\PassOptionsToPackage{unicode}{hyperref}
\PassOptionsToPackage{hyphens}{url}
\PassOptionsToPackage{dvipsnames,svgnames,x11names}{xcolor}
\documentclass[
  11pt,
]{article}
\usepackage{xcolor}
\usepackage[margin=1in]{geometry}
\usepackage{amsmath,amssymb}
\usepackage{iftex}
\ifPDFTeX
  \usepackage[T1]{fontenc}
  \usepackage[utf8]{inputenc}
  \usepackage{textcomp} % provide euro and other symbols
\else % if luatex or xetex
  \usepackage{unicode-math} % this also loads fontspec
  \defaultfontfeatures{Scale=MatchLowercase}
  \defaultfontfeatures[\rmfamily]{Ligatures=TeX,Scale=1}
\fi
\usepackage{lmodern}
\ifPDFTeX\else
\fi
\IfFileExists{upquote.sty}{\usepackage{upquote}}{}
\IfFileExists{microtype.sty}{% use microtype if available
  \usepackage[]{microtype}
  \UseMicrotypeSet[protrusion]{basicmath} % disable protrusion for tt fonts
}{}
\makeatletter
\@ifundefined{KOMAClassName}{% if non-KOMA class
  \IfFileExists{parskip.sty}{%
    \usepackage{parskip}
  }{% else
    \setlength{\parindent}{0pt}
    \setlength{\parskip}{6pt plus 2pt minus 1pt}}
}{% if KOMA class
  \KOMAoptions{parskip=half}}
\makeatother
\usepackage{longtable,booktabs,array}
\usepackage{calc} % for calculating minipage widths
\usepackage{etoolbox}
\makeatletter
\patchcmd\longtable{\par}{\if@noskipsec\mbox{}\fi\par}{}{}
\makeatother
\IfFileExists{footnotehyper.sty}{\usepackage{footnotehyper}}{\usepackage{footnote}}
\makesavenoteenv{longtable}
\usepackage{graphicx}
\makeatletter
\newsavebox\pandoc@box
\newcommand*\pandocbounded[1]{% scales image to fit in text height/width
  \sbox\pandoc@box{#1}%
  \Gscale@div\@tempa{\textheight}{\dimexpr\ht\pandoc@box+\dp\pandoc@box\relax}%
  \Gscale@div\@tempb{\linewidth}{\wd\pandoc@box}%
  \ifdim\@tempb\p@<\@tempa\p@\let\@tempa\@tempb\fi% select the smaller of both
  \ifdim\@tempa\p@<\p@\scalebox{\@tempa}{\usebox\pandoc@box}%
  \else\usebox{\pandoc@box}%
  \fi%
}
\def\fps@figure{htbp}
\makeatother
\providecommand{\tightlist}{%
  \setlength{\itemsep}{0pt}\setlength{\parskip}{0pt}}
\usepackage[sort&compress]{natbib}
\usepackage[LGR,T2A,T1]{fontenc}
\usepackage[greek,russian,english]{babel}
\usepackage{CJKutf8}
\DeclareUnicodeCharacter{2212}{\ensuremath{-}}
\DeclareUnicodeCharacter{2248}{\ensuremath{\approx}}
\DeclareUnicodeCharacter{2264}{\ensuremath{\leq}}
\DeclareUnicodeCharacter{221A}{\ensuremath{\surd}}
\DeclareUnicodeCharacter{03C3}{\ensuremath{\sigma}}
\DeclareUnicodeCharacter{03C4}{\ensuremath{\tau}}
\DeclareUnicodeCharacter{03B2}{\ensuremath{\beta}}
\DeclareUnicodeCharacter{207B}{\ensuremath{^{-}}}
\DeclareUnicodeCharacter{2074}{\ensuremath{^{4}}}
\DeclareUnicodeCharacter{2075}{\ensuremath{^{5}}}
\DeclareUnicodeCharacter{2076}{\ensuremath{^{6}}}
\DeclareUnicodeCharacter{2077}{\ensuremath{^{7}}}
\usepackage{float}
\usepackage[font=small,labelfont=bf]{caption}
\usepackage{newtxtext,newtxmath}
\setcitestyle{numbers,square,comma}
\AtBeginDocument{\setlength{\bibsep}{1pt plus 0.3ex}}
\usepackage{bookmark}
\IfFileExists{xurl.sty}{\usepackage{xurl}}{} % add URL line breaks if available
\hypersetup{
  pdftitle={Detection of dark compute via grid-side power monitoring},
  pdfauthor={Dan Parshall},
  colorlinks=true,
  linkcolor={Maroon},
  filecolor={Maroon},
  citecolor={Blue},
  urlcolor={Blue},
  pdfcreator={LaTeX via pandoc}}

\title{Detection of dark compute via grid-side power monitoring}
\author{Dan Parshall\\ \normalsize Canary Institute\\ \normalsize \texttt{dan@canaryinstitute.ai}}
\date{November 2026}

\begin{document}
\maketitle
\begin{abstract}
Distributed training is a problem for AI compute governance
\citep{rahman2026distributed}. A single large datacenter is easy to
identify via national technical means (NTM), e.g.~satellites. However a
distributed run can be spread across hundreds of small sites, which
might allow covert training by private but well-funded actors.

Here we show that it is possible to discover small training islands by
analyzing electrical load throughout the national grid. As a lower
threshold, we examine training islands at the smallest scale proposed
for governance (16 H100e chips), as well as larger islands (up to 2048
chips).

Because of its high average load (roughly 20x that of a typical
household) we find that even the smallest island can be ruled out of
operating on a residential meter, leaving only commercial and industrial
meters as possible sites. Even amongst those, we find that a naive
operator with a bare island is easily detected due to its high average
load and round-the-clock profile.

We also consider `cloaked' islands which are behind the meter of a
larger host. Such islands are not directly detectable, but by examining
the distribution of commercial and industrial meters and their load
profiles, we set bounds on how many sites would need to be physically
inspected under a compute-regulation regime such as the MIRI proposal
\citep{scher2025agreement}. We find that in many scenarios the
inspection load is on the same order as current OSHA inspections
(\textasciitilde30k / year).
\end{abstract}

\section{Background}\label{background}

Within the arms control framework, there are two primary things which
must be moniored for dangerous material or equipment:

\begin{itemize}
\tightlist
\item
  completeness: that all GPUs that could train a model are accounted for
\item
  correctness: that the identified chips are being used consistently
  with declaration
\end{itemize}

Throughout this work we address solely the question of
\emph{completeness}; we assume that any declared datacenters are subject
to some treaty-specified inspection regime to certify correctness.
Instead, we explore what limits we can place on undeclared illicit
training by performing monitoring and accounting of electricity
consumption for the national power grid. We set aside the issue of
off-grid power production; we also note that the grid-side accounting
examined here, when combined with the power-monitoring \emph{within}
declared facilities \citep{kimpson2026detecting}, could place tight
bounds on how much undeclared compute could be hidden within a declared
facility (e.g.~that no power is diverted to a hidden island on-site).
Both possibilities are avenues for future work.

We now provide a brief review of distributed training, as well as an
overview of the power grid.

\subsection{Distributed training}\label{distributed-training}

\subsubsection{DiLoCo and its relatives}\label{diloco-and-its-relatives}

Conventional pretraining takes place inside of a single datacenter.
After each batch of data, the improvements (gradients) to the model
weights are calculated, and all machines within the datacenter update
their weights. At each step, all machines are working from and updating
single version of the model.\footnote{Albeit there are multiple copies
  of the model spread across the various machines.} It is because
machines need to exchange updates so often that it's desirable to place
them all in a single building, to minimize transfer time.

Distributed training, as described by the ``DiLoCo'' model in
\citep{douillard2023diloco}, takes a different approach. Instead of
creating a single large datacenter all training a single replica of a
model, the developers split the training amongst ``islands'' of compute,
each of which works with a single copy of the model. Each island runs
several iterations of the process-calculate-update loop, and then
periodically those islands exchange their separate updates to the other
islands. This splitting and exchange process adds overhead to the
training run.

The precise amount of overhead is not known at frontier scale, so we use
the simulator of \citet{rahman2026distributed} as the basis for our
calculations.\footnote{Our changes (a search fix and the replica penalty
  on reducible loss only) are documented in the reproduction package
  (\url{https://doi.org/10.5281/zenodo.24485749}).} We find that the
memory and bandwidth terms only have a significant impact when there are
many islands, restricted memory, and little bandwidth. Otherwise
multiple islands makes only trivial changes to the overall compute
required.

\subsubsection{Number of islands versus memory versus
link}\label{number-of-islands-versus-memory-versus-link}

The DiLoCo method requires each island to maintain a full copy of the
model within the GPU's RAM. This sets a maximum size on the model
itself, which in turn means that in order to achieve the same level of
performance, the model must be overtrained, which increases the amount
of compute required. Precisely for this reason, the MIRI proposed treaty
\citep{scher2025agreement} specifies a maximum VRAM of 1,280 GB without
registration. The treaty also specifies a maximum network connection of
25 Gb/s, which does little to constrict an evader (except at the
smallest island sizes), but enables standard commercial use of servers.

\subsection{Treaty thresholds for an
island}\label{treaty-thresholds-for-an-island}

There are a number of proposals for a compute-governance regime. Here we
use the MIRI draft agreement \citep[Art. II def.
15--16]{scher2025agreement}, which sets the largest cluster that can go
unregistered as one that stays below \emph{either} of:

\begin{itemize}
\tightlist
\item
  16 H100-equivalents of compute (15,824 TFLOP/s at 989 TFLOP/s FP16 per
  H100)
\item
  1,280 GB of total accelerator memory (no restriction on CPU RAM)
\end{itemize}

We refer to this as a ``treaty-compliant'' island. Under the proposed
MIRI treaty, this is the largest size that would be unregistered, and so
we use this as our minimal island since it is least likely to generate
3rd-party reporting requirements. We ask under what conditions we could
identify treaty-compliant islands from grid data.

Strictly speaking the agreement also treats chips as one cluster if they
are co-located, networked to run workloads together, or joined by more
than:

\begin{itemize}
\tightlist
\item
  25 Gbit/s aggregate inter-node bandwidth
\end{itemize}

This restriction is of course unenforceable at the time of sale, so we
allow an evader to use a range of network speeds in order to set bounds
on the feasible space for distributed training.

We also consider ``compute limited'' islands, which consider whatever
VRAM is needed in order to hold a model in memory, thus setting training
time purely at the level of chip calculations, plus the overhead
introduced by training in a distributed fashion. Such a configuration
might be used by a state-level actor willing to go to great lengths to
evade, and so is willing to e.g.~desolder high-capacity VRAM from other
devices and resolder onto a chip board.

Finally we consider ``absolute floor'' islands, which lets us set the
limiting case, by calculating training time as if all of the islands
were colocated in a conventional datacenter. We find that except for
low-bandwidth scenarios, the compute-limited configurations are
typically quite close to this limit.

For concreteness we consider specifically H100 chips. The H100 chip
itself is rated at 700 W, but a server also powers CPUs, memory, fans
and power conversion: a DGX H100 server is rated at 10.2 kW for 8 GPUs,
about 1.28 kW per GPU. In practice jobs run at about 80\% of that rating
(the heaviest measured jobs average at most 76\%), so about 1 kW per
GPU, meaning \textasciitilde16 kW at the wall.\footnote{Node rating 10.2
  kW and measured peak 8.48 kW for one 8-GPU H100 node under training
  \citep{latif2024empirical}; heaviest jobs averaging at most 76\% of
  the 10.2 kW rating \citep{newkirk2025empirically}.} Cooling increases
that by another 30--110\% (we use 60\%), so an island of 16 chips would
be around 26 kW of total load at the meter. For comparison, a typical
American household averages about 1.2 kW, so an island uses about 20
households' worth of electricity. Newer chips use less power per H100e,
but overall provide little benefit to an evader (the most efficient
designs require liquid cooling, which requires specialized equipment,
rather than ordinary HVAC).\footnote{On the same assumptions as for the
  H100 (80\% of rating, 60\% for cooling): an air-cooled 8-GPU DGX B200,
  rated at about 14.3 kW \citep{nvidia2026dgxb200} and about 18 H100e,
  is a 16-H100e island by itself at about 18 kW at the meter. NVIDIA's
  liquid-cooled four-GPU GB200 NVL4 is rated at just over 6 kW per
  server \citep{techradar2024gb200nvl4}, about 0.66 kW per H100e,
  roughly half the H100's 1.0--1.3 kW; a 16-H100e island of two such
  boxes would be about 15 kW at the meter. Single-GPU desktops such as
  NVIDIA's DGX Station (one GB300, 1,600 W \citep{nvidia2026dgxstation})
  are sold individually, but use only 10--25\% less power per H100e than
  an 8-GPU Blackwell server.}

\subsection{Overview of power
distribution}\label{overview-of-power-distribution}

Electricity is generated at power plants, of which there are roughly
12,000--15,000 utility-scale plants in the USA
\citep{eia2025epatable41}.\footnote{Plants of 1 MW and up. The Electric
  Power Annual's Table 4.1 counts 15,254 plant-by-fuel entries for 2025,
  so there are fewer unique sites.} There are 3 sections to the national
grid: eastern, western, and Texas. Each grid can be considered a
directed acyclic graph (DAG) which has power plants as sources, and
meters as sinks. Energy flows along the edges (power lines) and is
conserved (minus transmission losses of a few percent) across switchgear
which splits the power into successively smaller amounts (and generally
lower voltage) at substations. While the lines leaving a power plant
transmit hundreds of megawatts (MW) or more and operate at hundreds of
kilovolts (kV) potential, the power is split as it gets closer to the
end users, and at ``distribution'' substations the exiting lines run
\textasciitilde15 kV potential and carry roughly 6 MW (so 2 MW per
phase).\footnote{We estimate roughly 60k distribution substations
  \citep{nas2009energyfuture} and some 200k feeder lines (204,813
  distribution circuits at 35 kV or below \citep{eia2025form861}).}

\begin{figure}
\centering
\includegraphics[width=1\linewidth,height=\textheight,keepaspectratio,alt={Nodes and edges of the power grid hierarchy. Energy is generated at the power plant and sent via high-voltage transmission lines (1), then split at a transmission substation (2), and at the distribution substations (3), which typically serve multiple neighborhoods. Feeder lines consist of 3 phase wires (4), and are further split closer to the final consumer (5a/b). A 16-chip island (about 26 kW at the meter) is half or more of a home's service transformer, about 1-5\% of one phase of a typical feeder depending on time of day, about 10⁻³ of a 115 kV line, and under 10⁻⁴ only at 345 kV and above. Photo credits in Image credits.}]{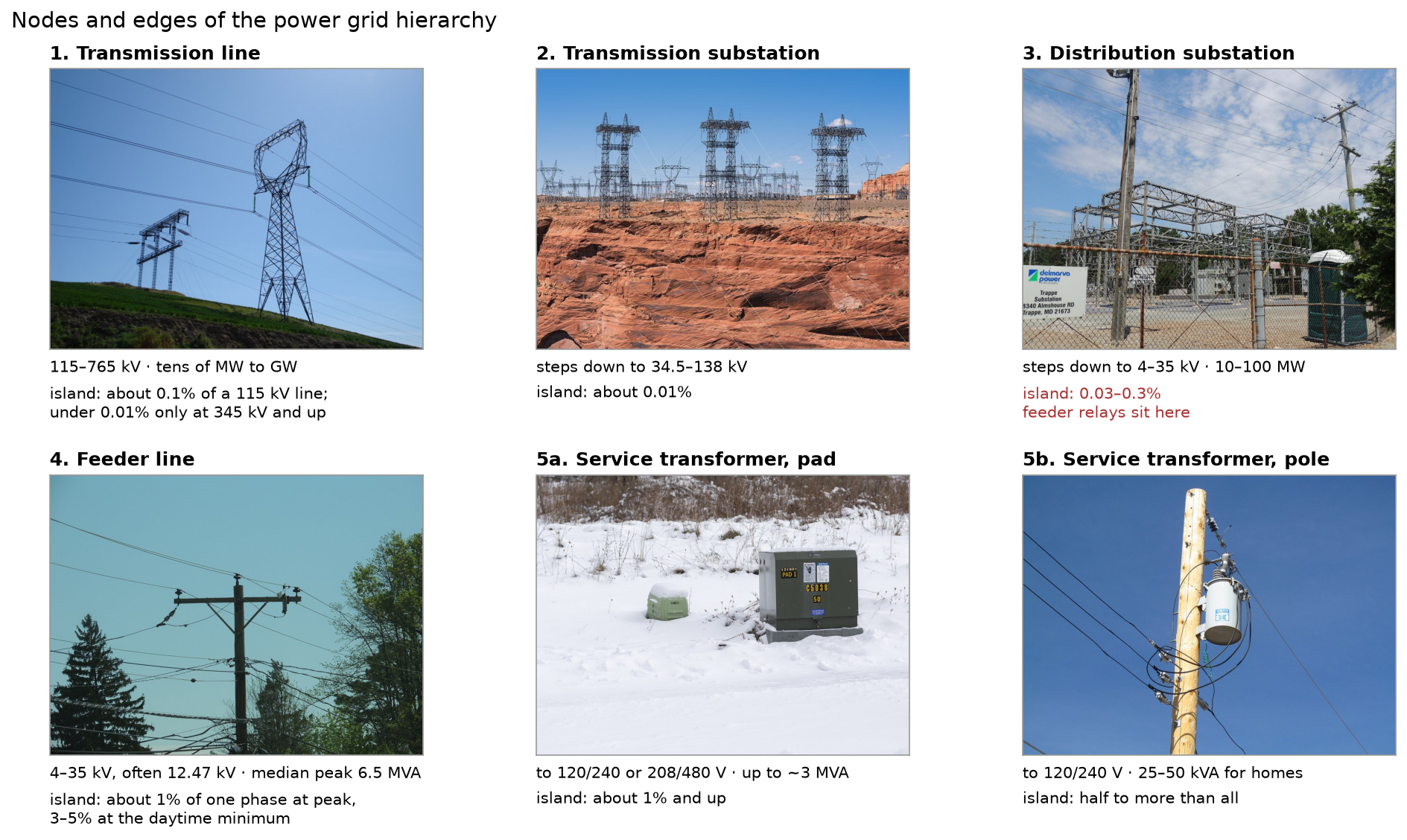}
\caption{Nodes and edges of the power grid hierarchy. Energy is
generated at the power plant and sent via high-voltage transmission
lines (1), then split at a transmission substation (2), and at the
distribution substations (3), which typically serve multiple
neighborhoods. Feeder lines consist of 3 phase wires (4), and are
further split closer to the final consumer (5a/b). A 16-chip island
(about 26 kW at the meter) is half or more of a home's service
transformer, about 1-5\% of one phase of a typical feeder depending on
time of day, about 10⁻³ of a 115 kV line, and under 10⁻⁴ only at 345 kV
and above. Photo credits in \hyperref[image-credits]{Image credits}.}
\end{figure}

However a single feeder is actually composed of 3 separate wires, each
on a different phase\footnote{The grid carries three phases 120 degrees
  apart because they deliver constant power and create a rotating field
  that starts motors without extra windings.} carrying one-third of the
total load, i.e.~around 2 MW. A single island would thus draw about 1\%
of a phase's load at peak, and about 2\% of its average load. The
absolute accuracy varies based upon meter type. For billing meters at
the junction of regional grids (where different power plants exchange
energy) the meters can be as good as 0.1\% (other common classes are
0.2\% and 0.5\%). The meters for protection relays at feeder heads are
less sensitive, and typically have a fixed absolute tolerance in normal
operation
\citep{abb2009ref615, ge2018urspecs, energyqueensland2023stnw3114}.

The load for a particular site can obviously be read directly from that
site's meter. What is less obvious is that signatures of training (such
as a step or drop of \textasciitilde26 kW when an island switches on or
pauses) can be detected further upstream, at the level of the feeder
lines leaving a distribution substation. Feeders generally have a
monitor on the protection relay, which is designed to cut the circuit in
the event of a fault or overload. These meters normally have an accuracy
that's around half the load of a 16-chip island (we estimate 6--18 kW in
typical usage\footnote{Relay datasheets give a fixed tolerance in normal
  operation: ABB's REF615 measures current to ±0.5\% of reading or
  ±0.2\% of rated current, and GE's Multilin UR to ±0.25\% of reading or
  ±0.1\% of rated \citep{abb2009ref615, ge2018urspecs}; Energy
  Queensland's acceptance criteria for protection devices are ±1\% for
  current and ±5\% for power \citep{energyqueensland2023stnw3114}.
  ``Rated'' is the full current of the current transformer (CT) that
  feeds the relay, not the present load. ABB's worked feeder example
  assumes a 400:5 CT for a 13.2 kV feeder with a maximum load of 330 A
  \citep[slides 93--95]{erwin2013feeder}; there ±0.2\% is ±0.8 A, or
  about ±6 kW on one phase (7.6 kV phase to neutral). A CT sized to a
  1,200 A breaker, one of the standard 15 kV continuous ratings
  \citep[slide 44]{erwin2013feeder}, would give about ±18 kW. GE's
  tolerance halves both.}). Meters can have a \emph{dynamic} precision
that is considerably better than the absolute accuracy. For instance,
using data from the Caltech campus, we found that two class 0.5 meters
on either side of a transformer \citep{xie2025socal} agree to within 0.2
kVA, roughly 0.02\% of the load
(\hyperref[app-dynamic-precision]{Appendix A}). That is 25× better than
their rated accuracy\footnote{The Caltech measurements are with class
  0.5 revenue meters; we could not find a public dataset from a
  protection-relay meter, so we have not yet measured the dynamic
  precision at a utility feeder head.}, and the step from a 26 kW island
turning on or off is around 100× larger.

Detection ability is then ultimately limited by the background signal
produced by other loads. This varies throughout the day, e.g.~at a
\textasciitilde1 MW substation meter, ordinary load makes daytime steps
of 30--53 kW, larger than the island's 26 kW, but at night the island's
step is 7--8× the noise level, so a single turn-on step can often be
picked out at the feeder head. This background can be largely removed
when monitors are installed at multiple levels of the hierarchy. Since
the sum of power in downstream monitors must equal the upstream monitor
(an \emph{energy balance}), load on an unmetered or illicit branch can
be estimated from the residual difference. For example on the Caltech
campus data the noise level is 3.6--5.3 kW, and a 26 kW step stands out
at 5--7× that even during the day. These multiples are against the
typical background; Appendices A.4 and A.5 also compare the step with
the largest ordinary steps, and \hyperref[app-high-frequency]{Appendix
B} (B.3) gives turn-on detection rates on real recordings.

\section{Results}\label{results}

We find that a `bare' island, operating directly on its own meter, can
almost always be easily identified from feeder-level data, solely by
average load and load factor. We also consider `cloaked' islands which
are behind the meter of a host whose larger power consumption
potentially masks the increased load.

We restrict the discussion to a simple analysis of average load and load
factor (average load divided by peak demand). This data can be obtained
from power monitors installed at feeder heads (roughly 200k in the USA,
each of which has 3 phases) or service transformers. In theory they
could also be collected directly from the customer meter; around 85\% of
the commercial meters in the USA are smart meters that collect data on
15-minute or hourly frequencies, which is sufficient for most of our
analysis. In \hyperref[app-high-frequency]{Appendix B} we explore other
analyses that are possible with meters capable of high-frequency
sampling (broadly, our results are consistent with
\citet{kimpson2026detecting}, who find that training can be told apart
from inference by the cadence of its power draw at a facility meter; we
can do so further upstream).

Note that gathering data directly from the meter raises legal and
privacy concerns in some jurisdictions. Our focus in this paper is
primarily about what is possible on a technical basis, and we leave
concerns about how this impacts data collection out of scope for this
work. However we note that certainly in the USA, the legal scrutiny for
search-and-seizure concerns (i.e.~4th Amendment) is explicitly lower for
commercial customers than residential ones, and e.g.~administrative
warrants are possible.

Residential islands are very unlikely: in addition to the extreme power
requirements relative to normal usage, they would also have extreme
cooling requirements. A hidden 16-chip island would probably demand a
second air-conditioning unit, and would need to run year round.

\subsection{Bare islands}\label{bare-islands}

In the USA there are approximately 20 million commercial meters, and
roughly another 1 million industrial meters.\footnote{Definitions vary
  by utility. In Texas, one retailer (TXU Energy Retail) reports 315,197
  industrial customers averaging 17 kW, up from 32,595 in 2018, while
  the Texas wires companies report 24,776 industrial accounts in all
  \citep{eia2025form861}. Without it the national industrial count is
  about 0.8 million.} There seems to be no standardized definition, so
we lump commercial and industrial customers together.

There is not a perfect one-to-one relationship between buildings and
meters. At some commercial properties each customer within a building
will have their own meter; likewise some large campuses have a single
meter but multiple buildings. Nationally there are about 1.06 meters per
commercial account \citep{eia2025form861}; at the loads that matter here
we estimate about 1.17 accounts per site (range 1.0--2.0), and equate
accounts and sites unless stated.\footnote{There can also be sites such
  as pumping stations which have a meter even though they aren't
  buildings; our analysis is restricted to buildings.}

\begin{figure}
\centering
\includegraphics[width=1\linewidth,height=\textheight,keepaspectratio,alt={US commercial buildings by average electric load. The dashed line at 25 kW is about one 16-chip island: about 943,000 buildings, 16\% of the stock, average more than that. The dotted line at 128 kW marks the smallest building that could cloak such an island under a 20\% energy screen. Data: {[}@eia2018cbecs{]}.}]{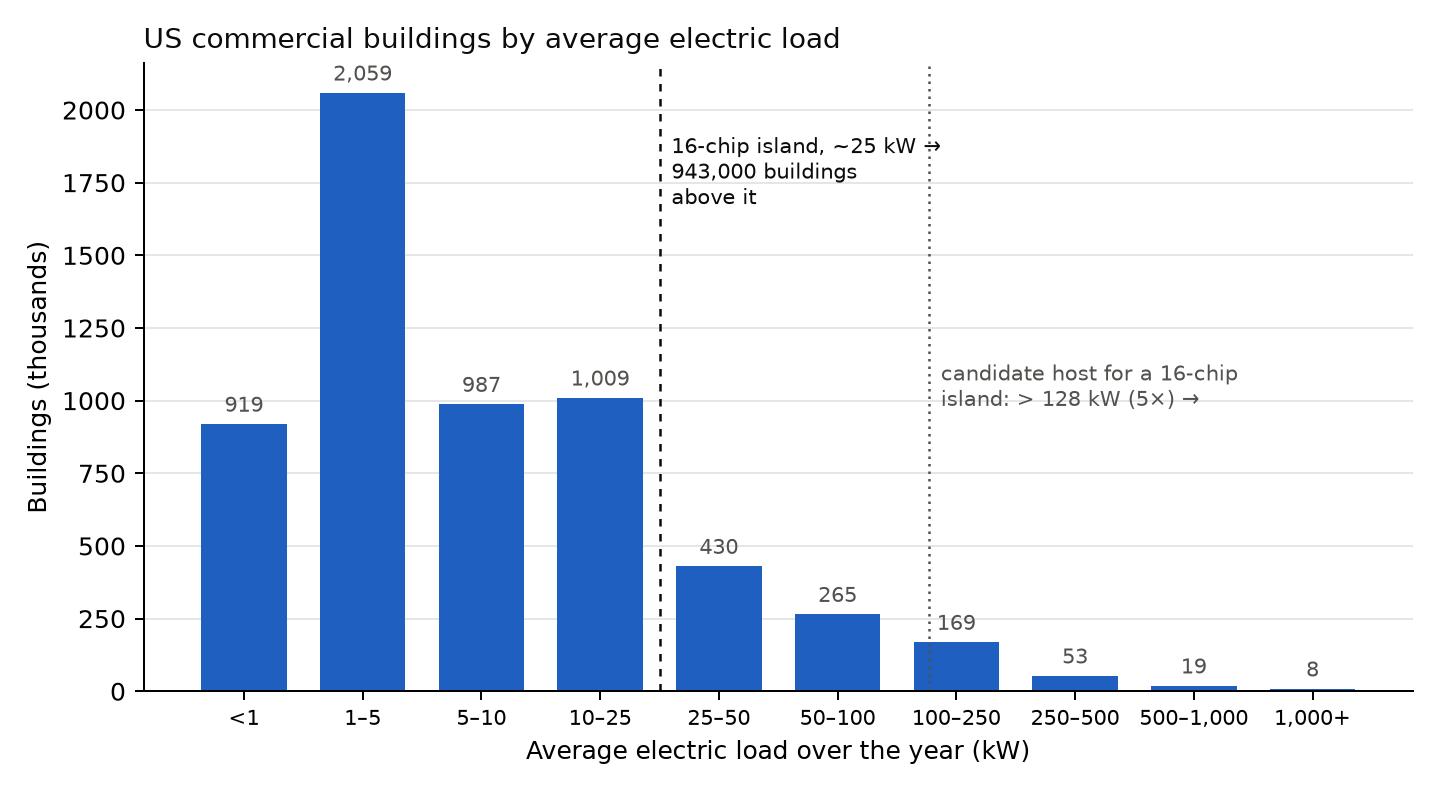}
\caption{US commercial buildings by average electric load. The dashed
line at 25 kW is about one 16-chip island: about 943,000 buildings, 16\%
of the stock, average more than that. The dotted line at 128 kW marks
the smallest building that could cloak such an island under a 20\%
energy screen. Data: \citep{eia2018cbecs}.}
\end{figure}

Since most businesses are open in the day and closed at night, the
typical load factor is \textasciitilde0.5, whereas an island runs
around-the-clock. An evader who wishes to disguise their presence by
only running half time has reduced their effective number of chips by a
factor of two. That in turn means they either compute half as much, or
they obtain extra chips and need twice as many sites (which increases
the probability of detection even when cloaked, see below).

Using data from CBECS, we estimate that only some \textbf{943,000 ±
39,000} of the USA's 5.9 million commercial buildings have an average
load above 25 kW.\footnote{Average load is annual electricity ÷ 8,760 h
  \citep{eia2018cbecs}.} Amongst those with average load \textgreater{}
50 kW, only 0.8\% of a measured sample\footnote{6 of 796 buildings in
  \hyperref[src-bdgp2]{BDGP2} \citep{miller2020bdg2}.} and effectively
none in the nationally representative \hyperref[src-comstock]{ComStock}
\citep{nrel2025comstock} model\footnote{About 7.5 of 492,000, or
  0.002\%.} have a median monthly load factor \textgreater{} 0.9; in a
treaty regime such a profile could easily be flagged for followup
in-person inspection (\hyperref[app-notes]{Appendix D}).

\begin{figure}
\centering
\includegraphics[width=1\linewidth,height=\textheight,keepaspectratio,alt={Median monthly load factor of buildings averaging 50 kW or more. Convenience sample of measured meters (blue, BDGP2 {[}@miller2020bdg2{]}) and representative sample (orange, ComStock {[}@nrel2025comstock{]}). Most buildings are unoccupied during the day, so their mean-divided-by-peak is relatively low, whereas AI islands run round-the-clock.}]{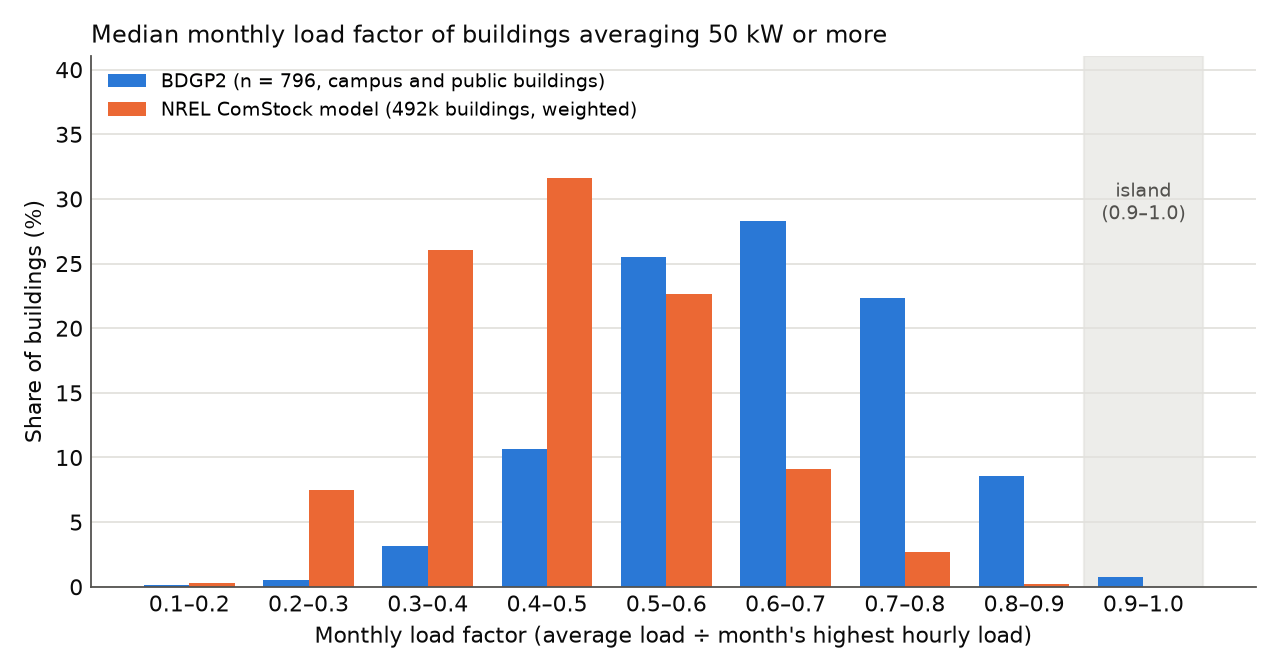}
\caption{Median monthly load factor of buildings averaging 50 kW or
more. Convenience sample of measured meters (blue,
\hyperref[src-bdgp2]{BDGP2} \citep{miller2020bdg2}) and representative
sample (orange, \hyperref[src-comstock]{ComStock}
\citep{nrel2025comstock}). Most buildings are unoccupied during the day,
so their mean-divided-by-peak is relatively low, whereas AI islands run
round-the-clock.}
\end{figure}

\subsubsection{Real-world discovery}\label{real-world-discovery}

Utilities already find hidden loads of smaller size by analyzing
customer meters. Sacramento's municipal utility identified those
customers with an average load greater than 3.8 kW
\citep{theregister2025smud}, and submitted lists to police for follow-up
inspection. Chelan County PUD in Washington State identified
high-consumption home crypto mines in 2018 by watching for sustained
spikes in use \citep{nwnews2018chelan, kpq2018chelan}, and disconnected
service or fined customers exceeding rated service capacity. Malaysia's
TNB detected 5,133 thefts through smart meters from 2021 to 2025
\citep{thestar2026smartmeters}, and in Russia a flat using more than
30--40\% of its apartment building's electricity brings an inspector
\citep{izvestia2022snikkars}. A 16-chip island uses about seven times
the Sacramento flag. \hyperref[app-notes]{Appendix D.2} lists further
cases, and D.3 describes how utilities find and trace stolen power.

Hidden load has also been found directly from the grid side at the
feeder heads (rather than solely customer meters). In British Columbia,
cannabis grow-ops were stealing about 850 GWh a year by the early 2010s.
BC Hydro used a combination of new and existing devices to analyze data
from feeders and meters, and perform energy-balancing across
\textasciitilde9,000 devices to estimate the location of illegal loads.
Based on police seizure data, we estimate each site was some 20-40 kW
(about the same as a 16-chip island), and at its peak the theft amounted
to some 4,000 sites. Ultimately BC Hydro was able to reduce annual theft
by over 80\%.

\subsection{Cloaked islands}\label{cloaked-islands}

An obvious method for an evader who wishes to run training full-time is
to cloak themselves behind the meter of a larger host. We consider here
the scenario in which an evader must find a willing host whose average
load is at least 5x the evader's own (under the logic that any smaller
level would be noticeable). We evaluate how many potential hosts exist
in the USA at a variety of island sizes, and provide estimates of how
many physical inspections would be needed to find at least one island of
the covert network with 95\% probability.

\begin{figure}
\centering
\includegraphics[width=1\linewidth,height=\textheight,keepaspectratio,alt={US industrial sites with average load of 150 MW or more. These sites could potentially cloak a 20k-chip island of \textasciitilde30 MW. Blue is the portion bought from the grid, grey is the portion generated on site; whiskers show estimated range. Data centers and crypto mines are excluded as declared compute within a treaty regime.}]{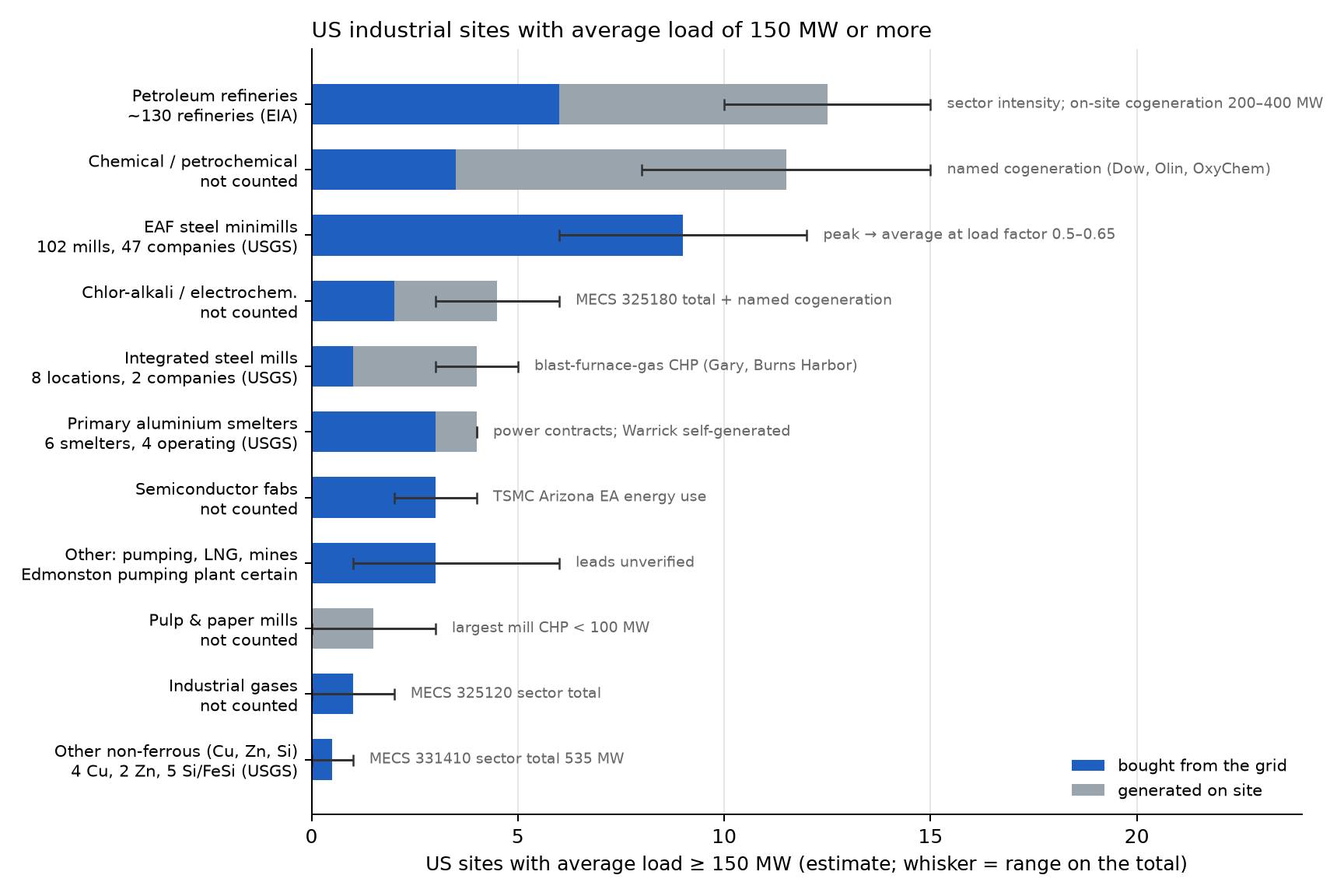}
\caption{US industrial sites with average load of 150 MW or more. These
sites could potentially cloak a 20k-chip island of \textasciitilde30 MW.
Blue is the portion bought from the grid, grey is the portion generated
on site; whiskers show estimated range. Data centers and crypto mines
are excluded as declared compute within a treaty regime.}
\end{figure}

At one extreme we consider what island size would be required to
complete pretraining of a model with 10\textsuperscript{27} FLOP within
the two-year window that we target for discovery. \footnote{The longest
  training run worth starting is T = 1/(g\textsubscript{H} +
  g\textsubscript{S} + g\textsubscript{I}), where g\textsubscript{H},
  g\textsubscript{S} and g\textsubscript{I} are the yearly growth rates
  of hardware price-performance, algorithmic efficiency and spending
  \citep{sevilla2022longest}; a longer run is overtaken by a later one
  on better hardware and methods. The same rates set how fast completed
  covert compute loses value, which is why we target discovery within
  this window. Under a MIRI-style regime Rahman takes 6\%, 50\% and 3\%
  a year \citep[App. A]{rahman2026distributed}: ln 1.06 + ln 1.5 + ln
  1.03 ≈ 0.49 a year, so T ≈ 740 days, about two years. At today's rates
  (1.37×, 3× and 3.5× a year) T is about 4.5 months.}

That works out to roughly 20,000 H100s, even under generous terms that
consider constant uptime.\footnote{10\textsuperscript{27} FLOP ÷ (1,979
  TFLOPS × 40\% utilisation × 740 days) ≈ 19,800 chips. NVIDIA lists the
  H100 SXM at 3,958 TFLOPS FP8 with sparsity \citep{nvidia2026h100}; the
  dense rate used here is half that.} Such a facility would consume an
average of 32 MW (the equivalent of 26,000 homes) around the clock;
facilities above about 10 MW are readily detectable via satellite
\citep{koopmanschap2026catch}. The physical footprint would require at
least 4000 square meters to hold the accelerators, and another 1000+
square meters of cooling. On the order of fifty industrial sites in the
USA draw power at that level, about half of them from their own
generation, and all in heavily documented industries: refineries,
chemical complexes, steel mills, metal smelters, chlor-alkali plants and
semiconductor fabs.\footnote{A site that generates its own power could
  absorb 32 MW by cutting exports or production by 20\%, but neither is
  quiet: exports are metered at the interconnection and settled by the
  grid operator, generation is reported monthly \citep{eia923}, and
  output appears in production statistics and emissions reports.
  Self-generation moves the evidence off the retail meter and into the
  plant's energy balance.}

\begin{figure}
\centering
\includegraphics[width=1\linewidth,height=\textheight,keepaspectratio,alt={Covert sites required versus island size. Sites a covert run needs to finish within two years by island size, against the candidate-host ceiling (every site whose average load exceeds 5× the island). Lines show final-compute training target for compute-bound evader; upper band is treaty-compliant units. Dots calculated for 1 Gbps per island; triangle is the treaty-compliant island at 25 Gbps.}]{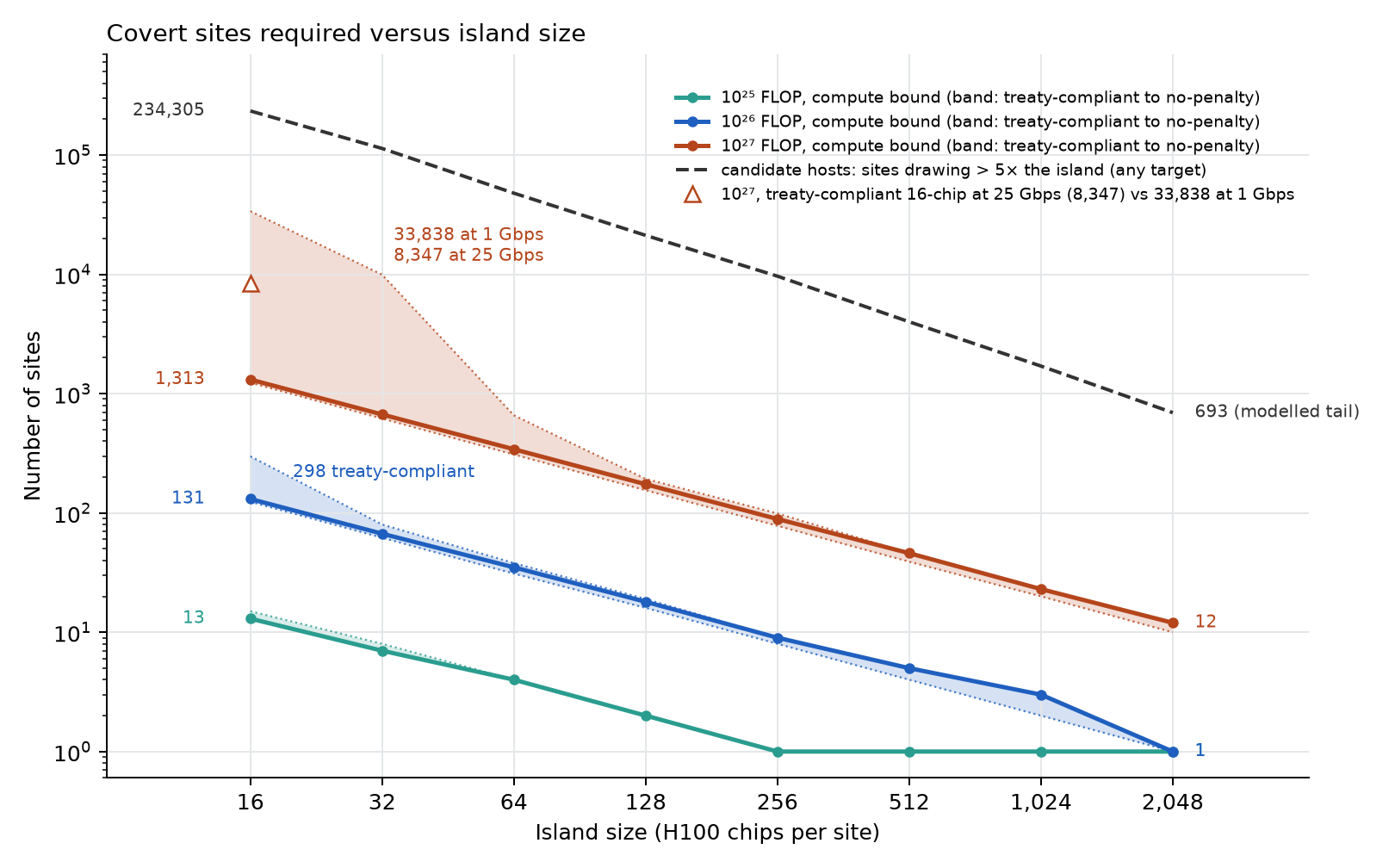}
\caption{Covert sites required versus island size. Sites a covert run
needs to finish within two years by island size, against the
candidate-host ceiling (every site whose average load exceeds 5× the
island). Lines show final-compute training target for compute-bound
evader; upper band is treaty-compliant units. Dots calculated for 1 Gbps
per island; triangle is the treaty-compliant island at 25 Gbps.}
\end{figure}

An evader might try to split the compute into smaller islands, and so we
consider everything from 16- to 2048-chip islands. Even a 2k-chip island
consumes 3.3 MW of power, and we estimate there are less than a thousand
sites where this much power would account for \textless{} 20\% of the
host. The smaller each individual island, the more sites are required in
order to achieve the same level of compute. To complete a training run
of 10\textsuperscript{26} FLOP using 16-chip islands, an evader would
need to have \textasciitilde130 hosts, each of which expands the size of
the conspiracy. Ultimately the evader's gains by moving to smaller
islands are limited: although there are more potential hosts, more hosts
are required. While there are roughly 235,000 sites that could
potentially host a small 16-chip island, the result is quite favorable
for the inspector: at 130 islands amongst 235k potential hosts, less
than 3,000 randomly-chosen inspections/year will find at least one
island with 95\% probability.\footnote{Assuming every visit finds a
  present island (d = 1); \hyperref[app-detection-probability]{Appendix
  C} gives d = 0.75 and 0.5.} For comparison, OSHA conducts
\textasciitilde30k inspections annually.

\begin{figure}
\centering
\includegraphics[width=1\linewidth,height=\textheight,keepaspectratio,alt={Random inspections required to catch a covert training run. Number of annual inspections required to discover at least one island within 2 years (with 95\% chance), when using 1 Gbps connection. Colored dashed lines show inspector's worst-case scenarios (which allows small islands hidden in large hosts). Black dashed line shows number of annual OSHA inspections.}]{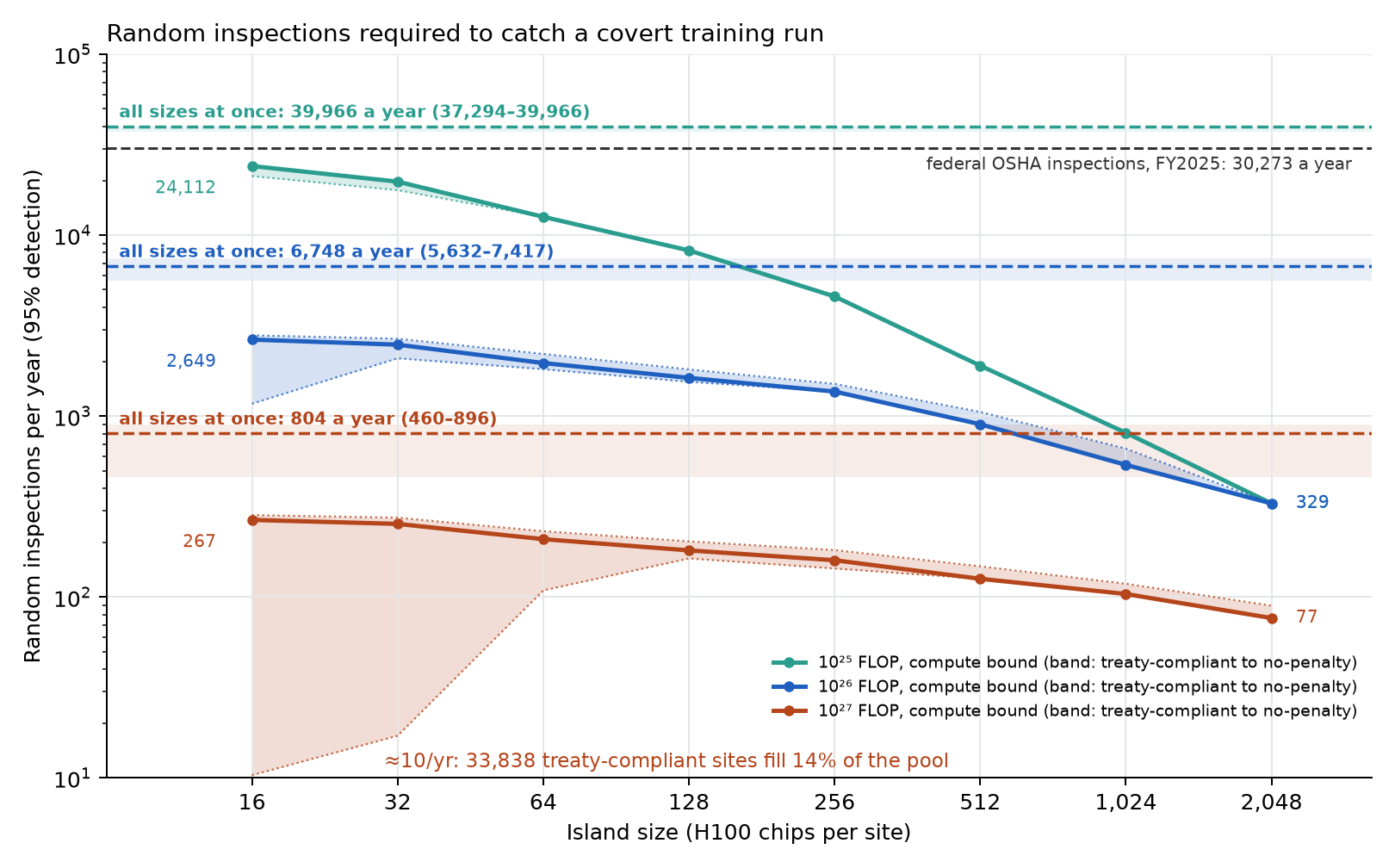}
\caption{Random inspections required to catch a covert training run.
Number of annual inspections required to discover at least one island
within 2 years (with 95\% chance), when using 1 Gbps connection. Colored
dashed lines show inspector's worst-case scenarios (which allows small
islands hidden in large hosts). Black dashed line shows number of annual
OSHA inspections.}
\end{figure}

The detection goal is largely a political decision, and should be based
upon what the \emph{evader} perceives as the risk-reward tradeoff. A
rational actor evades only when the probability of being caught times
the penalty is less than the probability of escaping detection times the
gain \citep[Theorem 9.5]{avenhaus1996compliance}, i.e.: \[
P(\text{caught}) \times \text{penalty} \;<\; \big(1 - P(\text{caught})\big) \times \text{gain}
\quad\Longleftrightarrow\quad
P(\text{caught}) \;<\; \frac{\text{gain}}{\text{gain} + \text{penalty}}
\]

The number of inspections required is not very sensitive to the goal;
relaxing from 95\% to 90\% reduces the inspections by
\textasciitilde1.25x, and tightening from 95\% to 99\% increases the
number of inspections by \textasciitilde1.4x (see Table 2). However an
evader who believes that a frontier model brings great benefits is much
more likely to attempt evasion, and so the inspector should increase the
detection target. The inspector can normally stay under 30k
inspections/year (the current number OSHA does annually), although they
may miss a run of size 10\textsuperscript{25} FLOP.

\begin{longtable}[]{@{}
  >{\raggedright\arraybackslash}p{(\linewidth - 14\tabcolsep) * \real{0.0667}}
  >{\raggedright\arraybackslash}p{(\linewidth - 14\tabcolsep) * \real{0.1667}}
  >{\raggedleft\arraybackslash}p{(\linewidth - 14\tabcolsep) * \real{0.1556}}
  >{\raggedleft\arraybackslash}p{(\linewidth - 14\tabcolsep) * \real{0.1000}}
  >{\raggedleft\arraybackslash}p{(\linewidth - 14\tabcolsep) * \real{0.1556}}
  >{\raggedleft\arraybackslash}p{(\linewidth - 14\tabcolsep) * \real{0.1000}}
  >{\raggedleft\arraybackslash}p{(\linewidth - 14\tabcolsep) * \real{0.1556}}
  >{\raggedleft\arraybackslash}p{(\linewidth - 14\tabcolsep) * \real{0.1000}}@{}}
\caption{Number of annual inspections required to find at least one
covert island within two years, depending on both the targeted final
compute, as well as the memory allotted. Here ``treaty compliant'' means
1,280 GB of VRAM per 16 H100e; ``no-penalty'' is with both unrestricted
memory and bandwidth. All values at 95\% detection, 1 Gbps per island,
and assuming 100\% chance that an inspection finds a hidden
island.}\tabularnewline
\toprule\noalign{}
\begin{minipage}[b]{\linewidth}\raggedright
FLOP
\end{minipage} & \begin{minipage}[b]{\linewidth}\raggedright
Evader scenario
\end{minipage} & \begin{minipage}[b]{\linewidth}\raggedleft
16-chip sites
\end{minipage} & \begin{minipage}[b]{\linewidth}\raggedleft
insp./yr
\end{minipage} & \begin{minipage}[b]{\linewidth}\raggedleft
128-chip sites
\end{minipage} & \begin{minipage}[b]{\linewidth}\raggedleft
insp./yr
\end{minipage} & \begin{minipage}[b]{\linewidth}\raggedleft
2,048-chip sites
\end{minipage} & \begin{minipage}[b]{\linewidth}\raggedleft
insp./yr
\end{minipage} \\
\midrule\noalign{}
\endfirsthead
\toprule\noalign{}
\begin{minipage}[b]{\linewidth}\raggedright
FLOP
\end{minipage} & \begin{minipage}[b]{\linewidth}\raggedright
Evader scenario
\end{minipage} & \begin{minipage}[b]{\linewidth}\raggedleft
16-chip sites
\end{minipage} & \begin{minipage}[b]{\linewidth}\raggedleft
insp./yr
\end{minipage} & \begin{minipage}[b]{\linewidth}\raggedleft
128-chip sites
\end{minipage} & \begin{minipage}[b]{\linewidth}\raggedleft
insp./yr
\end{minipage} & \begin{minipage}[b]{\linewidth}\raggedleft
2,048-chip sites
\end{minipage} & \begin{minipage}[b]{\linewidth}\raggedleft
insp./yr
\end{minipage} \\
\midrule\noalign{}
\endhead
\bottomrule\noalign{}
\endlastfoot
10²⁵ & no-penalty & 13 & 24,112 & 2 & 8,239 & 1 & 329 \\
10²⁵ & compute bound & 13 & 24,112 & 2 & 8,239 & 1 & 329 \\
10²⁵ & treaty-compliant & 15 & 21,209 & 2 & 8,239 & 1 & 329 \\
10²⁶ & no-penalty & 124 & 2,796 & 16 & 1,812 & 1 & 329 \\
10²⁶ & compute bound & 131 & 2,649 & 18 & 1,627 & 1 & 329 \\
10²⁶ & treaty-compliant & 298 & 1,172 & 19 & 1,548 & 1 & 329 \\
10²⁷ & no-penalty & 1,235 & 284 & 155 & 203 & 10 & 90 \\
10²⁷ & compute bound & 1,313 & 267 & 174 & 181 & 12 & 77 \\
10²⁷ & treaty-compliant & 33,838 & 10 & 193 & 163 & 12 & 77 \\
\end{longtable}

In addition, servers require maintenance: roughly one hands-on
intervention from a hardware technician per 16-chip island/year. So for
the 130 islands needed for a 10\textsuperscript{26} run, the evader's
team would at a minimum consist of one or two people whose sole time is
dedicated to keeping the equipment running. A lazy evader who allows the
island to stay out of commission risks giving away their usage when the
island goes offline and subsequently returns a few days later. A more
proactive one needs a larger conspiracy.

\begin{longtable}[]{@{}
  >{\raggedright\arraybackslash}p{(\linewidth - 6\tabcolsep) * \real{0.0909}}
  >{\raggedright\arraybackslash}p{(\linewidth - 6\tabcolsep) * \real{0.3030}}
  >{\raggedright\arraybackslash}p{(\linewidth - 6\tabcolsep) * \real{0.3030}}
  >{\raggedright\arraybackslash}p{(\linewidth - 6\tabcolsep) * \real{0.3030}}@{}}
\caption{Number of annual inspections required when allowing small
islands to hide within much larger hosts. Central number is
compute-bound evader, with the treaty-compliant to no-penalty range in
parentheses.}\tabularnewline
\toprule\noalign{}
\begin{minipage}[b]{\linewidth}\raggedright
FLOP
\end{minipage} & \begin{minipage}[b]{\linewidth}\raggedright
99\% detection
\end{minipage} & \begin{minipage}[b]{\linewidth}\raggedright
95\% detection
\end{minipage} & \begin{minipage}[b]{\linewidth}\raggedright
90\% detection
\end{minipage} \\
\midrule\noalign{}
\endfirsthead
\toprule\noalign{}
\begin{minipage}[b]{\linewidth}\raggedright
FLOP
\end{minipage} & \begin{minipage}[b]{\linewidth}\raggedright
99\% detection
\end{minipage} & \begin{minipage}[b]{\linewidth}\raggedright
95\% detection
\end{minipage} & \begin{minipage}[b]{\linewidth}\raggedright
90\% detection
\end{minipage} \\
\midrule\noalign{}
\endhead
\bottomrule\noalign{}
\endlastfoot
10²⁵ & 52,943 (49,437--52,943) & 39,966 (37,294--39,966) & 33,142
(30,942--33,142) \\
10²⁶ & 9,780 (8,094--10,661) & 6,748 (5,632--7,417) & 5,350
(4,486--5,905) \\
10²⁷ & 1,220 (693--1,357) & 804 (460--896) & 621 (357--693) \\
\end{longtable}

The geographic dispersion matters as well: when the sites are physically
nearby, a single tech might be able to cover the 130 islands needed for
a 10\textsuperscript{26} run, but at the cost of putting thousands of
households worth of load on a single subsection of the grid; spreading
them out reduces the signal in the power grid, but increases the number
of conspirators. The hardware techs comprise a bare minimum; in reality
there would be electricians, internet providers, facilities workers who
all might be aware of an unusual installation.

\section{Conclusion}\label{conclusion}

We have explored the use of power-grid monitoring to detect dark
compute. In the domestic case where measurements from the grid and
customer meters are trusted, it can often detect illicit compute for
islands of 25 kW and up. We've estimated the number of locations and
inspection regime required in this scenario, and find that it is
tractable if not perfect. This would obviously become more difficult
with more efficient chips, but the broad result holds.

AI statement: the main body of the paper is artisanal handcrafted prose;
the appendices and most of the footnotes are all mass-produced
claudeslop (but have many technical details that some readers will
nevertheless find useful).

\section{Figure sources and credits}\label{image-credits}

Figure 1 photographs are from Wikimedia Commons, downscaled and cropped:
transmission line, Stefan Andrej Shambora (CC BY 2.0); transmission
substation, Christian David (CC BY-SA 4.0); distribution substation and
feeder, Tonyglen14 (CC BY 2.0); pad-mounted transformer, Tessa Bury (CC
BY 4.0); pole-mounted transformer, Kkiefuik (CC BY-SA 4.0). The combined
figure is CC BY-SA 4.0. Feeder sizes in Figure 1 are from Portland
General Electric's published feeder peak loads \citep{pge2026feeders}
(559 feeders, median peak 6.5 MVA). Figure 2 is drawn from EIA CBECS
2018 public-use microdata \citep{eia2018cbecs}; Figure 3 from the
Building Data Genome Project 2 (CC BY-SA 4.0) \citep{miller2020bdg2} and
NREL ComStock 2025 release 3 (CC BY 4.0) \citep{nrel2025comstock};
Figures 4--6 from this project's own results. Data includes information
from the ComStock dataset developed by the National Laboratory of the
Rockies (NLR) with funding from the U.S. Department of Energy (DOE).

\renewcommand\refname{References}
\bibliography{refs.bib}

\clearpage

\section{Appendices}\label{appendices}

\subsection*{Appendix A. Dynamic precision of installed grid
instruments}\label{app-dynamic-precision}
\addcontentsline{toc}{subsection}{Appendix A. Dynamic precision of
installed grid instruments}

This appendix asks whether the instruments already installed on the grid
are good enough to see a 16-chip island (about 26 kW) switch on or off.
A.1 separates the two properties of a meter that matter, accuracy and
precision. A.2 measures precision on two pairs of installed instruments.
A.3 tests a rule of thumb that would let us estimate the precision of
instruments we could not measure. A.4 compares the instruments' noise
with the natural ups and downs of the load and finds that the load, not
the meter, sets the limit. A.5 tries energy balance, which removes most
of that load.

\subsubsection*{A.1 Static accuracy versus dynamic
precision}\label{a.1-static-accuracy-versus-dynamic-precision}
\addcontentsline{toc}{subsubsection}{A.1 Static accuracy versus dynamic
precision}

A revenue meter's accuracy class says how far its reading may stray from
a reference standard: a class 0.5 meter must stay within ±0.5\% of
reading across the test conditions the standard sets. ANSI C12.20 sets
classes \citep{nema2015c1220} of 0.5\%, 0.2\% and, since 2015, 0.1\%,
and the revenue meters made for large commercial and industrial
customers come in the tighter two: for example the Alpha 4 C\&I meter at
class 0.2, and the SEL-735 and the ION9000 at class 0.1.\footnote{\citep{nema2015c1220, honeywell2020alpha4, sel2024sel735, schneider2025ion9000}.
  We found no figure for what share of installed meters falls in each
  class. The class covers the meter alone; the current and voltage
  transformers that feed it add their own error.}

Detecting an island asks something different of the meter. It need not
report the true load, only report the same load the same way from one
interval to the next, so that a 26 kW step shows up as a 26 kW change.
We call this dynamic precision: the scatter of an instrument's own
reading of a steady load over averaging windows of seconds to minutes.
We measure it as an Allan deviation σ\textsubscript{A}: average the
reading over successive windows of length τ, take the difference between
each window and the next, and σ\textsubscript{A} is the square root of
half the mean squared difference. This is the natural measure for a step
detector, because the change a detector sees when a load switches on is
exactly such a difference of adjacent window averages, and the
root-mean-square size of that difference for a steady load is √2
σ\textsubscript{A}.

Dynamic precision can be far better than the accuracy class. The class
bounds an error that varies slowly, if at all, over a run: a gain error,
an offset, and how the gain changes with load. A step detector is hurt
only by the fast scatter, which A.2 measures. A gain error of 1\% turns
a 26 kW step into a 25.7 kW step; it does not hide it. A meter can
therefore be precise and inaccurate at once, and for picking out loads
that switch on and off, precision is the property that counts (A.4 shows
that in practice even precision is not the limit).

\subsubsection*{A.2 Measured noise
floor}\label{a.2-measured-noise-floor}
\addcontentsline{toc}{subsubsection}{A.2 Measured noise floor}

We found no published measurement of the dynamic precision of grid
instruments (A.3), so we measured it ourselves. The method uses two
installed instruments that meter the same power flow. We fit one reading
to the other on one stretch of data, then apply that fit to a different,
held-out stretch and look at what is left over (the pair residual). The
true load cancels in the residual, so its Allan deviation bounds the two
instruments' independent noise taken together; one instrument alone has
about 1/√2 of it. Unmetered loads between the two instruments, losses
that the fit does not capture, and clock skew between the two also add
to the residual, so the bound is conservative: the instruments are at
least this good.

The first pair is two eGauge meters on either side of substation
transformer 21/17 on the Caltech campus: meter 21 on the 16.5 kV side
and meter 17 on the 2.4 kV side. The transformer carries about 1.2 MVA.
This is a net flow: the buildings on the 2.4 kV side draw about 1.7 MW,
and on-site generation supplies about 0.7 MW of it. The fit has three
terms: a constant (73 kVA), a gain, and a quadratic term for transformer
losses, which grow with the square of the current. It was fitted on the
week of 2025-03-10 and scored on two held-out weeks, 2025-03-17 and
2025-07-14; no time lag between the meters was needed.

The second pair is two PSL µPMUs (micro phasor measurement units,
research-grade instruments that record voltage and current waveforms
with precise timestamps) on the Lawrence Berkeley National Laboratory
(LBNL) 12.5 kV bus, which carries about 0.7 MVA. The pair was fitted on
days 0--5 of an 11.6-day record from October 2015 and scored on days
6--11. The two LBNL sensors turn out not to see the same current: one
reads a steady 52--59\% of the other, phase by phase, either because the
load splits between two parallel paths or because a transformer ratio
was applied wrongly (the data cannot tell which). The share holds steady
over the whole record, so the fit absorbs it and the residual is still a
valid bound. We convert the residual to the larger sensor's side by
dividing by that share, which is the conservative direction.

\begin{longtable}[]{@{}
  >{\raggedright\arraybackslash}p{(\linewidth - 8\tabcolsep) * \real{0.3506}}
  >{\raggedright\arraybackslash}p{(\linewidth - 8\tabcolsep) * \real{0.1688}}
  >{\raggedleft\arraybackslash}p{(\linewidth - 8\tabcolsep) * \real{0.1818}}
  >{\raggedleft\arraybackslash}p{(\linewidth - 8\tabcolsep) * \real{0.1688}}
  >{\raggedleft\arraybackslash}p{(\linewidth - 8\tabcolsep) * \real{0.1299}}@{}}
\caption{Allan deviation of each pair's held-out residual against the
ambient load scatter at the same point (the Allan deviation of the load
itself), at τ = 8, 16, 64 and 128 s. The pair values bound the
independent noise of both instruments together. Caltech values are
apparent power, with the ambient taken at meter 21 over the scored week;
LBNL values are real power on the bus, with the ambient taken over the
whole 11.6-day record.}\tabularnewline
\toprule\noalign{}
\begin{minipage}[b]{\linewidth}\raggedright
Pair
\end{minipage} & \begin{minipage}[b]{\linewidth}\raggedright
Scored on
\end{minipage} & \begin{minipage}[b]{\linewidth}\raggedleft
Pair residual σ\textsubscript{A}
\end{minipage} & \begin{minipage}[b]{\linewidth}\raggedleft
Ambient σ\textsubscript{A}
\end{minipage} & \begin{minipage}[b]{\linewidth}\raggedleft
Ambient ÷ pair
\end{minipage} \\
\midrule\noalign{}
\endfirsthead
\toprule\noalign{}
\begin{minipage}[b]{\linewidth}\raggedright
Pair
\end{minipage} & \begin{minipage}[b]{\linewidth}\raggedright
Scored on
\end{minipage} & \begin{minipage}[b]{\linewidth}\raggedleft
Pair residual σ\textsubscript{A}
\end{minipage} & \begin{minipage}[b]{\linewidth}\raggedleft
Ambient σ\textsubscript{A}
\end{minipage} & \begin{minipage}[b]{\linewidth}\raggedleft
Ambient ÷ pair
\end{minipage} \\
\midrule\noalign{}
\endhead
\bottomrule\noalign{}
\endlastfoot
Caltech eGauge, transformer 21/17 & 2025-03-17 & 0.13--0.26 kVA &
4.6--5.5 kVA & 21--36 \\
Caltech eGauge, transformer 21/17 & 2025-07-14 & 0.11--0.16 kVA &
5.0--6.1 kVA & 32--45 \\
LBNL µPMU, 12.5 kV bus & days 6--11, Oct 2015 & 0.30--0.35 kW & 3.6--7.9
kW & 11--23 \\
\end{longtable}

Across the transformer the two meters agree to within 0.11--0.26 kVA at
τ = 8, 16, 64 and 128 s, 21--45× below the ambient scatter at meter 21
(20--45× against meter 17). The fit from March scores better in July
than in the following March week, so it did not fail when the season
changed, though one held-out week per season is a small test. The
held-out test was planned before its results existed. It replaced a
first pass, fitted and scored on two halves of the single week of
2025-03-10, which gave a looser 0.22--0.72 kVA (8--23× below ambient),
with a bump at 300--500 s that did not recur in the held-out weeks.

The LBNL pair bounds the larger sensor's own noise at 0.033--0.041 A on
phase 1 at 8--128 s, about 0.11--0.14\% of its 29.4 A; the other two
phases are quieter (0.027--0.035 A). This comes from fitting the full
current phasor (magnitude and phase) on one phase. A separate fit of the
in-phase current summed over the three phases, converted to power, gives
about 0.33 kW across the bus, 11--23× below the ambient load.

In both chains the residual is smallest at τ of 16--32 s and rises
slowly at longer windows, from drift or from a misfit of the losses;
March's top value of 0.26 kVA is the 128 s point, twice the 16 s value.
Either way the bound loosens, which is the safe direction.

\begin{figure}
\centering
\includegraphics[width=1\linewidth,height=\textheight,keepaspectratio,alt={Instrument noise against load scatter, two metering chains. Allan deviation σA against averaging time τ for the ambient load (solid) and for the held-out residual of a same-flow instrument pair (dashed and dotted), at the Caltech substation transformer 21/17 (blue, kVA) and on the LBNL 12.5 kV bus (orange, kW; the pair residual is converted to the larger sensor's side). The shaded band marks the 8--128 s range from which the values in the text are taken (at 8, 16, 64 and 128 s). Over that range the instruments sit one to two orders of magnitude below the load.}]{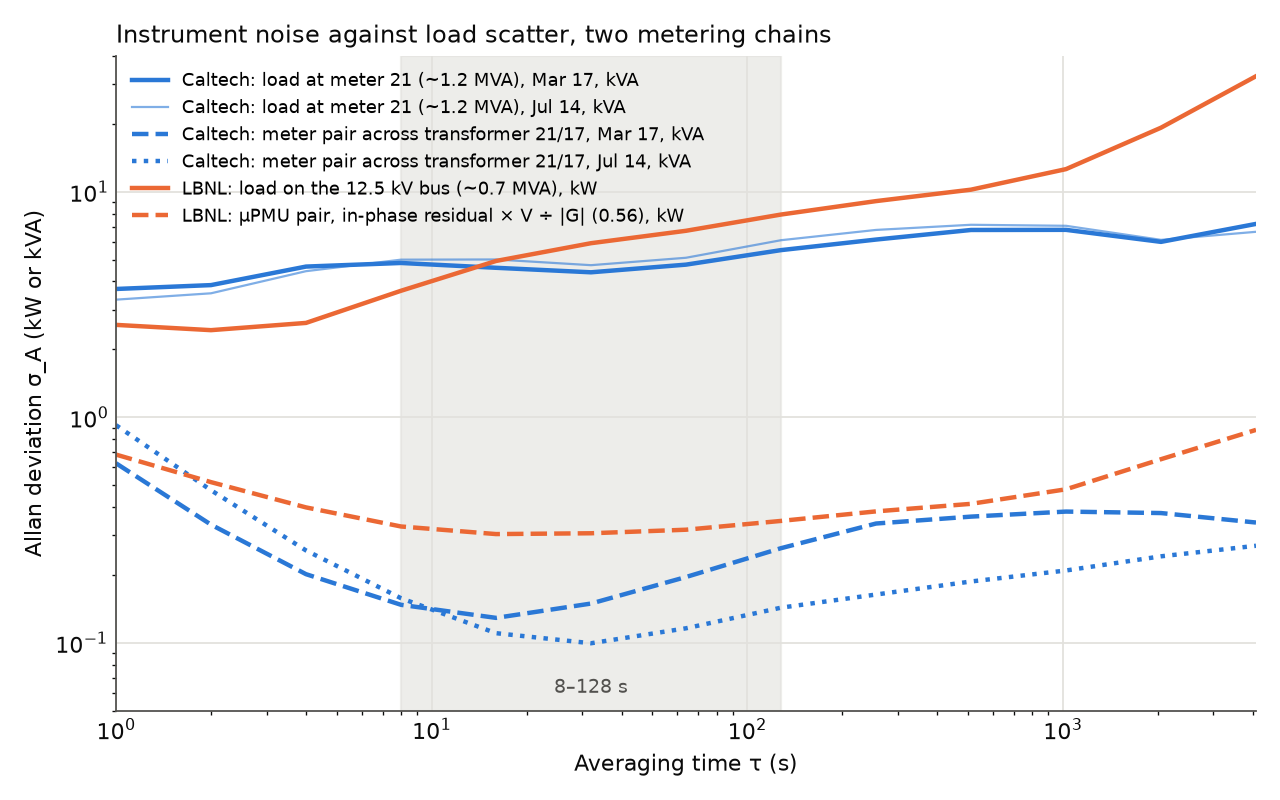}
\caption{Instrument noise against load scatter, two metering chains.
Allan deviation σ\textsubscript{A} against averaging time τ for the
ambient load (solid) and for the held-out residual of a same-flow
instrument pair (dashed and dotted), at the Caltech substation
transformer 21/17 (blue, kVA) and on the LBNL 12.5 kV bus (orange, kW;
the pair residual is converted to the larger sensor's side). The shaded
band marks the 8--128 s range from which the values in the text are
taken (at 8, 16, 64 and 128 s). Over that range the instruments sit one
to two orders of magnitude below the load.}
\end{figure}

These bounds cover independent noise only. Errors common to both
instruments of a pair (shared firmware averaging, matching transformer
non-linearity at the same current, a shared clock) cancel in the
residual and are not seen, and fixed ratio or phase errors are absorbed
by the fit. Fixed errors do not matter to a step detector, which sees
changes; but common errors that vary over time would, and the pair
cannot see them. So the bound covers the independent part of what a step
detector sees, and it says nothing about accuracy. Both chains are
revenue or research instruments, not protection relays, and LBNL is one
pair at one site, with its current transformers running at 1.3--2.9\% of
rating.

\subsubsection*{A.3 Precision against
rating}\label{a.3-precision-against-rating}
\addcontentsline{toc}{subsubsection}{A.3 Precision against rating}

Each instrument reads current or voltage through an instrument
transformer (a current or voltage transformer) that scales the line
quantity down to what the instrument can take; the transformer has a
rating, the largest current or voltage it is built to carry. Expressed
as a share of the reading, the measured precision differs widely between
chains. Expressed as a share of the rating of the transformer that feeds
the instrument, it lines up. We noticed this with the two current chains
and then tested it on a third, the LBNL voltage pair, after writing down
the prediction: if precision is a fixed share of the transformer's
rating, the voltage chain (whose transformer runs at 103\% of rating)
should come out at 0.001--0.003\% of its reading. It came out at
0.0009--0.0018\%.

\begin{longtable}[]{@{}
  >{\raggedright\arraybackslash}p{(\linewidth - 8\tabcolsep) * \real{0.0986}}
  >{\raggedleft\arraybackslash}p{(\linewidth - 8\tabcolsep) * \real{0.2817}}
  >{\raggedleft\arraybackslash}p{(\linewidth - 8\tabcolsep) * \real{0.1268}}
  >{\raggedleft\arraybackslash}p{(\linewidth - 8\tabcolsep) * \real{0.2535}}
  >{\raggedleft\arraybackslash}p{(\linewidth - 8\tabcolsep) * \real{0.2394}}@{}}
\caption{Dynamic precision at τ = 8--128 s for three instrument chains,
as a share of the reading and of the transformer's rating. The values
are pair bounds and cover independent noise only (A.2). The Caltech
per-phase range spans the cases of independent and fully correlated
noise across the three phases; the Caltech residual bounds meters 21 and
17 together and is charged here to meter 21's 600 A current transformer
(meter 17's is rated 2,000 A). The LBNL current row comes from a fit of
current magnitude only, as for the voltage row; the full-phasor fit of
A.2 gives 0.11--0.14\% of reading on phase 1.}\tabularnewline
\toprule\noalign{}
\begin{minipage}[b]{\linewidth}\raggedright
Chain
\end{minipage} & \begin{minipage}[b]{\linewidth}\raggedleft
Transformer rating
\end{minipage} & \begin{minipage}[b]{\linewidth}\raggedleft
Loading
\end{minipage} & \begin{minipage}[b]{\linewidth}\raggedleft
Share of reading
\end{minipage} & \begin{minipage}[b]{\linewidth}\raggedleft
Share of rating
\end{minipage} \\
\midrule\noalign{}
\endfirsthead
\toprule\noalign{}
\begin{minipage}[b]{\linewidth}\raggedright
Chain
\end{minipage} & \begin{minipage}[b]{\linewidth}\raggedleft
Transformer rating
\end{minipage} & \begin{minipage}[b]{\linewidth}\raggedleft
Loading
\end{minipage} & \begin{minipage}[b]{\linewidth}\raggedleft
Share of reading
\end{minipage} & \begin{minipage}[b]{\linewidth}\raggedleft
Share of rating
\end{minipage} \\
\midrule\noalign{}
\endhead
\bottomrule\noalign{}
\endlastfoot
Caltech eGauge, current (meter 21) & 600 A & 6.9\% & 0.009--0.038\% &
0.0006--0.0026\% \\
LBNL µPMU, current & 1,200 A & 1.3--2.9\% & 0.07--0.09\% &
0.0011--0.0026\% \\
LBNL µPMU, voltage & 7,200 V & 103\% & 0.0009--0.0018\% &
0.0009--0.0019\% \\
\end{longtable}

So across three chains dynamic precision comes to about 0.001--0.003\%
of the transformer's rating, while the share of reading spans 100×. How
the Caltech residual is split between its two meters changes its row:
charged to meter 17 alone it would be up to 0.0055\% of rating, above
the band; split between the two at the same share of each meter's own
rating, as the rule itself assumes, it is 0.0006--0.0024\%, inside the
band.

As a rule of thumb, the floor in power is about 2×10⁻⁵ of the
transformer's rated apparent power. For Caltech meter 21 (600 A at 16.5
kV, 17.1 MVA) the band gives 0.17--0.51 kVA, against the 0.11--0.26 kVA
measured. This is a consistency check, not an independent test: the band
was drawn from the same three chains, Caltech among them. We treat the
rule as a hypothesis, not a law. It rests on three chains at two sites
from two makers, two of the chains share the same µPMUs, and the match
holds against the transformer's rating but not against the µPMU's own
input ranges, so we do not know the mechanism.

Both chains are also one to two orders of magnitude more precise than
their rated accuracy. The Caltech chain is rated at about ±1--1.5\% of
reading (meter plus instrument transformers). The LBNL chain's class 1.2
transformers are, we infer, guaranteed nothing below 10\% of rating (we
read that floor for class 0.3 and assume it holds for class 1.2), and
the LBNL transformers run at 1.3--2.9\%.

The rule matters most for the instruments we could not measure. We found
no published dynamic precision for any protection relay, fault recorder,
power-quality recorder or PMU current channel: no Allan deviation, no
step-repeatability test, no same-flow comparison. No vendor specifies it
either. The nearest figure is on PSL's µPMU data sheet
\citep{psl2020micropmu}, which gives a ``typical short-term TVE
stability for differential measurements'' of ±0.002\% for its voltage
inputs (TVE, total vector error, combines magnitude and phase error).
That is the same size as our measured band, but it is a voltage figure,
not a current one. Relay datasheets give only accuracy, as a share of
reading or a fixed share of rated current, and the fixed term (±0.2\% of
rated current for the ABB REF615, ±0.1\% for the GE UR family
\citep{abb2009ref615, ge2018urspecs}) is about two orders of magnitude
above our measured floor. Feeder-head relays digitise through protection
transformers sized for fault current, which is many times normal load,
so if the rule of thumb holds their floor will be higher than on these
revenue chains. Their dynamic precision remains unmeasured; a relay
recording beside a revenue meter on the same flow is the cheapest way to
measure it.

\subsubsection*{A.4 The load, not the meter, sets the
limit}\label{a.4-the-load-not-the-meter-sets-the-limit}
\addcontentsline{toc}{subsubsection}{A.4 The load, not the meter, sets
the limit}

What a single turn-on step has to clear is the largest step that
ordinary load makes in the same window, not the meter's noise. We
measure this as the step exceeded once a day: the size of change between
adjacent window averages that the ordinary load exceeds, on average,
once per 24 hours of record.

Caltech substation meter 21 carries about 1 MW of real power (1,014 kW;
the 1.2 MVA of A.2 is the apparent power of the same net flow). Over
whole days in the week of 2025-03-10, the step exceeded once a day is
30--53 kW at 8--128 s averaging. The largest of these come in the
daytime, since night-time steps are smaller. They are larger than the
16-chip island's 26 kW, so in the daytime a single turn-on step is hard
to pick out on its own.

At night (00:00--05:00 local time) the load is quieter. In the same week
the night-time scatter falls to σ\textsubscript{A} = 3.1--3.9 kW at
64--128 s, so a 26 kW step is about 7--8× the scatter. It is about
100--200× the transformer pair's noise at the same windows (0.12--0.26
kVA, A.2), and an order of magnitude above a class 0.2 meter's absolute
band of ±2 kW at 1 MW, which in any case scales the step rather than
hiding it. The margin over the largest natural steps is thinner. In the
night-time data the step exceeded once per 24 hours of night-time record
(about once in five nights) is 18--20 kW at 64--128 s, so the island
clears it by only 1.3--1.4×. At shorter windows the margin is gone: the
same step is 25 kW at 16 s (a margin of 1.04×) and 31 kW at 8 s, above
the island.

The LBNL bus is noisier. There the step exceeded once a day at 8--128 s
is 52--57 kVA over the whole day and 33--43 kVA at night. In real power
the night figure is 26--30 kW, so the island does not clear it at any
window.

So at Caltech meter 21, a substation meter of about 1 MW, a single
turn-on step is visible at night to a detector that averages over a
minute or two, and is often lost in the daytime. At the LBNL bus it is
not visible on its own even at night. In every case the meter is not the
limit; the ordinary load is. The limit only moves to the instrument on a
very quiet line, behind a heavily oversized transformer, or for a
detector that cancels most of the ambient load, which is what energy
balance does (A.5).

\subsubsection*{A.5 Energy balance}\label{a.5-energy-balance}
\addcontentsline{toc}{subsubsection}{A.5 Energy balance}

Energy balance subtracts the readings of the meters downstream of a
point from the reading at that point, interval by interval. Whatever
load is metered cancels, and what is left (the balance residual) is
unmetered load, losses and instrument noise.

We tried it on the Caltech 2.4 kV network. The inflow is measured on the
16.5 kV side of the two substation transformers. From it we subtract the
metered 480 V boards below: the ten building mains, the inverter board
that feeds on-site generation into the network, and the solar and plant
board, which enters at a fitted weight of 0.58 that we cannot explain
(without the weight, the residual swings by about ±30 kW). We also
subtract a fitted constant of about 77 kW, roughly what the transformers
between the measuring points should lose, plus any steady unmetered
load. Fitted on the week of 2025-03-10 and scored on the same two
held-out weeks as in A.2, the residual has σ\textsubscript{A} = 3.6--4.4
kW (2025-03-17) and 3.7--5.3 kW (2025-07-14) at 8--128 s, so a 26 kW
step stands out at 5--7× the residual. After the constant, the
residual's mean moves from about 2 kW in March to about 6 kW in July;
that is a slow offset, not scatter, and a step detector does not see it.

Because the fitted constant absorbs steady losses and steady unmetered
load, the balance sees changes rather than levels. An island already
running when monitoring starts sits inside the constant and shows only
when it stops or restarts (see the restart test in Appendix B.5).

The balance also lowers the largest natural steps. The step in the
balance residual exceeded once a day is 19--24 kW at 64--128 s in the
two held-out weeks, against about 30 kW at meter 21 alone in the week of
2025-03-10 (meter 21's scatter is similar in all three weeks), and still
47--51 kW at 8 s. In real power, the balance cuts these largest steps by
about a fifth to a third, a little more than it cuts the scatter (by up
to 29\% at 128 s, and hardly at all at 8--16 s). At 64--128 s the
balance brings the whole day down to about where the head meter sits at
night, so a 26 kW step clears the daily extreme by only 1.1--1.3×, a
much thinner margin than its 5--7× over the scatter. Tests that use more
than one step (does the load stay up, does it recur) are not tried here.

This is one campus network, metered at 1 s by meters that report
honestly, and the fitted weight on one board is not explained. It has
not been tried on a utility feeder, where the downstream meters are
hundreds to thousands of customer meters (about 1,300 on an average BC
Hydro feeder, Appendix D.2) read at 15-minute or hourly intervals with
their own clock skew. The instrument test worth doing next is one
utility feeder's customer interval data together with its head-meter
data, to measure the balance residual at the cadence utilities actually
collect.

\subsection*{Appendix B. High-frequency
detection}\label{app-high-frequency}
\addcontentsline{toc}{subsection}{Appendix B. High-frequency detection}

\subsubsection*{B.1 Low-frequency versus high-frequency
data}\label{b.1-low-frequency-versus-high-frequency-data}
\addcontentsline{toc}{subsubsection}{B.1 Low-frequency versus
high-frequency data}

A revenue meter already measures the training pulse; it just does not
record it. Inside the meter, voltage and current are sampled thousands
of times a second (up to 32 kHz on Itron's grid-edge meters
\citep[p.~2]{itron2023dimeters}), multiplied, and added into an energy
register, which runs continuously like an odometer. The fast samples are
used up in that sum and never stored. At the end of each interval,
usually 15 minutes for commercial customers, the meter logs the energy
used in it. That energy divided by the interval is the ``15-minute
average'' demand: 10 kWh in 15 minutes is 40 kW. The ``peak'' or maximum
demand on a bill is the highest of those averages over the billing
period (PG\&E Schedule B-20 bills the ``highest 15-minute average'' in
the month \citep{pge2026scheduleb20}; Georgia Power bills the highest
30-minute demand of the current and preceding eleven months
\citep{georgiapower2025pll18}), not the peak of the fast samples, so a
one-second spike barely moves it. The logged intervals are uploaded in
batches; PG\&E posts non-residential 15-minute data daily
\citep{pge2020eamds}. Many meters also keep instantaneous
``instrumentation'' readings (present kW, volts, amps) that can be read
on demand or logged at a faster interval, but the sources do not say
that utilities send them anywhere: the Alpha 4 C\&I meter can log them
once a minute \citep{honeywell2020alpha4}, and the SEL-735 trends its
load-profile channels every three seconds \citep{sel2024sel735} and can
stream its waveform at 3,000 samples a second.

Interval data at 15 minutes is enough for everything in the main text.
It gives average load, load factor (the average over the highest
interval), energy steps that last weeks or months, and outages of more
than an interval. Most commercial and industrial meters are advanced
(AMI) meters that can collect it: in 2025, 84.7\% of the 20.66 million
US commercial meters and 80.9\% of the 1.04 million industrial meters
were AMI, up from 61.6\% and 58.1\% in 2020.\footnote{\citet[Table
  10.05]{eia2025epatable1005}. The shares are AMI as a fraction of all
  meters in each sector, from the same table; the table does not report
  what interval a utility configures.} What interval data cannot show is
anything faster than the interval. The training pulse of B.3 repeats
every 10--16 s, so seeing it at all needs a reading at least every 5--8
s, and seeing its shape needs one every second or faster. Turn-on edges,
the short drops of B.5 and the harmonics of B.8 need faster still. The
rest of this appendix therefore uses research recordings, not utility
data: one-second readings from the Caltech eGauge meters, 120 samples
per second from the LBNL µPMUs (micro phasor measurement units), and, in
B.8, the Caltech meters' 2.5 kHz waveform captures.

Whether utilities could collect such data from their existing equipment
depends on two things: whether the device can compute something
sub-second, and whether its link can carry the result. At the customer
meter, the newer ``grid-edge'' meters (Itron Riva, Landis+Gyr Revelo)
compute per-second power and capture waveforms on the meter, and run
apps that the utility installs remotely. They were about 4\% of US smart
meters in 2021 by one industry estimate, forecast to pass 25\% by 2030;
the same estimate counts more than 4 million Itron Riva meters deployed,
and Landis+Gyr reports more than 6 million Revelo meters under contract
in North America.\footnote{Guidehouse Insights for Sense (2022)
  \citep{guidehouse2022insidethemeter}; Landis+Gyr press release, 29 Oct
  2025 \citep{landisgyr2025rappahannock}. Both are industry sources, and
  neither splits commercial from residential.} The first-generation
meters that make up most of the installed fleet record nothing faster
than their profile interval, a minute at best, without a hardware
change. The link is the tighter limit. Itron's own figure for its mesh
network is that ``1kb packets can typically be transmitted once every
few seconds'' per device, shared with every meter that relays through it
\citep[p.~3]{itron2023networks}. That is enough for a few meters per
mesh to stream one channel at a few readings a second, or for every
meter to send a summary feature (a count of steps, the strength of a
spectral line) every few minutes; it is not enough for every meter to
stream. Itron makes the same point \citep{itron2023dimeters}, keeping
analytics on the meter because backhauling the raw data ``is an enormous
IT challenge.'' We found no US utility streaming sub-second data from
commercial meters outside pilots.

At the feeder head the situation is better. Microprocessor feeder relays
sample at kHz rates, and many can be set to report synchrophasors
(time-stamped measurements of voltage and current) at 1--60 times a
second; older units cap at once a second.\footnote{SEL synchrophasor
  product table \citep{sel2026synchrophasors}.} A cooperative that added
this function to its installed relays found it ``in many relays at no
extra cost.'' \citep{hataway2013synchrophasors} What limits it is again
the link and the configuration. SCADA (the utility's supervisory control
system) usually polls the feeder head every few seconds, which is
borderline for a 10 s pulse and too slow for its shape. At 60 reports a
second, each synchronized measurement loads the network at about 29
kbps, and the DOE's distribution roadmap judges that ``access to
high-speed communication is limited in most areas'' of the distribution
system \citep{quanta2021roadmap}. A relay's full stream of eight phasors
and a few other values at 60 a second needs 61.2 kbps, more than the
57.6 kbps of a typical serial port, so it has to go out over Ethernet
\citep{hataway2013synchrophasors}. Substations on fibre or microwave
could carry a reduced stream (one power channel at 10 a second is a few
kbps) for every feeder, and a summary feature computed in the relay fits
within the existing SCADA poll almost everywhere SCADA reaches. We found
no national count of feeder heads with microprocessor relays, or of the
substation links, so we give no fraction.

So the sub-second signal is already measured at both ends of the
distribution system. Recording it, and sending either the raw data or a
feature computed from it, is a question of configuration and
communications, not of new sensing hardware, wherever a grid-edge meter
or a modern relay is installed. Where neither is, a dedicated monitor is
needed.

Better data matters because a screen's false negatives feed straight
into the inspection count. A host with an island that the screen fails
to flag stays in the pool of accounts from which random inspections are
drawn. The naive step screen of B.4 (a rise in monthly energy, with no
high-rate sampling or waveform analysis) misses 17--26\% of the islands
behind accounts whose load is between 1 and 5 times the island's load,
depending on the island's size. Those accounts outnumber the candidate
hosts (accounts above 5 times the island's load) by 3--7 to one, so the
misses roughly double the pool. This is not the same as d, the chance
that an inspection of the right host finds the island (Appendix C). A
low d leaves the pool alone and makes each inspection less likely to
succeed; d = 0.5 also roughly doubles the inspections when a run is
spread over many sites. The two multiply, so for a run spread over many
sites the naive screen's false negatives and an inspection that succeeds
half the time together cost about four times the paper's counts. For the
runs on two sites and on one site (C.3) the combined cost is about 6 and
10 times; those runs need the fewest inspections in any case.

\subsubsection*{B.2 Data}\label{b.2-data}
\addcontentsline{toc}{subsubsection}{B.2 Data}

The high-rate tests add logged training jobs (the signal) to real grid
recordings (the background). The backgrounds come from two places. The
LBNL open µPMU data set \citep{stewart2016openupmu} gives 11.57 days
(2015-10-01 to 10-12) of one 12 kV bus at 120 samples per second, mean
load 724 kVA; we analysed it mainly as 1-second means. The Caltech SoCal
28-bus data set \citep{xie2025socal} records eGauge meter waveforms at
2.5 kHz; we used the one-second magnitudes that its client software
serves (documented in the client's README). We used three of its meters:
a 252 kVA building main at 480 V; a 1,160 kVA 2.4 kV section, which is
the low side of the substation transformer 21/17 of Appendix A and
carries a net flow (the buildings on it draw about 1.66 MW, about 0.7 MW
of which comes from on-site generation); and the 5,489 kVA 16.5 kV main
supply into the campus substation. The first Caltech tests used one week
(2025-03-10 to 03-17). The later re-test in B.3 used three blocks of
four weeks in 2025, one each in winter, March and summer.

The signal comes from two sources. The NLR data set
\citep{vercellino2026measurement} logs GPU power every 0.2 s on 16 H100s
in 4 nodes: five replicate runs of a Llama-2 70B LoRA fine-tune, and
five Stable Diffusion runs, of which we used one.
\citet{elsayed2026characterization} publish 15-minute recordings of
per-GPU power from single 8-GPU cloud nodes, of which we used two H100
sessions, a Llama-8B run and a small diffusion run (data licensed CC BY
4.0). The Caltech data are licensed CC BY 4.0 (IEEE DataPort,
doi:10.21227/hsfq-fq78); the full set needs a GitHub login approved by
the maintainers, and its meters carry anonymised names. All of the job
logs are software readings of GPU power: the NLR readings are, and the
Elsayed readings behave like, a mean over the previous second. So they
show the step frequency but not the edges. The NLR GPU readings include
the GPUs' own memory and leave out host memory, switches, network cards,
storage, fans and power-supply losses.

\subsubsection*{B.3 The training pulse}\label{b.3-the-training-pulse}
\addcontentsline{toc}{subsubsection}{B.3 The training pulse}

Synchronous training draws power in a regular pulse. Each training step
has a compute-heavy part, when the GPUs run at high power, and a lighter
part (communication between chips, and in the NLR fine-tune also traffic
to the CPU memory that holds part of the model), when they draw less.
The whole island moves from one part to the other at the same moment,
every step, so its power rises and falls with the step time. Grid
operators already run forced-oscillation detectors, built to find loads
or generators that push the grid at a steady frequency, and a periodic
load like this is the kind of thing they look for. In the NLR fine-tune
the step is a near-square pulse every 10.1--10.3 s: 5.6--5.8 s high at
about 11 kW, then 4.4--4.6 s low at about 8.6 kW. We measure the size of
the pulse by its \emph{swing}, the 95th percentile of the island's GPU
power minus its 5th percentile. The 16 GPUs move together: the total
swing is 0.96--0.97 of the sum of the GPUs' own swings. That swing,
2.32--2.45 kW, is 21--22\% of GPU power, but only about 9\% of the
island's 26 kW at the meter, since the rest of the island's load (CPUs,
fans, cooling) does not follow the step.

\begin{figure}
\centering
\includegraphics[width=1\linewidth,height=\textheight,keepaspectratio,alt={The two logged workloads. Top: two minutes of the Llama-2 70B LoRA fine-tune on 16 H100s; the step is a near-square pulse. Bottom: six minutes of the NLR Stable Diffusion run; long blocks are flat at one-second resolution, and the deep swings come in other blocks, which the source paper attributes to evaluation (the logs carry no labels for the parts of a run). This is not the Elsayed diffusion run in the job table below. Both panels are software GPU readings, each a mean over the previous second, not wall power. Source: NLR data set {[}@vercellino2026measurement{]}.}]{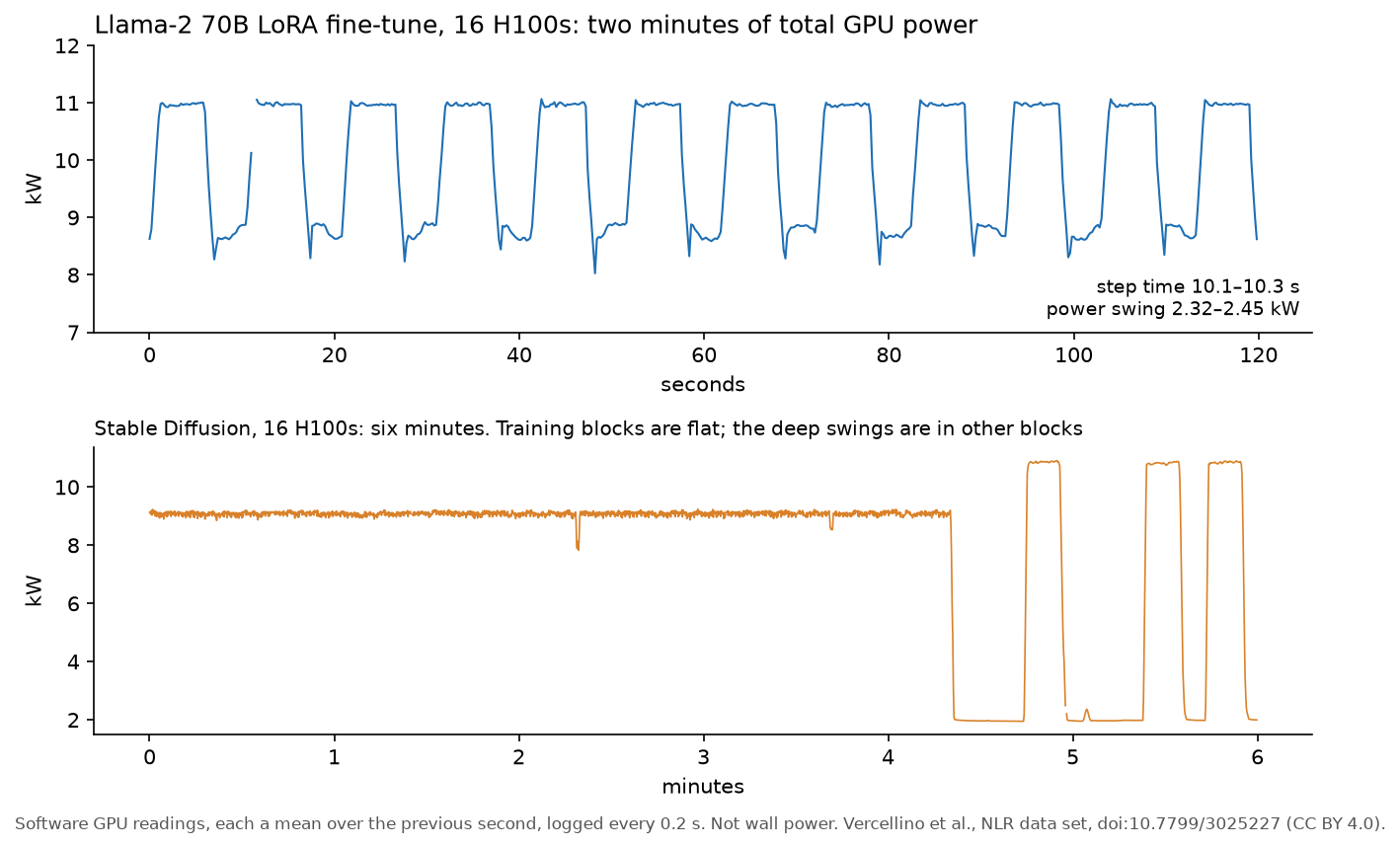}
\caption{The two logged workloads. Top: two minutes of the Llama-2 70B
LoRA fine-tune on 16 H100s; the step is a near-square pulse. Bottom: six
minutes of the NLR Stable Diffusion run; long blocks are flat at
one-second resolution, and the deep swings come in other blocks, which
the source paper attributes to evaluation (the logs carry no labels for
the parts of a run). This is not the Elsayed diffusion run in the job
table below. Both panels are software GPU readings, each a mean over the
previous second, not wall power. Source: NLR data set
\citep{vercellino2026measurement}.}
\end{figure}

\begin{longtable}[]{@{}
  >{\raggedright\arraybackslash}p{(\linewidth - 8\tabcolsep) * \real{0.1765}}
  >{\raggedright\arraybackslash}p{(\linewidth - 8\tabcolsep) * \real{0.1765}}
  >{\raggedleft\arraybackslash}p{(\linewidth - 8\tabcolsep) * \real{0.2353}}
  >{\raggedright\arraybackslash}p{(\linewidth - 8\tabcolsep) * \real{0.1765}}
  >{\raggedleft\arraybackslash}p{(\linewidth - 8\tabcolsep) * \real{0.2353}}@{}}
\caption{The three jobs added to the grid recordings, as a 16-chip
island. The fine-tune was logged on 16 GPUs and is used as logged; the
two Elsayed jobs were logged on one 8-GPU node and doubled, with the two
copies stepping together.}\tabularnewline
\toprule\noalign{}
\begin{minipage}[b]{\linewidth}\raggedright
Job
\end{minipage} & \begin{minipage}[b]{\linewidth}\raggedright
Source
\end{minipage} & \begin{minipage}[b]{\linewidth}\raggedleft
Step time
\end{minipage} & \begin{minipage}[b]{\linewidth}\raggedright
Pulse shape
\end{minipage} & \begin{minipage}[b]{\linewidth}\raggedleft
Swing, 16 chips
\end{minipage} \\
\midrule\noalign{}
\endfirsthead
\toprule\noalign{}
\begin{minipage}[b]{\linewidth}\raggedright
Job
\end{minipage} & \begin{minipage}[b]{\linewidth}\raggedright
Source
\end{minipage} & \begin{minipage}[b]{\linewidth}\raggedleft
Step time
\end{minipage} & \begin{minipage}[b]{\linewidth}\raggedright
Pulse shape
\end{minipage} & \begin{minipage}[b]{\linewidth}\raggedleft
Swing, 16 chips
\end{minipage} \\
\midrule\noalign{}
\endhead
\bottomrule\noalign{}
\endlastfoot
Llama-2 70B LoRA fine-tune & NLR \citep{vercellino2026measurement} &
10.1--10.3 s & high 5.6--5.8 s of each step & 2.32--2.45 kW \\
Llama-8B training run & \citet{elsayed2026characterization} & 13.4 s &
low about 14\% of each step & 7.4--8.1 kW \\
Diffusion (107M parameters) & \citet{elsayed2026characterization} & 16.1
s & high about a third of each step & 4.5 kW \\
\end{longtable}

An island makes two signals: its average load, which appears as a step
in the line's load when it turns on or off, and the pulse on top of that
load. This section tests both. Whether the pulse can be found depends
mostly on how the island is wired. On one phase, the pulse was found in
almost every hour at every meter we tried. Spread evenly over three
phases, the pulse was mostly lost in the background, but the turn-on
step was still found at the building main at both alarm thresholds, and
in about a fifth of runs at the 1.2 MVA section. False alarms stayed
near their target in most tests, but ran far over it in a few, because
the background changed from week to week.

\emph{How a job is added.} A \emph{logged job} is a real training run
whose GPU power was recorded by software while it ran (the three in the
table above). Each test adds a logged job's power trace to a real grid
recording and asks whether a detector can find it. For the pulse, the
trace is cut to the job's busy stretch, its mean is removed, and it is
repeated to fill each hour, as a 16-chip island (32 chips where stated).
Removing the mean does not change the pulse test: a constant load sits
at zero frequency, outside both of the detector's bands (below). The
average load is tested separately, as a step. The meters record apparent
power (kVA), the product of voltage and current magnitudes. The rest of
the load on a line draws its current somewhat out of step with the
voltage, by an angle set by its power factor, while the island is
modelled as drawing in step. The two currents therefore do not add in
full: to first order, the island raises the apparent power by its own
real power times the power factor of the rest of the load. In the pulse
re-test (the table below and the three-phase re-test figures after it)
we count 90\% of the swing, a power factor of 0.90. Measured on these
meters, the power factor of the rest of the load is 0.855 at the
building main and the 1.2 MVA section and 0.87--0.90 at the campus
supply, so at the first two meters the island adds slightly less than we
counted. The earlier test, the turn-on test and the test of offset
copies (all below) count the whole swing or step.

\emph{Phase unbalance and three-phase total.} Most commercial lines
carry three phases. An island wired to one phase is read on the
\emph{phase unbalance}: that phase minus the average of the other two.
The unbalance cancels most of the variation that all three phases share,
so the island's pulse stands out. An island split evenly across the
three phases leaves the unbalance unchanged, so it has to be read on the
\emph{three-phase total}, where the whole load of the line varies around
it. A monitor would not know which phase an island is on. So each test
put the island on each phase in turn, and we report the phase where it
was hardest to find (the \emph{worst phase}).

\emph{The pulse detector.} The detector looks for a repeating pulse. For
each hour it measures how much power the recording carries in a band of
frequencies around the expected step rate, and compares that with the
site's usual level in the same band. An hour whose score passes a
threshold raises an alarm. We tried two bands, both fixed before any
logged job was added. The \emph{short-segment} setting covers 0.08--0.5
Hz, which holds steps of about 2--12 s, and works on 100 s stretches of
data; the Llama-8B run's 13.4 s step and the diffusion run's 16.1 s step
fall just below this band and are seen in it through their harmonics.
The \emph{long-segment} setting covers 0.016--0.3 Hz, steps of about
3--60 s, on 500 s stretches. Every site also has frequencies of its own
at which some equipment pulses all the time (\emph{spectral lines}); the
detector learns these and ignores them (B.6).

\emph{The re-test.} The table below comes from a re-test at three
Caltech meters: the 252 kVA building main, the 1.2 MVA section of the
campus network, and the 5.5 MVA main supply into the campus substation.
Each meter was tested on three four-week blocks of 2025 data, in winter,
March and summer. Each block is used in three parts. The first week
teaches the detector the site's usual level and its spectral lines. The
second week sets the threshold at the score of its seventh-loudest hour,
so that about one hour a day would pass it: 13.9 alarms expected in the
336 scored hours (336 × 7/169). The last two weeks, 336 hours, are
scored. A \emph{false positive} is an alarm in an hour without the job.
One combination of meter, setting and season is a \emph{cell} of the
table.

\emph{The earlier test.} An earlier version of the test used one Caltech
week and the LBNL record, each split into a half that the detector
learned on and set its threshold from, and a half that was scored. It
had two thresholds: the score of the loudest hour without the job in the
learning half (the lower), and twice that (the higher). We quote it
below for LBNL, which the re-test did not include, and for how much
larger an island must be to be found on the three-phase total, which the
re-test's tables do not give.

\begin{longtable}[]{@{}
  >{\raggedright\arraybackslash}p{(\linewidth - 8\tabcolsep) * \real{0.2000}}
  >{\raggedright\arraybackslash}p{(\linewidth - 8\tabcolsep) * \real{0.2000}}
  >{\raggedright\arraybackslash}p{(\linewidth - 8\tabcolsep) * \real{0.2000}}
  >{\raggedright\arraybackslash}p{(\linewidth - 8\tabcolsep) * \real{0.2000}}
  >{\raggedright\arraybackslash}p{(\linewidth - 8\tabcolsep) * \real{0.2000}}@{}}
\caption{Single-phase 16-chip island on the phase unbalance, Caltech
re-test, worst phase, power factor 0.90. Each cell gives the share of
the 336 scored hours in which the job was found, for the fine-tune / the
Llama-8B run, and in brackets the false positives in the same hours
without the job (13.9 expected at one a day). ``Found'' counts the hours
in which the island's own phase passes the threshold; the bracket counts
the hours in which any of the three phases passes it without the job,
which is the false-alarm cost of a monitor that does not know the
island's phase. So a found rate can sit below its
bracket.}\tabularnewline
\toprule\noalign{}
\begin{minipage}[b]{\linewidth}\raggedright
Meter
\end{minipage} & \begin{minipage}[b]{\linewidth}\raggedright
Setting
\end{minipage} & \begin{minipage}[b]{\linewidth}\raggedright
Winter
\end{minipage} & \begin{minipage}[b]{\linewidth}\raggedright
March
\end{minipage} & \begin{minipage}[b]{\linewidth}\raggedright
Summer
\end{minipage} \\
\midrule\noalign{}
\endfirsthead
\toprule\noalign{}
\begin{minipage}[b]{\linewidth}\raggedright
Meter
\end{minipage} & \begin{minipage}[b]{\linewidth}\raggedright
Setting
\end{minipage} & \begin{minipage}[b]{\linewidth}\raggedright
Winter
\end{minipage} & \begin{minipage}[b]{\linewidth}\raggedright
March
\end{minipage} & \begin{minipage}[b]{\linewidth}\raggedright
Summer
\end{minipage} \\
\midrule\noalign{}
\endhead
\bottomrule\noalign{}
\endlastfoot
252 kVA & short-segment & 100 / 100 (13) & 100 / 100 (122) & 100 / 100
(23) \\
252 kVA & long-segment & 100 / 100 (9) & 100 / 100 (99) & 100 / 100
(11) \\
1.2 MVA & short-segment & 100 / 100 (31) & 100 / 100 (1) & 100 / 100
(178) \\
1.2 MVA & long-segment & 100 / 100 (20) & 100 / 100 (16) & 100 / 100
(18) \\
5.5 MVA & short-segment & 0 / 100 (41) & 1 / 100 (16) & 1 / 100 (43) \\
5.5 MVA & long-segment & 100 / 100 (19) & 100 / 100 (23) & 100 / 100
(4) \\
\end{longtable}

On one phase, the pulse was found almost everywhere. The Llama-8B run
was found in every hour in every cell. The fine-tune was found in every
hour at the 252 kVA and 1.2 MVA meters in every season, and at 5.5 MVA
on the long-segment setting. Its one miss is the short-segment setting
at 5.5 MVA, where the result depends on the phase. The table shows the
worst phase. On its best phase the fine-tune is found in every hour in
March and in 67\% of hours in summer, where that phase alarms in 4 and 0
hours without the job; in winter no phase does better than its own
alarms without the job. At that meter and setting the fine-tune is close
to the limit: a job with a much smaller swing would be missed.

The earlier test agrees. At LBNL all five fine-tune runs and the Stable
Diffusion run were found in 100\% of hours on the phase unbalance, with
0--2 false positives in 121 scored hours, and both Elsayed jobs, even at
their logged 8-GPU size, were found in every hour on both settings. A
32-chip fine-tune island was found in every hour at all four meters,
including the 5.5 MVA short-segment cells, and the re-test found it in
every hour at 5.5 MVA on that setting in every season.

Spread evenly over three phases, the island is much harder to see. The
fine-tune was found clearly more often than the false positives in one
cell only (252 kVA, long-segment, March: 32\% of hours against 3\%). The
Llama-8B run has about 3.3 times the swing. It was found in every hour
at the 252 kVA building main on the long-segment setting in winter and
summer, but in 3\% in March, when the learning week marked 63\% of the
band as site lines and set it aside (18\% in winter). At 5.5 MVA it was
found in no more hours than the false positives. At LBNL, in the earlier
test at its lower threshold, the fine-tune on the total was found in 6\%
of hours against 3 false positives in 121, no different from background.
In the earlier test, at the higher threshold, a three-phase fine-tune
island had to be about 2.2--2.3 times larger to be found in 80\% of
hours at the 724 kVA LBNL bus, 3.2--7.5 times at 1.2 MVA and 9--11 times
at 5.5 MVA; each range spans the two ways of splitting the record
(learning on the first half and scoring the second, and the reverse).

\emph{The turn-on step.} The turn-on test asks whether an island can be
seen arriving, as a step up in the line's mean load, where its pulse
cannot be seen. It adds the island's whole load: each logged job's GPU
power, topped up to the island's full load at the meter of 1.6 kW per
chip (25.6 kW for 16 chips). This assumes the island has its own cooling
unit, which starts with the chips. The load was added in full, with no
power-factor scaling, as runs of 1 hour (84 runs) or 6 hours (82 runs)
at random start times in the two scored weeks, with nothing before or
after; only the fine-tune and the Llama-8B run were used. A second
detector looks for a change in the mean load over windows of 2 to 60
minutes. It has two alarm thresholds, both set on the calibration week
without the job: the lower is the largest such change seen there, and
the higher is twice that. The higher threshold is the headline, since it
keeps false positives to a few a week.

On one phase, read on the phase unbalance, the turn-on was found at the
252 kVA and 1.2 MVA meters in 98--100\% of runs at the higher threshold,
with no false positives in the scored weeks, and in 99--100\% at the
lower threshold (3.5--10 false positives a week), in every season. At
5.5 MVA the higher threshold was mostly too high: it found 2--10\% of
turn-ons on the worst phase and up to 46\% on the best, with 0--2.5
false positives a week. At the lower threshold (1--7 false positives a
week) it found every turn-on on every phase in winter and summer. In
March it found none at either threshold, because one switching step in
the calibration week set both far too high (B.9).

The three-phase total sees the island's whole load however it is wired
across the phases, so the results there hold for any wiring. At the 252
kVA building main the turn-on was found in 98--100\% of runs and the
turn-off (the island switching off) in 99--100\%, at the higher
threshold in every season, with 0.5--2.5 false positives a week. At the
lower threshold both were found in 99--100\% of runs, with 8--9 false
positives a week. At the 1.2 MVA section a 16-chip island's turn-on was
found in 18--24\% of runs at the lower threshold in March and summer,
against 5--6\% and 0\% of runs without the job, with 21 and 5 false
positives a week; its turn-off was found in 35--87\%. In winter neither
was found, and at the higher threshold neither was found more often than
without the job. A 32-chip island (51.2 kW) was found there in 91--99\%
of runs at the lower threshold in March and summer. At 5.5 MVA no island
was found on the total more often than without the job. The 724 kVA LBNL
bus was not in the re-test; in the earlier test, with the island topped
up to 1 kW per chip, the turn-on on its total was no different from
background.

\emph{False positives.} In most cells of the pulse re-test, false
positives came close to the one a day the threshold aims at: the median
cell had 1.2 times the 13.9 expected. A few cells ran far above it:
7--13 times in 3 of 18 phase-unbalance cells, and 6--7 times on the 1.2
MVA total in March. The cause was the background changing between weeks,
so that a threshold set on one week did not fit the next. On the 1.2 MVA
total, the median hourly score rose from 1.2--1.6 in the first two March
weeks to 2.7--5.9 in the last two. In summer, a new spectral line
appeared on the 252 kVA phase unbalance in the final week. Setting the
threshold more often does not fix this. We scored each cell's last week
against a threshold set on the week before and against one set two weeks
before. The older threshold did worse in 16 of 34 cells and better in
16.

\emph{Two simplifications in the test.} They push the results in
opposite directions. The first makes the island easier to find than a
real one would be; the second hides part of its signal.

First, some islands are built from copies that step in exact unison. The
two Elsayed jobs, logged on one 8-GPU node, became 16- and 32-chip
islands as two and four copies of that node, and the 32-chip fine-tune
is two copies of the logged 16-chip trace. (The 16-chip fine-tune is not
a copy: NLR logged its 16 GPUs directly, and they move together on their
own.) Real nodes would drift slightly out of step, which smears the
pulse. To test this we spread the copies evenly across a window of ±0.3
s, in the 34 usable cells for each of the five islands built from copies
(170 cells). Found rates changed by a median of 0 points and at most 10,
and moving the copies apart never raised a found rate. The change is
small because most of the pulse's power sits at 0.10--0.15 Hz, where
even the worst arrangement of that spread, half the copies early and
half late, costs only 4--8\% of it; the evenly spread copies we tested
lost 1--4\% of the power in the band.

Second, the logs are one-second means, so nothing above about 0.5 Hz has
been measured. A real pulse has sharp edges, which put power into
harmonics of the step frequency that one-second means smooth away. A
simulation that rebuilt the fine-tune's pulse with sharp edges and added
it to the LBNL record suggests a µPMU at 120 samples per second would
see harmonics of the step up to about 1.8 Hz in a quiet hour (B.8).
Testing that needs a trace of a real training job measured at the wall.

\subsubsection*{B.4 Where and how fast you look decides what you can
see}\label{b.4-where-and-how-fast-you-look-decides-what-you-can-see}
\addcontentsline{toc}{subsubsection}{B.4 Where and how fast you look
decides what you can see}

There is no single probability that ``the grid'' detects an island. It
depends on the level at which the load is measured (customer meter,
service transformer, building main, feeder head, substation), and on how
fast the instrument samples. Each level sees the island against a
different background: the higher the level, the more ordinary load
varies around the island. Each sampling rate sees a different part of
the island's signal: monthly or 15-minute energy shows its average load,
one-second readings show its training pulse, and faster readings show
the pulse's harmonics (B.8).

The high-rate tests of B.3 show the effect of level, and of how the
island is wired. A 16-chip island on one phase, read on the phase
unbalance, was found in every hour at the 252 kVA building main and at
the 1.2 MVA section of the campus network (the 2.4 kV side of the
transformer whose two meters are compared in
\hyperref[app-dynamic-precision]{Appendix A.2}) in every season, and at
the 5.5 MVA supply into the campus substation on the long-segment
setting. Spread evenly over three phases, the same island has to be read
on the three-phase total, where the whole load of the line varies around
it, and its pulse is lost much sooner. The logged fine-tune was found
clearly more often than false positives in only one cell, at the
building main. The Llama-8B job, with about three times the swing, was
found in every hour at the building main on the long-segment setting in
two seasons of three. In the earlier test, an island spread over three
phases had to have 2.2--2.3 times the fine-tune's logged swing to be
found in 80\% of hours at the higher threshold on the 724 kVA LBNL bus,
3.2--7.5 times on the 1.2 MVA section and 9--11 times at 5.5 MVA (each
range spans the two ways of splitting the record into a learning half
and a scoring half). So an island that is plain on the meter of the
building it sits in can be invisible a few levels up, unless it sits on
one phase.

The other end of the range of sampling rates is the one screen we ran on
measured meter data. The main text's tests need only interval data; this
screen uses only monthly energy. It is the simplest screen possible: it
flags a meter whose monthly energy rises by 20\% or more and stays up
for six months. It uses no high-rate sampling, no waveform or pulse
analysis, no energy balance and no comparison across meters. We ran it
on 804 meters at 14 sites from the Building Data Genome Project 2
(\hyperref[src-bdgp2]{BDGP2} \citep{miller2020bdg2}), adding to each
meter a constant load the size of one 16-chip island (25.6 kW, about 26
kW at the meter) from January 2017; larger islands are represented by
the same ratio of island load to the meter's own load. Its
\emph{false-negative rate} is the share of meters carrying an island
that it fails to flag. Weighted to the US mix of accounts whose load is
between 1 and 5 times the island's load, it missed 17--26\% of islands,
more for smaller islands.\footnote{The host's own year-over-year fall
  and its seasonal high months break the six-month run.} Single groups
of BDG2 meters, some of only 5--8 meters, ran higher, depending on the
meter's load factor: 55--75\% for accounts whose load is 3--5 times the
island's, and up to 50\% at 1--2 times.

These numbers come from the most naive method possible: a single
threshold on a rise in monthly energy, held for six months, on one meter
at a time. The base data are also thin: one pair of years at 14 mostly
campus sites, and almost no industrial meters (BDG2 holds 6). The rates
are a floor for what grid-side screening can do, not an estimate of it.

The main text's 5× rule grants the evader, a priori, that an island
under 20\% of its host's load is not noticeable, even though a better
screen might sometimes catch it; at 20\% and above it becomes
detectable, often but not always. Even this naive screen flags most
islands there, so the signal is present. The inspection counts, though,
treat every account whose load is under 5 times the island's as screened
out, which assumes a false-negative rate of zero. If the naive screen's
measured false negatives are added back to the pool of accounts that
random inspections are drawn from, random inspections at 95\% come out
1.9 times the main text's counts at 16 chips and 2.2 times at 2,048
chips.\footnote{At 1,024 and 2,048 chips only 1--4 BDG2 meters fall
  between 1 and 5 times the island's load, so the false-negative rate
  there is carried over from smaller islands; at 512 chips the band
  holds 31 meters, which is thin.} Stronger methods are what would close
that gap.

The improvement we expect comes from different data and methods, not
from tuning this screen. The one variant we tried also required the
meter's load factor to rise, since an island runs around the clock. With
no island added, the plain screen flags 17 of the 804 meters and the
variant flags 4 of those 17, so false positives fell about fourfold, but
on so few meters that the figure is rough. The variant also missed
46--69\% of the islands that the plain screen caught, depending on
island size. The candidate methods are energy balance, which cancels the
metered load around a point (\hyperref[app-dynamic-precision]{Appendix
A.5}); the high-rate pulse, read at a level where it is visible (B.3);
and correlation of steps and pulses across meters (B.5). None of these
has yet been measured as a screen with false-positive and false-negative
rates, and we give none.

The screen's false-negative rate at each level decides how far the grid
data can narrow the pool of accounts that inspections are drawn from;
the main text's 5× rule stands in for it. It does not set d, the chance
that an inspection of a host holding an island finds the island (the
main text takes d = 1). Both the right screening threshold and the
number of inspections depend on d
(\hyperref[app-detection-probability]{Appendix C}).

\subsubsection*{B.5 Open instrument
questions}\label{b.5-open-instrument-questions}
\addcontentsline{toc}{subsubsection}{B.5 Open instrument questions}

An evader can try to flatten or hide the pulse of B.3. Three ways have
been described: software filler, power smoothing on the chip, and an
energy buffer at the rack. A crash or stall also interrupts the pulse,
whether or not the evader wants it to. None of these has been measured
against a real installation. Each leaves something behind, and each is a
question about what an instrument has to see. Filler and chip smoothing
leave features smaller and faster than the pulse; a crash leaves a brief
drop as deep as the island's whole load; a rack buffer leaves slower
ones (the turn-on and turn-off, the extra energy, the load factor). We
treat all of them as open research questions for better instruments, not
as settled defeats or settled detections.

\emph{Software filler.} Filler gives the GPUs junk work in the low phase
of each step, so power stays near the high level.
\citet{choukse2025power} describe one such tool at under 5\% of
throughput, plus dedicated CPU cores and host bandwidth for fast power
polling. On the logged fine-tune, holding GPU power at the high level
adds 10.9--11.7\% to GPU energy, about 4\% at the wall. What remains is
the extra energy and one short drop in power per step, lasting as long
as the filler takes to notice the low phase and react. For a 10.2 s
step, that drop keeps about 0.8\% of the pulse's spectral line (the
power at the step frequency that the detector of B.3 looks for) at a 25
ms reaction, 3.1\% at 100 ms and 31\% at 1 s; faster steps keep more.
The drop is only as deep as the step's swing, about 2.4 kW for 16 chips.
The counts we have of natural drops that could be mistaken for it are
for deeper drops. On the LBNL bus at 120 samples per second over 3.1
days, sub-second drops (8--500 ms) of 15 kVA or more ran to hundreds or
thousands a day on the three-phase total and none on the phase
unbalance. How often the grid makes a sub-second drop as small as 2.4 kW
on one phase has not been counted.

\begin{figure}
\centering
\includegraphics[width=0.8\linewidth,height=\textheight,keepaspectratio,alt={Software filler on a square step. Illustration, not a measurement: a square step at the logged fine-tune's levels and timing. Once the filler reacts it holds GPU power at the high level, leaving one short drop per step, as long as its reaction time; the shaded area is the extra energy it burns.}]{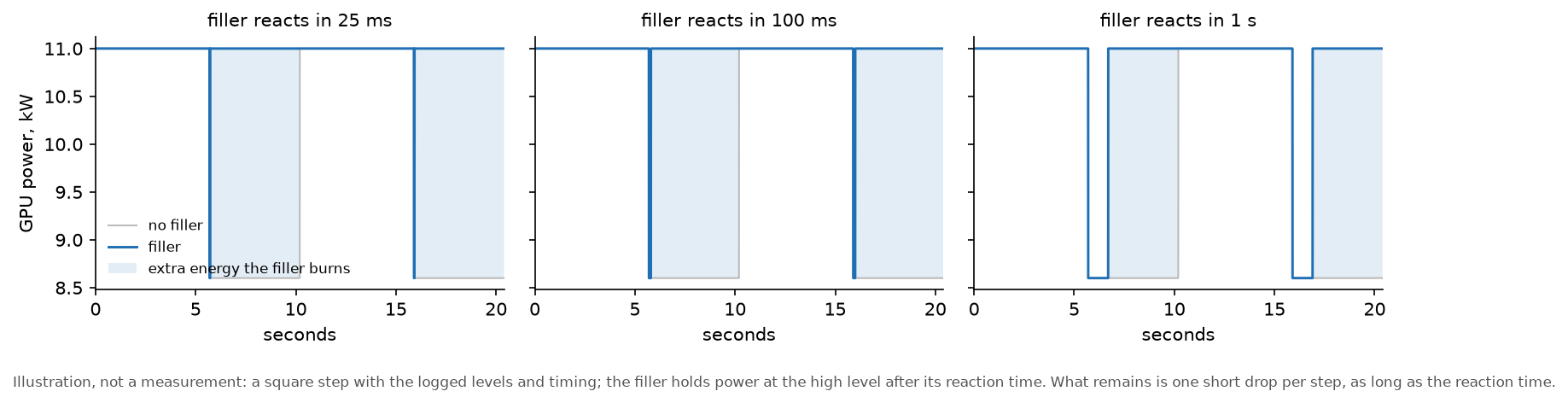}
\caption{Software filler on a square step. Illustration, not a
measurement: a square step at the logged fine-tune's levels and timing.
Once the filler reacts it holds GPU power at the high level, leaving one
short drop per step, as long as its reaction time; the shaded area is
the extra energy it burns.}
\end{figure}

\emph{Smoothing on the chip.} Built-in smoothing on GB200-generation
chips holds GPU power between a floor of at most 90\% of rated power and
a short-peak ceiling of 1.1×, at a simulated energy cost of 10.5\% and a
limited lifetime \citep{choukse2025power}. It is not shown on H100s.
With the floor at its highest setting and a job that peaks at rated
power, the leftover swing at one-second resolution is about 10\% of
rated power, and a lower floor leaves more; a job that uses the full
ceiling leaves at least 20\%, the figure the source gives. Whether the
detector finds that smaller pulse has not been measured. Ramps of about
a second would also weaken the step's harmonics above about 0.5 Hz
(B.8); the source does not give the ramp rates. During a stall the
feature holds the floor for a programmable delay and then ramps down at
a set rate; the longest delay and its energy cost are not published.

\emph{A buffer at the rack.} A capacitor bank or battery behind a
bidirectional converter charges in each low phase and discharges in each
high phase, so the line sees only the mean. The energy moved per step is
small: 5.3 kJ for a 16-chip island with the fine-tune's swing and 11.5
kJ with the Llama-8B job's, integrated from the logged traces. That is
3--8\% of one rated charge and discharge of a 48--51 V supercapacitor
module (up to 13\% for a 32-chip island). What it leaves is the turn-on
and turn-off (to hide those it would have to carry the island for the
whole run), the energy and its conversion losses, and the load factor.
On GB300-class hardware, a power floor plus the 65 J per GPU of storage
built into the GB300 NVL72 power shelves would cover a filler's short
drop at reaction times of 25--100 ms, by our arithmetic on NVIDIA's
figures. How flat a real converter's control loop holds the line current
at 0.1--1 Hz, and what it leaves at the step's harmonics and at 60 Hz,
has not been measured, and we found no peer-reviewed method, with error
rates, that recognises a battery from upstream.

\emph{A crash or stall.} A crash drops the island's load and then
restores it. For one 16-chip island this is rare: Llama 3's 419
unexpected interruptions in a 54-day snapshot on up to 16K GPUs
\citep{grattafiori2024llama3}, scaled per GPU, give about one in 132
days, closer to a lower bound for less carefully managed hardware. At
LBNL, natural drops of 15 kVA or more lasting 8--17 ms occurred 5--7
times a day on one phase (1--3 at 20 kVA), never on the phase unbalance,
and about 3,000 times a day on the three-phase total (about 1,000 at 20
kVA; 3.1 days, one 0.7 MVA bus). The same applies to the step when an
island restarts after pausing for an exchange of updates with the other
islands of a run (the outer exchange of DiLoCo-style training, main
text): the edge can be found, but on its own it does not mark a restart.
A synthetic 20 kVA island that pauses at each exchange had 89\% of its
restart edges found within ±2 s, at a 15 kVA threshold fixed in advance,
against 258 natural up-steps a day of that size on the total (LBNL, 3.2
days); this test ran on one-second means of the LBNL record. A sharper
detector, which counts rises completed within 1/6 s, could not be tried
on the logged jobs, because their one-second means smear every edge.

The candidate answer to all of these is high-rate sampling correlated
across sites: a phase-unbalance monitor at 120 samples per second for
the sub-second drops, and timing coincidence across meters for the
exchange and restart steps of islands that share a run. The baseline for
coincidence is not clean: unrelated loads on nearby meters already step
together more often than chance. On the three-phase total, load steps of
5 kVA or more on neighbouring Caltech building meters fall within ±2 s
of each other about twice as often as they would if the meters were
independent (1,773 against 995 in a week). On the phase unbalance, steps
of 2 kVA coincide 1.3 times as often (31 against 23), and steps of 5 kVA
are too rare to test. The measurements that would settle these questions
are: a wall-side trace of a training job run with filler or smoothing,
per phase at 100 samples per second or faster (the filler's drop lasts
25--100 ms); a count of sub-second drops of about 2.4 kW on the phase
unbalance of a real feeder; the input current of a rack buffer under a
training load; and the timing of exchange steps across the islands of
one run, which no one has recorded in power.

\subsubsection*{B.6 Where the pulse sits and how the detector looks for
it}\label{b.6-where-the-pulse-sits-and-how-the-detector-looks-for-it}
\addcontentsline{toc}{subsubsection}{B.6 Where the pulse sits and how
the detector looks for it}

The detector has to look at the right frequencies, so the first question
is what step rates real training produces. Reported step cadences for
synchronous training run from about once a second to once every tens of
seconds (about 0.02--1 Hz). For one at-scale DGX-H100 job,
\citet{choukse2025power} report spectral energy concentrated between 0.2
and 3 Hz, mostly above the 0.5 Hz top of both bands used here, and they
say the period depends on the scale of the job. Grid modelling studies
assume a training cadence of 0.5--1.5 Hz \citep{kimpson2026detecting}.
The three logged jobs here step every 10.1--16.1 s, about 0.06--0.10 Hz,
and we found no pre-training cadence measured on a 16-chip island. The
sources do not say what sets the step time beyond each job's
configuration. The fine-tune runs a batch of one sample per GPU under
DeepSpeed ZeRO-3 with CPU offload (part of the training state is kept in
host memory and moved to and from the GPUs each step), so its low phase
includes offload traffic; the Llama-8B run (ZeRO-3, no offload) is low
for only about 14\% of each step. Inside each island, DiLoCo-style
training is ordinary synchronous training \citep{douillard2023diloco},
so each island keeps its own step, and outer exchanges that pause
training add a slower on/off swing. No power measurement of a
DiLoCo-style run has been published.

The detector scores each hour by the spectral power above the site's
background in one band of frequencies. It estimates the spectrum by
Welch's method: the hour is cut into overlapping segments, each
segment's spectrum is computed, and the spectra are averaged; longer
segments resolve finer frequencies but give fewer segments to average.
Both settings were fixed before any logged job was added. The
long-segment setting (500 s segments, 0.016--0.3 Hz) comes from the
earlier synthetic-pulse tests at LBNL. The short-segment setting (100 s
segments, 0.08--0.5 Hz) was set for the Caltech data, whose one reading
a second cannot show anything faster than 0.5 Hz. The short setting
holds the fundamental (the step frequency itself) of steps of about
2--12.5 s. The 13.4 s Llama-8B step and the 16.1 s diffusion step fall
below it, and are seen there only through leakage from just outside the
band and through their harmonics. Which setting works therefore depends
on the step time, which a monitor does not know in advance. Running both
settings, with an hour alarming if either passes its own threshold, did
not help in the earlier test: above the building main it added 0--3
false positives in 84 hours and found nothing the long setting missed.
In the earlier test, which setting to headline at each Caltech meter was
chosen after scoring, forced by false alarms on the long-segment setting
at the 252 kVA building main (74 of 84 hours); the re-test table of B.3
reports both settings at every meter.

Each site has \emph{spectral lines} of its own: frequencies at which
some piece of equipment pulses all the time. The LBNL campus has
periodic loads at 1.00 Hz, 3.17 Hz, 1.17 Hz moving to 1.19 Hz, and 10
Hz, which come and go. The detector learns a site's lines while it
learns the background (at LBNL, from the first half of the record; at
Caltech in the re-test, from the first week) and \emph{notches them
out}, ignoring those frequencies when it scores an hour. That can cost
the island if its own step falls on a site line. At LBNL, on the
three-phase total of a 32-chip fine-tune island on the long-segment
setting, the notches removed the job's 0.098 and 0.100 Hz frequency
bins, 51\% of its in-band power, and the job was found in every hour
only through the neighbouring 0.096 Hz bin. A version of the test with
four times as many learning windows notched that bin too, and found the
job in 1\% of hours. A job running about 2\% faster would have sat in
the notched line. The fine-tune's own steps are not perfectly regular.
In its main mode (about 80\% of steps), successive step times are only
weakly correlated (correlation between one step time and the next, −0.19
to 0.06) and spread by 2.7--3.9\%; part of that may be noise in reading
step times from the log. Whether natural or deliberate jitter of that
size costs detection has not been measured.

What else on a line pulses near 0.1 Hz has been checked only on modelled
loads. In currents drawn from appliance models, the training job is the
one load whose comb of lines (the fundamental and its harmonics) stays
fixed for the whole hour. A printer's fuser makes a comb at 1.4 Hz only
while it prints, motor loads put their energy below 0.05 Hz, and a
charger, a crypto mine and a serving load make no line. A measured
feeder-head spectrum for 0.01--60 Hz would replace the models; a limited
search found none.

\begin{figure}
\centering
\includegraphics[width=1\linewidth,height=\textheight,keepaspectratio,alt={Modelled currents of a training job and six ordinary loads, as spectrograms. Illustration, not a measurement: one hour at 10 ms of the RMS current a single-phase 240 V service monitor would read, in ten-minute Hann windows at a five-minute stride, 0.01--50 Hz. The training panel uses the logged fine-tune's step, levels and swing over a 5 kW rest-of-node, so about 16 kW at the wall with no cooling, which makes its swing a larger share of the load than the 9\% of 26 kW in B.3; the other six are appliance models. Only the training job has a comb of lines fixed across all eleven windows; the fuser's comb at 1.4 Hz comes and goes with print jobs.}]{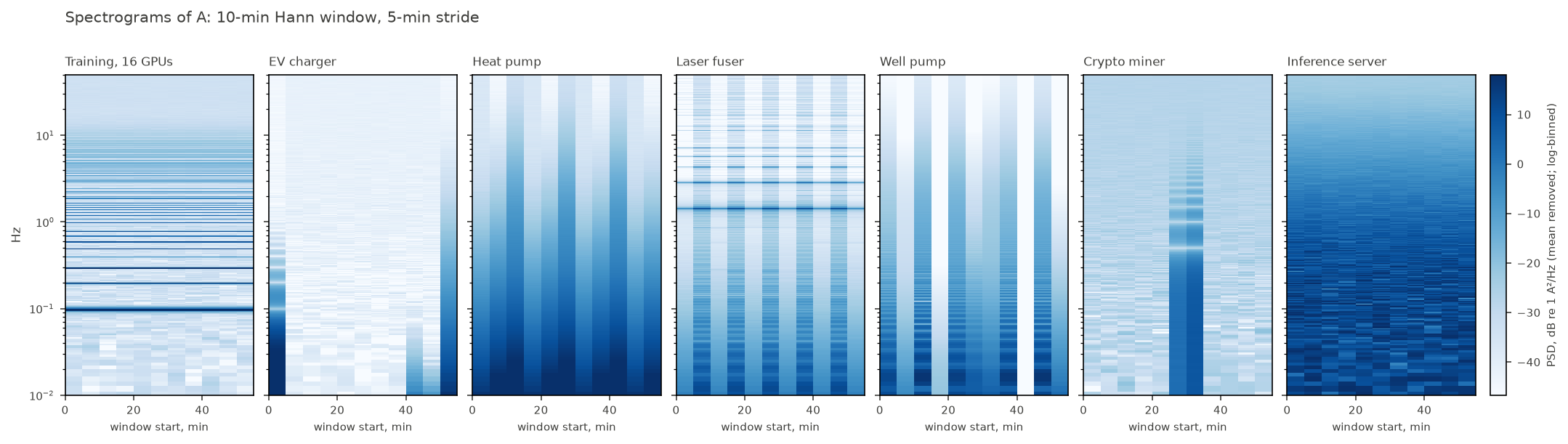}
\caption{Modelled currents of a training job and six ordinary loads, as
spectrograms. Illustration, not a measurement: one hour at 10 ms of the
RMS current a single-phase 240 V service monitor would read, in
ten-minute Hann windows at a five-minute stride, 0.01--50 Hz. The
training panel uses the logged fine-tune's step, levels and swing over a
5 kW rest-of-node, so about 16 kW at the wall with no cooling, which
makes its swing a larger share of the load than the 9\% of 26 kW in B.3;
the other six are appliance models. Only the training job has a comb of
lines fixed across all eleven windows; the fuser's comb at 1.4 Hz comes
and goes with print jobs.}
\end{figure}

\subsubsection*{B.7 Pulse shape, other workloads and the smallest swing
found}\label{b.7-pulse-shape-other-workloads-and-the-smallest-swing-found}
\addcontentsline{toc}{subsubsection}{B.7 Pulse shape, other workloads
and the smallest swing found}

A job's \emph{swing} is how far its power drops and recovers in each
step, measured as the 95th minus the 5th percentile of the island's GPU
power; the logged fine-tune's is 2.32--2.45 kW for 16 chips. The
detector keys on the step frequency, so what counts is how much of the
swing lands in the fundamental. A square pulse keeps about 80\% of its
variance there, and a pulse with a short dip keeps less. To compare jobs
we ask how large a swing each needs to be found in 80\% of hours. (The
80\% level is a convention, used here only to compare jobs.) On the LBNL
bus, at the setup of the table below, the Llama-8B run, low for only
about 14\% of each step, needs 2.0--2.5 times the fine-tune's swing in
kW, close to the 2.3 its shape predicts. Measured against its own logged
swing it needs a little more than the fine-tune (0.26--0.37 against
0.20--0.24).

The table gives, for each job, the swing needed to be found in 80\% of
hours, as a fraction of the job's own logged swing. It is from the
earlier test, which split the LBNL record into a learning half and a
scoring half, read on the phase unbalance on the long-segment setting,
at the higher threshold (twice the loudest no-job hour of the learning
half). A fraction below 1 means the job as logged has room to spare.

\begin{longtable}[]{@{}
  >{\raggedright\arraybackslash}p{(\linewidth - 4\tabcolsep) * \real{0.2727}}
  >{\raggedleft\arraybackslash}p{(\linewidth - 4\tabcolsep) * \real{0.3636}}
  >{\raggedleft\arraybackslash}p{(\linewidth - 4\tabcolsep) * \real{0.3636}}@{}}
\caption{Swing needed to find each logged job in 80\% of hours, as a
fraction of the job's own logged swing, LBNL 12 kV bus, phase unbalance,
long-segment setting, higher threshold of the earlier
test.}\\*
\toprule\noalign{}
\begin{minipage}[b]{\linewidth}\raggedright
Job
\end{minipage} & \begin{minipage}[b]{\linewidth}\raggedleft
As logged
\end{minipage} & \begin{minipage}[b]{\linewidth}\raggedleft
At 16 chips
\end{minipage} \\*
\midrule\noalign{}
\endfirsthead
\toprule\noalign{}
\begin{minipage}[b]{\linewidth}\raggedright
Job
\end{minipage} & \begin{minipage}[b]{\linewidth}\raggedleft
As logged
\end{minipage} & \begin{minipage}[b]{\linewidth}\raggedleft
At 16 chips
\end{minipage} \\*
\midrule\noalign{}
\endhead
\bottomrule\noalign{}
\endlastfoot
Llama-2 70B LoRA fine-tune (logged at 16 GPUs) & 0.20--0.24 &
0.20--0.24 \\*
Llama-8B training run (logged at 8 GPUs) & 0.26--0.37 & 0.13--0.18 \\*
Diffusion, 107M parameters (logged at 8 GPUs) & 0.29--0.42 &
0.15--0.21 \\*
\end{longtable}

The smallest swing found in 80\% of hours in these tests is the
fine-tune's on the LBNL bus: 0.50--0.59 kW, about a fifth of its logged
swing and about 2\% of the island's 26 kW at the meter. The swing needed
in kW barely changes with island size, so the fraction halves from 8 to
16 chips, and a larger island is only easier to find. The weak case is
the diffusion run on the short-segment setting, at 0.59--0.96 of its
8-GPU swing (0.29--0.48 of its 16-chip swing), because its 16 s step
falls below that band. Two workloads from one lab are not independent
tests: the five NLR fine-tune runs are near-identical, and the NLR
Stable Diffusion run is presumably found through its swinging blocks,
which the source paper attributes to evaluation, since its long blocks
are flat at one-second resolution.

How the pulse would look at other island sizes rests on scaling. Every
island here except the 16-chip fine-tune, which NLR logged directly, is
a logged trace multiplied, with the same per-GPU swing, step and duty
cycle (the share of each step spent in the high phase). That linear
scaling is not expected to hold for large islands, where
\citet{choukse2025power} report the low phase can fall close to idle and
the period changes with scale. The data hold no language-model
pre-training job, no DiLoCo-style job and nothing faster than about 1
Hz. The test that would settle this is wall-side traces of several
training styles on 8--64 chips, per phase, at 10 samples per second or
faster, with step and exchange times logged.

\subsubsection*{B.8 Harmonics, and what 120 samples per second
adds}\label{b.8-harmonics-and-what-120-samples-per-second-adds}
\addcontentsline{toc}{subsubsection}{B.8 Harmonics, and what 120 samples
per second adds}

A step that switches sharply between two levels puts power not only at
its own frequency (the fundamental) but at whole multiples of it (the
harmonics). The job logs are one-second means, so above about 0.5 Hz
they are attenuated and cannot show the pulse as it is at the wall. A
µPMU at 120 samples per second can see the harmonics, and we tested what
that adds in two ways.

The first is to sum harmonics. On the LBNL record, with 600 s segments,
summing ten harmonics of the step frequency gains at most about 2 times
in amplitude over the best single frequency bin, and most of that gain
comes from the fundamental's power spilling into a second, adjacent bin.
A harmonic-sum (comb) detector, which adds the power at the step
frequency and its multiples, was run on the LBNL three-phase total at
120 samples per second, with 60 s segments and 5 or 10 harmonics. It did
worse than the one-second long-segment detector: it needed 6.5 times the
fine-tune's logged swing and 7.7 times the Llama-8B run's 8-GPU swing at
the higher threshold, against 2.3 times for the one-second detector on
the fine-tune. Part of the loss is the 60 s segments: on them, the plain
best-bin score needed 4.5 times. A comb on 500 s segments has not been
run.

The second is the simulation that B.3 mentions. We rebuilt the
fine-tune's pulse as a two-level wave with instant edges at its own 578
logged edge times, which keeps its step-to-step jitter, and added it to
five minutes of the LBNL phase-1 unbalance. Averaged as the logger
averages, the rebuilt pulse reproduces the logged spectrum up to about 1
Hz, so the simulation is anchored to the data below that. On the 60 s
windows of the figure below, harmonics up to about 1 Hz stand clear of
the background in a busy weekday hour (7 of the first 10 multiples), and
up to about 1.8 Hz in a quiet hour. Above about 2 Hz the jitter smears
the lines, and longer windows stop helping. Some of that jitter is error
in estimating edge times from one-second means; if the real edges are
steadier, the higher harmonics would do better. Edge shape matters
little: a 30 ms rise filters the pulse above about 5 Hz (one over 2 pi
times the rise time), so millisecond rises weaken only harmonics above
about 5 Hz (this was not simulated), and only ramps of about a second,
as built-in smoothing might impose, would weaken the comb above about
0.5 Hz.

\begin{figure}
\centering
\includegraphics[width=1\linewidth,height=\textheight,keepaspectratio,alt={The rebuilt fine-tune pulse on the LBNL bus at 120 samples per second. Simulation on real background: the logged fine-tune's step rebuilt with instant edges and added to the LBNL phase-1 unbalance from 150 to 450 s, as a µPMU's current magnitudes would show it. Rows: a quiet hour (05:54) and a busy weekday hour (09:11). Columns: background only, with the job, and the difference in dB. Bottom: the spectrum over the 300 s the job is on, with and without it; orange lines mark multiples of the step frequency. The background's own lines at about 1.17 and 2.3 Hz are not the job.}]{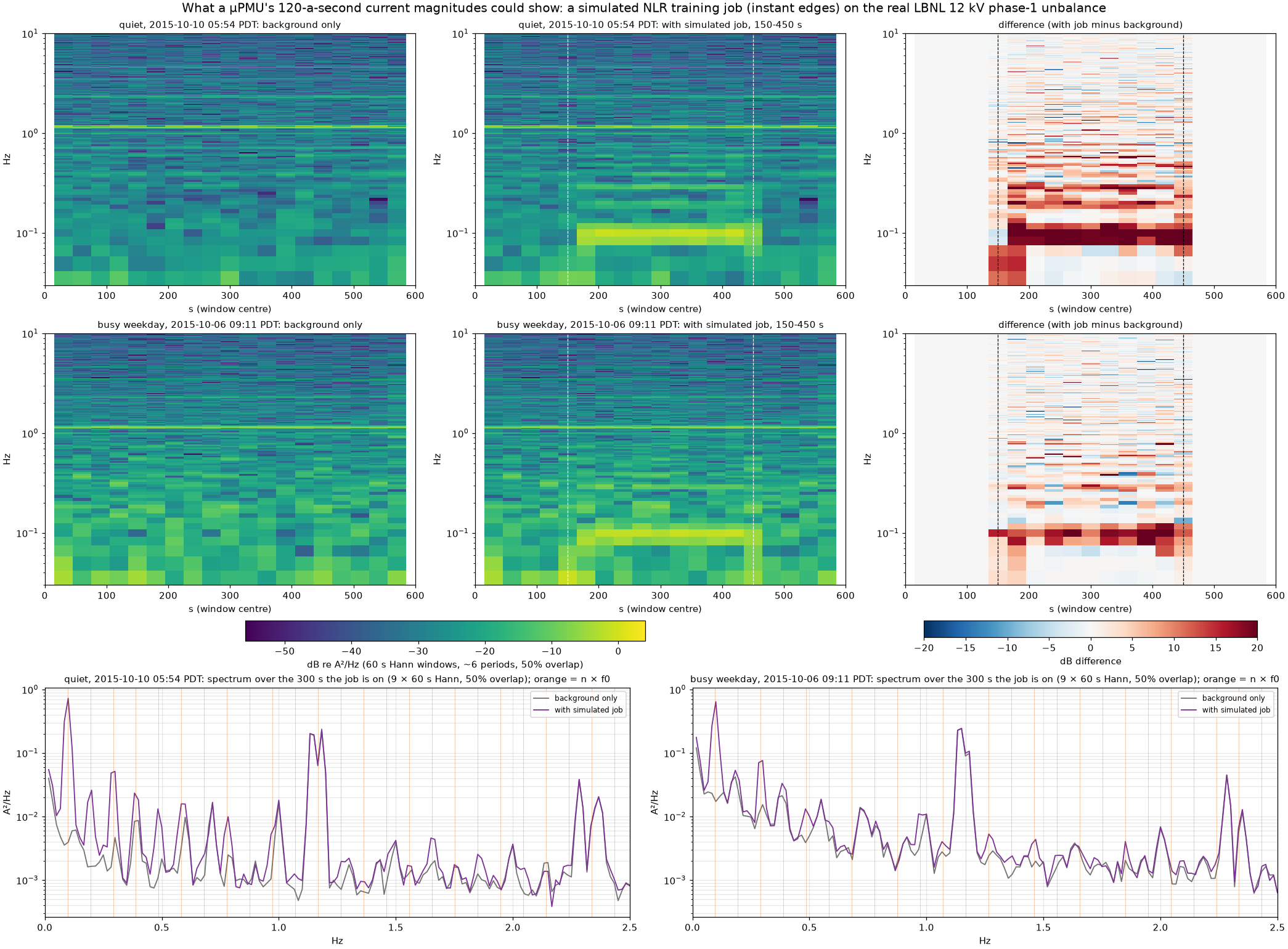}
\caption{The rebuilt fine-tune pulse on the LBNL bus at 120 samples per
second. Simulation on real background: the logged fine-tune's step
rebuilt with instant edges and added to the LBNL phase-1 unbalance from
150 to 450 s, as a µPMU's current magnitudes would show it. Rows: a
quiet hour (05:54) and a busy weekday hour (09:11). Columns: background
only, with the job, and the difference in dB. Bottom: the spectrum over
the 300 s the job is on, with and without it; orange lines mark
multiples of the step frequency. The background's own lines at about
1.17 and 2.3 Hz are not the job.}
\end{figure}

High-rate data also reach the harmonics of 60 Hz, at 180 Hz and above.
These are a separate signal: they come from the island's power supplies,
which draw current in a distorted wave, rather than from its training
step. The Caltech meters capture the waveform at 2.5 kHz, which resolves
frequencies up to about 1.25 kHz, for one second in every ten. One
capture resolves a current harmonic to about 0.04--0.10 A (a spread
between captures, so an upper bound). At the 252 kVA building main the
fifth harmonic is already 13\% of the 60 Hz current and varies by about
3.4 A (one standard deviation) across 60 one-second captures, so an
island has to clear that capture-to-capture spread, not the total. A
supply with power-factor correction shapes its current toward a sine
wave: one 1 kW reference design has total harmonic distortion of 1.4\%
at full load and about 2--7\% at a third to half load
\citep{ti2016tida00707}. We found no measurement of how much a server
supply puts at each harmonic, so whether a corrected island clears the
spread at the 11th and 13th harmonics is open. The same supplies switch
at 100--140 kHz in that reference design, which puts their distinctive
emissions in the 2--150 kHz band, out of reach of the meters in hand; no
measurement of a server supply there was found either. No harmonic
injection was run, and the background is one ten-minute set on one
weekday.

\subsubsection*{B.9 Controls and false-positive
calibration}\label{b.9-controls-and-false-positive-calibration}
\addcontentsline{toc}{subsubsection}{B.9 Controls and false-positive
calibration}

Every detection rate is paired with a false-positive count on the same
hours without the job. The false positives therefore do not depend on
the pulse shape, and they come out identical for every job scored on a
meter, which guards against a fault in the computation. The re-test's
March data are a fresh download, checked identical to the stored copy by
its SHA-256 hash.

The textbook threshold failed. The Follum--Pierre periodogram test
\citep{follum2016detection} sets its threshold from a target
false-positive rate by assuming the ambient noise is stationary: that
its spectrum is the same every hour, so that each hour's measured
spectrum, divided by the long-run ambient spectrum, follows a chi-square
distribution. We used it in a Welch-averaged form, with the ambient
spectrum taken from the learning hours, aimed at 10⁻³ false positives an
hour. It predicts 0.12 false-positive hours in 120 at LBNL and 0.084 in
84 at Caltech, and gave 37--110 and 27--84. The same thresholds fail on
the hours their spectra came from (40--95 and 31--77), so learning on
hours other than those scored is not the main cause. The cause is that
the ambient noise changes from hour to hour. We measure that by the
excess log-variance: for each frequency bin, the variance across hours
of the log of the hour's spectrum over the ambient spectrum, minus the
part that chi-square sampling alone would give. A stationary background
gives 0 dB; on the three-phase total it is 3.0--4.0 dB at LBNL and
1.9--6.1 dB at Caltech. PNNL's correction for correlated channels
\citep{follum2015multichannel} lowers the count on the phase unbalance
by at most about 20--30 hours. Every rate in this appendix therefore
uses thresholds learned from each meter's own earlier hours. Only one
statistic was tested, and PNNL's online form, which estimates the
ambient spectrum from the hour under test with a median filter, was not
run; what failed is the stationary assumption with an ambient spectrum
estimated in advance, not forced-oscillation detection in general.

\begin{figure}
\centering
\includegraphics[width=1\linewidth,height=\textheight,keepaspectratio,alt={False positives against the stationary prediction. Left: false-positive hours that the stationary chi-square null predicts at 10⁻³ per hour (black dashes) against what each band and direction produced on hours without the job; the gap is two to three orders of magnitude at every meter. Right: the hour-to-hour change in ambient noise on the three-phase total; a stationary background would sit at 0 dB.}]{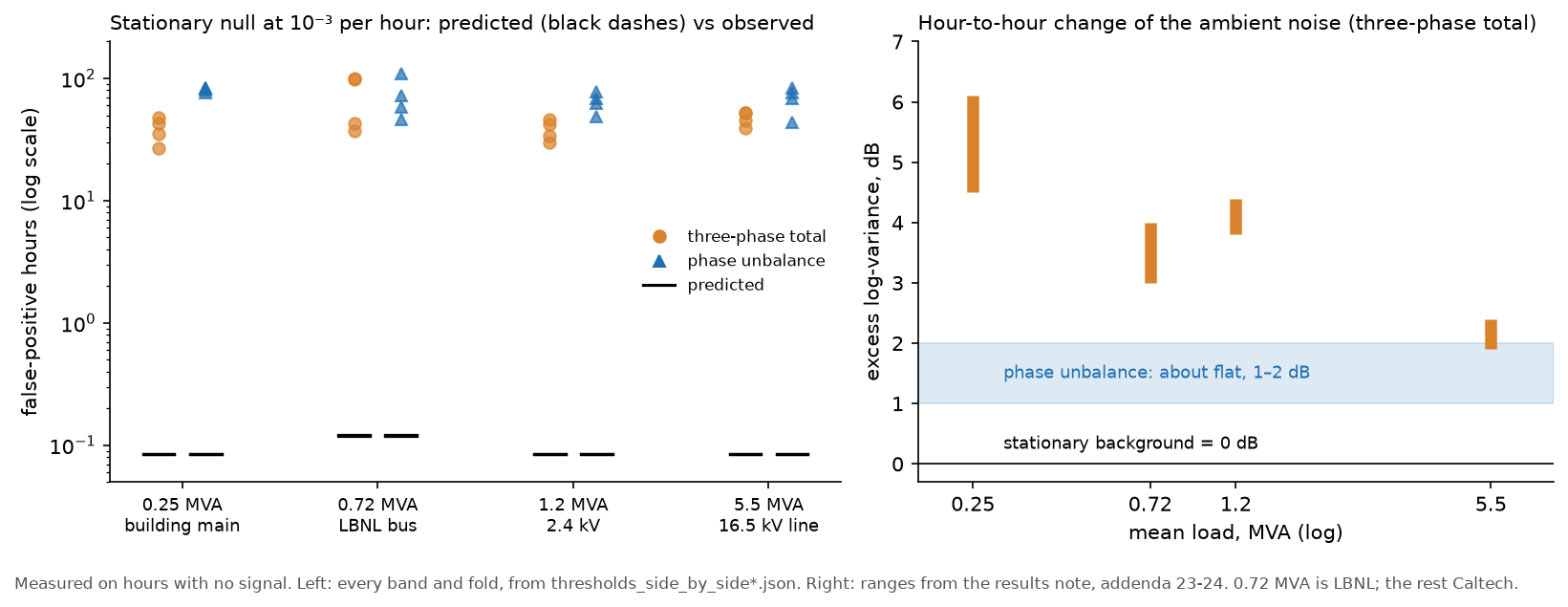}
\caption{False positives against the stationary prediction. Left:
false-positive hours that the stationary chi-square null predicts at
10⁻³ per hour (black dashes) against what each band and direction
produced on hours without the job; the gap is two to three orders of
magnitude at every meter. Right: the hour-to-hour change in ambient
noise on the three-phase total; a stationary background would sit at 0
dB.}
\end{figure}

Thresholds learned from the data have failure modes of their own, and
the tests were rebuilt twice to avoid them. The project's first
thresholds were set on the same 39 hours they were then scored on; on
202 new hours they raised false positives in 118, 121 and 41 hours, and
the figures built on them were withdrawn. The next design, the earlier
test quoted in B.3 and B.7, split the record into halves. At Caltech it
split one March week into halves of 3.5 days, which put the whole
weekend in one scoring half. It also set the threshold from the half
that had built the detector's baseline, which tends to set thresholds
low: its false-positive counts err high, and its detection rates err
high too. The re-test of B.3 replaced it with separate learning,
calibration and scoring weeks. A threshold set at the loudest
calibration hour can be thrown by a single event. One real switching
step of about 970 kVA at the 5.5 MVA meter on 2025-03-20 set it so high
that the turn-on was not found there in March. (An island split unevenly
across the phases, as NVIDIA's design guide lays out the power supplies
of a single rack, which balances the phases only across two racks, was
not found at 5.5 MVA in any season.) The threshold at the 7th-loudest
hour, used for the one-alarm-a-day target, is not thrown by one event.
In the re-test, 21 of 34 usable cells came within a factor of two of the
13.9 false positives expected, 8 ran well above (31--178) and 5 below
(0--6). Small counts bound little: no false positive in 336 hours bounds
the hourly rate below about 0.9\% at 95\% confidence.

A national screen needs a far lower rate than one a day. At one alarm a
day per line, 200,000 feeders raise about 6 × 10⁶ alarms a month. One
alarm a month in total needs 0.8--5 × 10⁻⁶ per line-month (from 1.22
million accounts averaging at least a 16-chip island's 25.6 kW,
i.e.~every account that could carry a bare island, not only the larger
hosts that could cloak one, down to 200,000 feeders), about seven to
seven and a half orders of magnitude below one a day. The plain detector
found the single-phase pulse at every meter, so method work there is
justified: a time-varying whitening model fitted before the spectral
test (one that tracks how the background spectrum changes through the
day), thresholds conditioned on hour of day, PNNL's online ambient
estimate, and evidence accumulated across hours. Accumulation helps only
if false positives are independent across hours and channels and their
rate is known, and our data show neither. A year of data per meter would
test both at the rates measured here. The far lower rate a national
screen needs cannot be checked directly: even 10⁻³--10⁻⁴ false positives
per line-month would take 80--800 meter-years of data to show a single
one, so that rate would rest on pooling meters or on extrapolation.

\subsection*{Appendix C. Inspections when a visit can miss the
island}\label{app-detection-probability}
\addcontentsline{toc}{subsection}{Appendix C. Inspections when a visit
can miss the island}

\subsubsection*{C.1 The visit-success probability
d}\label{c.1-the-visit-success-probability-d}
\addcontentsline{toc}{subsubsection}{C.1 The visit-success probability
d}

The main text assumes that an inspection of a host that holds an island
always finds it. Here we relax that assumption. Let d, the visit-success
probability, be the probability that an inspection of a host that does
hold an island discovers it. A visit can fail because the evader powers
the island down or moves it before the inspector arrives, because the
racks are disguised as ordinary servers, or because the inspector simply
misses them. In the language of inspection games, d is the detection
probability of a single inspection, 1 − β, where β is the probability
that an inspection misses a violation that is present: a false negative,
which \citet[ch.~1]{avenhaus1996compliance} and
\citet[ch.~6]{avenhaus2020inspection} call an error of the second kind.
d is not the detection goal of the main text (95\%). The detection goal
is the chance that the whole two-year campaign of inspections finds at
least one island of the run; d is the chance that one visit to the right
host succeeds.

Nor is d the catch rate of the grid screen. The screen decides which
hosts are worth visiting at all. The main text's rule that a candidate
host must average at least 5× the island's load stands in for that
screen, and Appendix B.4 shows that the share of islands a screen
catches depends strongly on where in the grid, and how fast, the load is
measured. d describes what happens once the inspector is at the door,
and the two effects multiply (B.1). Nothing in this paper measures d,
and nothing in the grid data sets it, so we take d = 1, 0.75 and 0.5 to
show how sensitive the counts are to it. d changes the number of
inspections (C.2 to C.4). It also changes how far the screen should
narrow the pool before inspectors are sent: checking a host the screen
has flagged costs a visit whatever d is, while the random visits needed
to cover the unflagged hosts grow as d falls, so the best balance
between flagging and random visits shifts with d.

Throughout, each visit is treated as an independent chance to find the
island, so a host visited twice hides its island with probability (1 −
d)\textsuperscript{2}. In practice misses at the same host are likely
correlated; C.3 says what that changes.

\subsubsection*{C.2 Runs spread over many
sites}\label{c.2-runs-spread-over-many-sites}
\addcontentsline{toc}{subsubsection}{C.2 Runs spread over many sites}

Suppose a run needs S island sites, one island per host, hidden among N
candidate hosts (the sites large enough to cloak an island of that
size). As in Table 1, the inspector is taken to know the island size,
and so which pool to draw from. Over the two-year run the inspector
makes k inspections, drawn at random from the pool without returning to
a host already inspected. Each island's host is then inspected with
probability k/N, and an inspected island is found with probability d, so
all S islands escape with probability about (1 −
d·k/N)\textsuperscript{S}. Setting that equal to 1 − g, where g is the
detection goal, gives

k = N·{[}1 − (1 − g)\textsuperscript{1/S}{]} ÷ d, inspections a year = k
÷ 2.

The bracket is the share of the pool that must be inspected for a chance
g of a find when every visit succeeds; at d = 1 the formula gives the
counts in Table 1. Each inspection of an island's host now succeeds only
with probability d, so the inspector needs 1/d times as many inspections
for the same chance of a find: ×1.33 at d = 0.75 and ×2 at d = 0.5, at
any detection goal.

This 1/d rule holds only while k ≤ N, that is, while a single pass over
the pool, visiting each host at most once, is enough. That requires 1 −
(1 − g)\textsuperscript{1/S} ≤ d.~At 95\% detection the rule therefore
needs more than two sites at d = 0.75 and more than four at d = 0.5. In
the table of C.4 it fails only for the three runs on one or two sites
(2,048-chip islands at 10\textsuperscript{25} and 10\textsuperscript{26}
FLOP, 128-chip islands at 10\textsuperscript{25}). There k exceeds N at
d \textless{} 1, the inspector has to return to hosts, and C.3 takes
over. Every other cell, including 16-chip islands at
10\textsuperscript{25} (13 sites) and 2,048-chip islands at
10\textsuperscript{27} (12 sites), is the d = 1 count times 1.33 or 2,
to within the rounding of the printed counts.

\subsubsection*{C.3 Runs on only a few
sites}\label{c.3-runs-on-only-a-few-sites}
\addcontentsline{toc}{subsubsection}{C.3 Runs on only a few sites}

The 1/d rule fails when a run needs so few sites that one pass over the
pool is not enough. In the table of C.4 these are the runs on one or two
sites (2,048-chip islands at
10\textsuperscript{25}--10\textsuperscript{26} FLOP, 128-chip islands at
10\textsuperscript{25}). Off the table, at d = 0.5, it also fails for
runs on three or four sites, such as 64-chip islands at
10\textsuperscript{25} (4 sites) and 1,024-chip islands at
10\textsuperscript{26} (3 sites).

The reason is the following. With many islands, each visit is a fresh
draw that lands on some island's host with a small probability, and d
simply scales that probability down. With a single island, the inspector
at d = 1 needs to visit only 95\% of the candidate hosts, once each, in
two years. At d \textless{} 1 a visit to the island's host can fail.
Since the inspector does not return to a host within a pass, that host
comes up again only on the next pass over the pool, so the inspector
must sweep the pool more than once. The chance of missing the island
after n full passes and a fraction f of the next is (1 −
d)\textsuperscript{n} (1 − f·d). Within the last, partial pass the
chance of a miss falls in proportion to f rather than geometrically, so
the number of passes needed is somewhat larger than the shortcut ln(1 −
g) ÷ ln(1 − d). For one site at 95\% detection the inspector needs 2.27
passes at d = 0.75 and 4.40 at d = 0.5. That is 2.39 and 4.63 times the
d = 1 count (0.95 of a pass), against the 1.33 and 2 times of the 1/d
rule. For two sites it needs 1.14 and 2.21 passes, 1.47 and 2.85 times
the d = 1 count.

Runs on very large islands nevertheless stay cheap to find, because few
hosts are large enough to hold them. At d = 0.5 the single-site
2,048-chip runs at 10\textsuperscript{25} and 10\textsuperscript{26}
FLOP need only 1,525 inspections a year, the fewest of the three island
sizes in the table. Off the table, at 10\textsuperscript{26} FLOP, a
1,024-chip run on three sites needs fewer once d \textless{} 1: 717 a
year at d = 0.75 and 1,300 at d = 0.5 (against 538 at d = 1, when the
2,048-chip run is the cheaper).

These figures treat each visit as an independent chance to find the
island. In practice misses at the same host are likely correlated, since
an island well hidden from one inspection is likely well hidden from the
next. That changes the game: repeat passes buy less than these figures
assume, so the counts for runs on a few sites are optimistic, and an
island hidden perfectly is not found by any number of passes.

\subsubsection*{C.4 Inspections by visit-success
probability}\label{c.4-inspections-by-visit-success-probability}
\addcontentsline{toc}{subsubsection}{C.4 Inspections by visit-success
probability}

The table collects the counts. Every column assumes the inspector knows
the island size, as Table 1 does. If the size is unknown, the inspector
must cover all sizes at once (Table 2); that schedule includes the one-
and two-site sizes that need repeat passes, so at 10\textsuperscript{25}
FLOP it grows by more than 1/d (×1.47 at d = 0.75 and ×2.42 at d = 0.5).
At 10\textsuperscript{25} FLOP with 16-chip islands, both d \textless{}
1 counts (32,149 and 48,224) are above the roughly 30,000 inspections
OSHA makes a year; every other cell stays below it.

\begin{longtable}[]{@{}lrrrr@{}}
\caption{Number of annual inspections required to find at least one
covert island within two years, by the chance d that a visit to an
island's host finds the island. Compute-bound evader, 95\% detection, 1
Gbps per island, island size known to the inspector. The d = 1 rows
equal the compute-bound rows of Table 1, and the number of sites per run
is the same in every row.}\\*
\toprule\noalign{}
FLOP & d & 16 chips & 128 chips & 2,048 chips \\*
\midrule\noalign{}
\endfirsthead
\toprule\noalign{}
FLOP & d & 16 chips & 128 chips & 2,048 chips \\*
\midrule\noalign{}
\endhead
\bottomrule\noalign{}
\endlastfoot
10²⁵ & 1 & 24,112 & 8,239 & 329 \\*
10²⁵ & 0.75 & 32,149 & 12,105 & 785 \\*
10²⁵ & 0.5 & 48,224 & 23,464 & 1,525 \\*
10²⁶ & 1 & 2,649 & 1,627 & 329 \\*
10²⁶ & 0.75 & 3,532 & 2,169 & 785 \\*
10²⁶ & 0.5 & 5,297 & 3,254 & 1,525 \\*
10²⁷ & 1 & 267 & 181 & 77 \\*
10²⁷ & 0.75 & 356 & 242 & 102 \\*
10²⁷ & 0.5 & 534 & 362 & 153 \\*
\end{longtable}

\subsection*{Appendix D. Building counts and real-world detection of
hidden loads}\label{app-notes}
\addcontentsline{toc}{subsection}{Appendix D. Building counts and
real-world detection of hidden loads}

This appendix supports two parts of the main text. D.1 gives the details
behind the bare-island screen: how the building counts and load factors
were computed. D.2 and D.3 support the claim that utilities already find
hidden loads of about an island's size: D.2 lists the cases we found,
with sources, and D.3 describes how utilities find and trace
non-technical losses (energy delivered but not billed) and how small a
loss they can see.

\subsubsection*{D.1 Bare islands: building counts and load
factors}\label{d.1-bare-islands-building-counts-and-load-factors}
\addcontentsline{toc}{subsubsection}{D.1 Bare islands: building counts
and load factors}

The building count matters because it is the population a bare island's
own meter has to blend into. An island on its own meter is a commercial
account averaging about 26 kW around the clock, so the question is how
many ordinary buildings look like that.

A building's average load is its annual electricity use divided by 8,760
hours, from the CBECS 2018 public-use microdata. CBECS is a sample
survey, and EIA publishes replicate weights with it (the jackknife
method: the estimate is recomputed many times with parts of the sample
reweighted, and the spread of those recomputations gives the standard
error); our standard errors come from them. The 25 kW threshold is the
round number just below the 16-chip island's 26 kW.

The main text counts buildings whose \emph{median} monthly load factor
exceeds 0.9, that is, buildings that are flat in a typical month. A
looser test asks whether a building is flat in even one month of the
year, by using each building's \emph{highest} monthly load factor
instead. On that test, 4.0\% of buildings averaging 50 kW or more pass
0.9 in BDGP2 (32 of 796) and 0.13\% in ComStock (about 640). An island
is flat in every month, so the looser test shows how many buildings
could pass for an island at some point in the year, not how many look
like one all year. CBECS reports annual and not hourly use, so it cannot
give load factors.

CBECS 2018 has no data-center building type. A standalone data center,
if the survey sampled one, would have to be coded under another type;
``Other'' is the likeliest, but we have not confirmed this. Data-center
rooms inside other buildings are recorded: 116,000 ± 15,000 buildings
report one. We found no EIA statement on whether standalone data centers
are sampled.

\subsubsection*{D.2 Real-world discovery: sources and
details}\label{d.2-real-world-discovery-sources-and-details}
\addcontentsline{toc}{subsubsection}{D.2 Real-world discovery: sources
and details}

This section lists the cases behind the main text's ``Real-world
discovery'' paragraphs: what was hidden, how large each site was, and
what found it. Power figures in the text and in the table at the end are
averages, from energy ÷ hours (730 h a month, 24 h a day). Where a
source gives energy only, the kW figure is our arithmetic, not the
source's.

British Columbia, in brief (details under Canada below). The site count
in the main text is our arithmetic. BC Hydro's 850 GWh a year of theft
divided by 20--32 kW of lighting per stealing site, from police seizure
files, gives about 3,000--4,800 sites; counting ventilation and cooling
as well would lower that to roughly 1,200--1,700. Provincial regulation
required a system that could ``estimate the amount of electricity
supplied from a portion of the electric distribution system to unmetered
loads that are not known to the authority and to estimate the location
of those loads'' \citep[s. 2(d)]{bcreg3682010}. BC Hydro built it from
9,116 measurement points by December 2015, mostly devices already on its
grid: SCADA devices (the switches, relays and reclosers that the
utility's supervisory control and data acquisition system reads
remotely), meters on distribution transformers, and check meters (meters
a utility installs on its own lines to compare against customer meters)
\citep[Table 1]{bchydro2016smiqr23}. The grid-side part came late in the
theft reduction: theft had been ``declining since 2012'', BC Hydro
``commenced integrating SCADA and check meter devices into the theft
solution'' in October 2014 \citep[p.~2]{bchydro2015smiqr20}, introduced
``Energy Balancing capabilities across distribution feeders'' in fiscal
2016 \citep{bchydro2016smiqr23}, and its vendor's release for ``full
energy balancing'' was due late in 2016 \citep{emc2016bchydro}. The 80\%
cut is BC Hydro's credit to its smart-meter programme as a whole, so the
feeder-level balance can claim little of the early reduction, which came
from the smart meters and from meter and transformer analytics. An
energy balance finds only energy that no meter bills
(\hyperref[app-notes]{Appendix D.3}), so it complements the meter
screens of the main text: an island that bypasses its meter shows up in
the balance, and one that pays on its own meter shows up on that meter.
Neither sees an island that pays through a larger host's meter, which is
the cloaked case.

Russia. The TASS quote \citep{tassndnorthcaucasus} is from Deputy Energy
Minister Pyotr Konyushenko at the Caucasian Investment Forum, who put
illegal mining in the North Caucasus at about 300 MW; the article is
dated 2 May with no year shown, and it mentions the 2025--2031 regional
bans, so it is from 2025 or later. The remarks of Deputy Energy Minister
Pavel Snikkars come from an Izvestia interview reported on 12 December
2022 \citep{izvestia2022snikkars}; Izvestia blocked our fetch, so we
read them in Habr's report of it \citep{habr2022miners}. Habr quotes him
as saying miners are found ``by growth in consumption and change in the
load profile'', that Irkutsk engineers began doing this ``from load
estimates at substations'', and that suspicion arises when a resident
``consumes more than 30--40\% of the apartment building's consumption''
or from ``a sharp rise in electricity use''; an inspector then visits,
and the miner is either rebilled at the business tariff or sued. This is
the only Russian detection criterion with a number that we found, and it
is relative, not a fixed kWh or kW level.

Kommersant \citep[8 December]{kommersant2021irkutsk} reports that
Irkutskenergosbyt, the Irkutsk regional electricity retailer, had found
more than 1,000 cases of ``grey'' (unregistered, household-tariff)
mining since the start of that year and filed 85 lawsuits worth 73.3
million roubles, of which it had won nine. It gives the Plishkino house,
148.5 m², at 193,000 kWh a month, ``about 22\% of the consumption of the
whole village''. Taken as a monthly average that would be about 264 kW,
above the 150 kW limit of the household connection category in Russia
\citep[para. 14]{russia2004decree861}; the article mentions no
connection, transformer or substation, so we do not use the figure. If
the figure is in fact annual, the average is about 22 kW. Kommersant
does not say how the cases were found. By August 2023 Irkutskenergosbyt
had filed 1,491 suits over five years, 855 of them granted \citep[citing
TASS]{forklog2023irkutsk}. Another Irkutsk cottage used up to 90,000 kWh
a month, ``20 times the settlement average'', and the court accepted a
site inspection and the usage statistics as enough evidence
\citep{fontanka2025energolimit}; as a monthly average that would be
about 123 kW, within the 150 kW household category, though the article
gives no connection details. On 11 August 2025 Irkutskenergosbyt
estimated ``more than 8,000'' suspected illegal mining points in the
region, about 1,500 in Irkutsk and 1,700 in Irkutsk district, and said
the real number is ``significantly higher'' \citep{ria2025irkutsk}; the
article does not say how they were counted.

The Dagestan figure is from ComNews \citep[24
July]{comnews2026dagestan}. In an experiment in Untsukul district,
Dagenergo's engineers cut the internet without warning in two villages,
and ``the load on the grid fell by 3.2 MW, which corresponded to about
900 devices''; 3.2 MW ÷ 900 ≈ 3.6 kW. The same article says cases are
usually found after ``an abnormal rise in load on a particular section
of the network'' or residents' complaints about outages, and that
finding one takes ``from several weeks to several months''. ForkLog
\citep[26 January]{forklog2024makhachkala} reports that Makhachkala
police and Dagenergo staff, ``working through complaints from residents
of an apartment building about voltage surges'', found 620 ASIC miners
(machines built only for cryptocurrency mining) in a garage in the
building's underground parking, connected around the meters; the owner
is accused of stealing 1.75 million kWh, worth more than 12.2 million
roubles, and faces up to five years in prison. No running time is given,
so the power cannot be worked out.

The MTS figures are from MTS's patent announcement
\citep{www12026energytool}: EnergyTool, running since 2021, uses ``a
statistical model'' on smart-meter data and ``since the beginning of
2024 \ldots{} the detection of several hundred illegal mining farms'',
with 15\% more farms recorded in 2025 than in 2024. ComNews \citep[17
October]{comnews2025mts} reports the agreement between Rosseti (the
state grid company) and MTS, announced on 16 October, and says the
system analyses smart-meter data in real time with a self-learning
anomaly model; neither source names the features it uses. Not every find
comes from grid data: in Buryatia a KamAZ truck holding 95 rigs and a
mobile transformer station was found on a routine inspection of a 10 kV
line \citep[citing TASS]{cointelegraph2025buryatia}, and the Cheremkhovo
case, 214 ASICs connected to a nearby transformer substation in early
2023, is reported without any detection method
\citep{forklog2024cheremkhovo}.

Two Russian limits are relevant, though neither is a detection rule.
Since 1 November 2024 individuals may mine without registering only
below 6,000 kWh a month, about 8.2 kW \citep{interfax2024miners}. Since
1 January 2025 the federal residential tariff's first band ends at 3,900
kWh a month, about 5.3 kW, with a second band up to 6,000 kWh
\citep{rg2024tariff}. Regions may set lower bands: St Petersburg's first
band has been 1,200 kWh a month, about 1.6 kW, since 1 July 2025
\citep{fontanka2025energolimit}. The 3,900 kWh band was introduced
partly to stop household tariffs being used for ``grey'' mining.

China. The NDRC notice \citep{ndrc2021mining} (\begin{CJK}{UTF8}{gbsn}发改运行〔2021〕1283号\end{CJK},
signed 3 September 2021, posted 24 September) orders analysis of
``grid-connected generation data and abnormal electricity use data''
(\begin{CJK}{UTF8}{gbsn}并网发电数据、异常用电数据分析\end{CJK}) with on-site checks of large
data-centre users. It also orders mines found by the crypto-asset
monitoring platform traced to ``IP address, specific enterprise and
physical address'', bans supply ahead of the meter or by direct line
from small hydro plants, and sets a surcharge of ¥0.30/kWh. The Sichuan
notice of 18 June 2021 told local governments, with State Grid, to shut
the 26 projects flagged by State Grid Sichuan by 20 June
\citep{coindesk2021sichuan}. The Jiangsu Communications Administration
found 4,502 mining IP addresses using about 260,000 kWh a day \citep[via
The Paper]{jiangsuca2021mining}; 260 MWh ÷ 24 h ÷ 4,502 ≈ 2.4 kW per IP.
About 21\% of those IPs belonged to public bodies, universities or firms
whose machines had been hijacked. These were found from network traffic
to mining pools, not grid data. Qihoo 360 reported about 109,000 mining
IPs active daily in November 2021 \citep{coindesk2021qihoo}. The ban did
not end mining: Cambridge's pool data put China at 22.3\% of Bitcoin
hashrate in September 2021 after showing almost none in August, which
Cambridge reads as ``significant underground mining activity''
\citep{cjbs2022bitcoin}.

Malaysia. On 27 November 2025 Deputy Prime Minister Fadillah Yusof told
the Dewan Rakyat (the lower house of parliament) that TNB, the national
utility, uncovered 14,489 premises from 2020 to 2025 and lost RM5.14
billion; TNB ``closely monitors suspicious electricity usage by tracking
demand spikes and checking meters for tampering'' and piloted
distribution transformer meters at substations \citep{thestar2025tnb}.
An earlier written reply (19 November 2025) gave 13,827 premises and
RM4.57 billion to August 2025 \citep{thevibes2025tnb}. In February 2026
he said 5,133 theft cases of all kinds were detected through smart
meters from 2021 to 2025, rising from 379 to 1,900 a year, and that
drones and handheld thermal devices are also used
\citep{thestar2026smartmeters}.

Iran and Kazakhstan. The CEO of Tavanir, Iran's state grid company, said
the 240,000 rigs seized over three years ``consumed an estimated 800
megawatts'', and a deputy estimated about 700,000 rigs still running at
about 2,000 MW \citep{iranintl2025rigs}; 800 MW ÷ 240,000 ≈ 3.3 kW. The
article names the Economic Security Police but not a detection method.
Kazakhstan's Financial Monitoring Agency reported 106 farms stopped (55
voluntarily, 51 by force), more than 67,000 devices seized and daily
consumption down 600 MWh \citep{cointelegraph2022kazakhstan}; 600 MWh ÷
24 h ÷ 106 ≈ 236 kW per farm. The release does not say how the farms
were found.

United States. A U.S. Attorney's press release, republished by the DEA
\citep[15 February]{dea2013elkgrove}, says SMUD (Sacramento's municipal
utility) investigators reported a residence using ``an unusually large
amount of electricity, commonly seen in indoor marijuana grows''; seven
houses and 2,912 plants followed. The Register \citep[21
July]{theregister2025smud} reports that SMUD sent address lists to
police on 90 occasions and that the threshold ``dropped from 7,000 kWh
per month to 2,800 kWh'' between 2014 and 2023; this figure comes from
court filings in a lawsuit against SMUD, and we found no utility
document stating it. 7,000 ÷ 730 ≈ 9.6 kW; 2,800 ÷ 730 ≈ 3.8 kW.

Chelan County PUD's miners were mostly paying customers running
high-density loads without the utility's approval: in March 2018 it put
a moratorium on applications for ``new high density loads'', and its
fees for ``unauthorized service'' included charges for ``equipment
degradation'' \citep[17 April]{kpq2018chelan}. NW News Network \citep[3
April]{nwnews2018chelan} says the PUD's steps ``resemble strategies used
in the past to detect illegal indoor marijuana grows'', including
``monitoring for extraordinary, sustained electricity usage spikes'',
and that it could report such loads as power theft ``because the
specific commercial use should have a higher rate''. By the time of the
moratorium the PUD ``had identified 28 unauthorized operations'', of
which ``19 were shut down'' \citep{mcdaniels2018bitcoin}. For the rate
after that, KPQ reports finds of ``two to three a day on average'' and
the Everett Herald ``about three more unauthorized operations each
week'', so the rate is uncertain.

Canada. BC Hydro's 2016 fact sheet \citep{bchydro2016smifactsheet} says
that ``more than 60 per cent of grow ops were stealing electricity'',
costing customers ``about \$100 million per year'', and that BC Hydro
responded with ``smart meters, distribution system inventory, analytics
and enhanced field inspection''. The distribution-system inventory is
the energy balance: sensors on the distribution grid compared against
the customer meters below them, with field crews then sent to the gap
\citep{cbc2016hydro, kellydetwiler2016bchydro}. It was built in 2015
(see the summary at the top of this section); before that, leads came
from ``highly specialized meter and transformer analytics''
\citep{emc2016bchydro}. The fact sheet credits smart meters with
``reducing theft by more than 80 per cent'', against a business-case
target of 75\% by fiscal 2016, and says theft detection paid for its
share of the programme within 12 months.

BC Hydro's 2016 letter \citep{reimer2016letter} gives the scale. A 2006
analysis ``using three independent methodologies, indicated that
grow-ops were stealing 500 gigawatt hours of electricity per year'', and
``a subsequent analysis indicated that theft had increased to 850
gigawatt hours'', which at \$124 per MWh ``equates to an annual cost of
\$105 million''; the letter puts the rise between 2006 and 2011. It
reports theft reduced by ``1,139 gigawatt hours of electricity from 2012
to 2015'', which is theft avoided against a baseline, not sites found.
For scale, BC Hydro's domestic sales in fiscal 2013/14 were 53,018 GWh,
with 5,900 GWh of line loss and system use
\citep[p.~15]{bchydroserviceplan201516}. Our arithmetic: 500 and 850 GWh
a year are average loads of about 57 and 97 MW, and 850 GWh is about
1.6\% of sales, or about a seventh of the line-loss figure it was hidden
in. An 80\% cut from 850 GWh leaves about 170 GWh a year, about 19 MW.
These are the utility's own figures; an opposition critic disputed them
at the time.

BC Hydro gives no count of sites, but University of the Fraser Valley
studies of police seizure files allow one. Grow-ops that stole power in
2006--2010 ran ``approximately 36 lights'', and BC Hydro's own data put
it at 36.5. A typical light was ``a 1000 W bulb'' on 18 hours a day for
a month and then 12 hours a day for two, ``approximately 5,040 kWh'' a
year per light, or 181,440 kWh a year per stealing site
\citep{diplock2011electrical}.\footnote{Regional figures are 37.1 lights
  in Mission and 48.9 in the Cariboo
  \citep{plecas2011mission, plecas2011cariboo}.} Ballasts (the lamp's
power-control gear) add about 10--13\%: 10\% on a magnetic ballast's
datasheet, 1,100 W of input per 1,000 W lamp
\citep{advance2019ballast71a8773}, and 13\% for the mix of magnetic and
digital ballasts in Mills's model \citep{mills2012carbon}. Our
arithmetic: at 20--32 kW of lighting per site, 500 GWh a year is roughly
1,800--2,800 stealing sites, 850 GWh is 3,000--4,800, and the 1,139 GWh
avoided in 2012--2015 is roughly 1,000--1,600 sites a year taken off
theft. Ventilation, dehumidification and cooling would cut these counts
by up to a factor of about 2.6 on Mills's US process model
\citep{mills2012carbon}, which puts whole sites at about 58--83 kW. That
factor rests on a US model, not on BC data, and it treats Mills's shares
of carbon emissions as shares of electricity. By comparison, BC Hydro
recorded only 430--594 theft cases a year in 2006--2010
\citep{bellett2010hydro}, and police found 2,113 indoor grow-ops in 2010
\citep[RCMP data,][]{diplock2011electrical}: before smart meters BC
Hydro was catching perhaps a tenth of the stealing sites each year. Each
site's lighting alone drew about as much as a small training island
(whole sites drew two to three times as much), and BC law already
flagged any home averaging ``93 kilowatt-hours per day or more''
\citep[s. 3.1]{bcreg1002004}, about 3.9 kW, or seven 1,000 W lamps on a
14-hour day.

The balance was resolved per feeder, not per transformer. BC Hydro first
planned 30,000 dedicated transformer and feeder meters. In October 2014
it decided instead that ``9,000 measurement points (compared to 30,000
originally planned) would be an appropriate number to perform
feeder-based energy balancing for all feeders'', built mostly from SCADA
devices and check meters already on the grid, because ``the actual level
of theft has been declining since 2012''
\citep[p.~2]{bchydro2015smiqr20}. By December 2015 it had 9,116. Of
these, 4,414 were SCADA reclosers, relays, switches and similar devices;
3,045 were transformer rebate meters; 1,000 were check meters ``for
temporary rotational purposes''; 548 were fixed check meters; and 109
were first-generation feeder and transformer meters \citep[Table
1]{bchydro2016smiqr23}. That covered 1,482 primary feeders and about
295,000 distribution transformers \citep{reimer2010costofservice}:
roughly six points per feeder, with about 1,300 customer meters below
each feeder on average (our arithmetic)
\citep[p.~7]{bchydro2016smiqr23}. The provincial regulation sets the
floor at ``at least 9 000 but no more than 35 000 system devices''
recording ``as frequently as smart meters'', that is, at least hourly.
It requires the system to ``estimate the amount of electricity supplied
from a portion of the electric distribution system to unmetered loads
that are not known to the authority and to estimate the location of
those loads'' \citep[ss. 1, 2(d), 4(1)]{bcreg3682010}. Earlier pilots
balanced ``at either the primary or secondary voltage level'' (the
medium-voltage feeder or the low-voltage side of a transformer) across
800 homes and found 22 thefts \citep[p.~23]{bchydro2011businesscase}. No
public filing we found states a minimum detectable gap.

UK and Netherlands. Ofgem, the GB energy regulator \citep[3
July]{ofgem2013pressrelease}, said ``up to one third of the volume of
electricity stolen each year is used to power cannabis farms'' and that
industry detects up to 25,000 theft cases a year; an Ofgem economist put
a cannabis farm at about 12,000 kWh a month against 300 kWh for a home
\citep{nbc2013marijuana}, or about 16 kW. Ofgem's 2023 open letter
\citep{ofgem2023openletter} gives 17,423 thefts confirmed by suppliers
in 2021/22 and 16,581 in 2022/23, against a target of 41,000. Liander, a
Dutch distribution company, working with police and municipalities,
closed 633 cannabis farms in its area in 2021 that had taken 31.9
million kWh \citep{nltimes2022liander}, about 50 MWh each; that would be
about 5.8 kW if each ran a full year, which the source does not say.

Almost all of these cases were theft, or a residential account doing
commercial work, which gave the utility its own reason to look. An
island that pays its bill on a commercial meter would be found only if a
regulator asks for the screen.

\begin{longtable}[]{@{}
  >{\raggedright\arraybackslash}p{(\linewidth - 8\tabcolsep) * \real{0.2000}}
  >{\raggedright\arraybackslash}p{(\linewidth - 8\tabcolsep) * \real{0.2000}}
  >{\raggedright\arraybackslash}p{(\linewidth - 8\tabcolsep) * \real{0.2000}}
  >{\raggedright\arraybackslash}p{(\linewidth - 8\tabcolsep) * \real{0.2000}}
  >{\raggedright\arraybackslash}p{(\linewidth - 8\tabcolsep) * \real{0.2000}}@{}}
\caption{Cases where hidden loads were found, and what found
them.}\tabularnewline
\toprule\noalign{}
\begin{minipage}[b]{\linewidth}\raggedright
Country
\end{minipage} & \begin{minipage}[b]{\linewidth}\raggedright
What was hidden
\end{minipage} & \begin{minipage}[b]{\linewidth}\raggedright
Power per site
\end{minipage} & \begin{minipage}[b]{\linewidth}\raggedright
How found
\end{minipage} & \begin{minipage}[b]{\linewidth}\raggedright
Source
\end{minipage} \\
\midrule\noalign{}
\endfirsthead
\toprule\noalign{}
\begin{minipage}[b]{\linewidth}\raggedright
Country
\end{minipage} & \begin{minipage}[b]{\linewidth}\raggedright
What was hidden
\end{minipage} & \begin{minipage}[b]{\linewidth}\raggedright
Power per site
\end{minipage} & \begin{minipage}[b]{\linewidth}\raggedright
How found
\end{minipage} & \begin{minipage}[b]{\linewidth}\raggedright
Source
\end{minipage} \\
\midrule\noalign{}
\endhead
\bottomrule\noalign{}
\endlastfoot
Russia (Irkutsk) & Home mining on household tariff & not stated & Growth
in consumption and change in load profile (deputy minister's general
account); \textgreater1,000 cases in 2021, \textgreater8,000 suspected
points by 2025 &
\citep{habr2022miners, kommersant2021irkutsk, ria2025irkutsk} \\
Russia (Dagestan) & \textasciitilde900 rigs in one district & 3.2 MW
total, \textasciitilde3.6 kW per rig & Internet cut by Dagenergo as a
test; load drop measured & \citep{comnews2026dagestan} \\
Russia (Makhachkala) & 620-device farm under an apartment block & not
stated (1.75 GWh stolen in total) & Residents' complaints of voltage
swings & \citep{forklog2024makhachkala} \\
Russia (national) & Mining farms behind smart meters & not stated &
Smart-meter anomaly model (MTS EnergyTool); ``several hundred'' since
2024 & \citep{www12026energytool, comnews2025mts} \\
China & Mining farms and home rigs & Jiangsu: 4,502 IPs,
\textasciitilde2.4 kW per IP & State Grid flagged, method not stated
(Sichuan); pool traffic (Jiangsu) &
\citep{ndrc2021mining, jiangsuca2021mining} \\
Malaysia & Crypto mining on stolen power & not stated & Demand spikes,
tamper checks, transformer meters, drones &
\citep{thestar2025tnb, thestar2026smartmeters} \\
Iran & 240,000 rigs seized, 3 years & \textasciitilde3.3 kW per rig &
Not stated (with economic police) & \citep{iranintl2025rigs} \\
Kazakhstan & 106 farms, 67,000 devices & \textasciitilde240 kW per farm
(average) & Not stated (reported by the Financial Monitoring Agency) &
\citep{cointelegraph2022kazakhstan} \\
USA (Chelan, WA) & Home crypto mines & not stated & Not stated
(usage-spike monitoring among the PUD's methods); 28 found, 19 shut down
by March 2018 & \citep{kpq2018chelan, mcdaniels2018bitcoin} \\
USA (Sacramento) & Cannabis grow houses & flag at 2,800 kWh/month
(\textasciitilde3.8 kW) & Utility usage lists &
\citep{dea2013elkgrove, theregister2025smud} \\
Canada (BC) & Cannabis grow-ops & \textasciitilde20--32 kW of lighting
(our estimate) & Smart meters; meter and transformer analytics; energy
balance on every feeder from 2015 (9,116 points) &
\citep{bchydro2016smifactsheet, bchydro2015smiqr20, bchydro2016smiqr23, emc2016bchydro} \\
UK & Cannabis farms & \textasciitilde16 kW (12,000 kWh/month) & Supplier
theft detection & \citep{ofgem2013pressrelease, ofgem2023openletter} \\
Netherlands & Cannabis farms (633 in 2021) & \textasciitilde5.8 kW if
run all year (\textasciitilde50 MWh each) & Liander with police &
\citep{nltimes2022liander} \\
\end{longtable}

\subsubsection*{D.3 Non-technical losses: how small, and how they are
found}\label{d.3-non-technical-losses-how-small-and-how-they-are-found}
\addcontentsline{toc}{subsubsection}{D.3 Non-technical losses: how
small, and how they are found}

Non-technical loss (NTL) is energy delivered but not billed: theft,
meter tampering, unmetered connections, and billing errors. Technical
loss, by contrast, is energy lost as heat in lines and transformers. An
island that bypasses its meter is NTL, so this section asks how small an
NTL utilities can see, and how they track one to a site.

Utilities find NTL in two ways. The first is to screen each customer's
own meter for volume, load shape or tamper alarms; this is how SMUD
found its cases, and TNB uses such screens alongside transformer meters
and drones. The second is an energy balance: subtract the downstream
customer meters from a meter at a supply point (a distribution
transformer, a feeder head or a substation), subtract an estimate of
technical losses, and treat what is left as NTL. An energy balance only
finds energy that no meter bills. A load that pays on its own meter,
however strange its shape, adds to both sides and cancels.

The sources in D.2 rarely say which method found a case. The UK is the
one place where lead sources are counted. In 2013 five suppliers' theft
leads came 26\% from data collectors (meter readers), 21\% from meter
operators, 14\% from revenue-protection services, 7\% from distribution
network operators, 8\% from data analysis, and 24\% from tip-offs,
police and others \citep[Table 1]{ofgem2013consultation}. In 2021/22
suppliers confirmed 17,423 thefts, and in 2022/23 16,581, each against
an incentive-scheme target of 41,000 \citep{ofgem2023openletter}.

\textbf{What limits a balance.} A balance that judges the \emph{level}
of loss (is this transformer losing more than it should?) cannot see NTL
smaller than its uncertainty in technical losses plus meter error. A
balance that looks for a \emph{step} (did the loss jump when something
switched on?) escapes fixed errors of that kind, because they are the
same before and after the step, but not clock misalignment between
meters (Appendix A.5). The figures below are all about the level.

\begin{itemize}
\tightlist
\item
  \emph{Size of technical losses.} In Great Britain, cumulative losses
  from the grid supply point are about 0.5\% at 132 kV, 1.5\% at 33 kV,
  3\% at 11 kV and 7\% at 400/230 V. Total distribution losses are ``of
  the order of 5\%''. The uncertainty in the annual figure is ``of the
  order of 0.15\% to 0.3\% of the units distributed -- that is about 3\%
  to 6\% of the losses themselves'' \citep{sohn2009losses}.
\item
  \emph{Interval averaging.} On 100 real UK smart meters at 1-minute
  resolution, placed on three modelled networks, low-voltage losses were
  about 4\% of energy supplied. Computing them from 30-minute data
  ``underestimated the losses by between 9\% and 24\%''
  \citep{poursharif2018smart}, because losses rise with the square of
  current and averaging smooths out the peaks. At 4\% losses that bias
  is 0.4--1\% of throughput (our arithmetic).
\item
  \emph{Brazil.} ANEEL, the national regulator, computes each
  distributor's technical losses by load flow (a network calculation) on
  its registered network data \citep{aneel2021prodistmod7}, and the NTL
  it reports is the residual. The same rules compute NTL feeder by
  feeder, as energy measured at the feeder head minus the customer
  meters and modelled downstream losses: a regulated energy balance on
  every medium-voltage feeder, used for tariff accounting rather than as
  a search tool. For 2023 it reports total losses of 14.1\% of injected
  energy: 7.4\% technical and 6.7\% non-technical
  \citep{aneel2024perdas}.
\end{itemize}

\textbf{Meter accuracy and clocks.} A meter's accuracy class is the
error it is allowed, in percent of reading (class 0.5S allows 0.5\%; the
S marks a version specified down to low currents).

\begin{itemize}
\tightlist
\item
  \emph{GB settlement meters \citep{sohn2009losses}.} Grid-supply-point
  meters are better than ±0.5\%. Half-hourly meters are ``generally
  accurate to within 1\%''. Non-half-hourly meters are within +2.5\% and
  −3.5\%.
\item
  \emph{India.} The metering regulations of the Central Electricity
  Authority (CEA) require energy-accounting meters in distribution to be
  no worse than class 0.5S where separate current transformers are used
  \citep{cea2020metering}.
\item
  \emph{China.} A 2023 Beijing draft specification for on-site
  verification of smart meters limits class 0.5S to meters connected
  through instrument transformers, and excludes gateway meters from its
  scope \citep{bjsamr2023smartmeterdraft}. That suggests a Chinese
  district master meter (the meter on a distribution transformer, see
  below) can be 0.5S while direct-connected customer meters are class 1
  or 2. This is our inference; no \begin{CJK}{UTF8}{gbsn}台区\end{CJK} rule we found states it.
\item
  \emph{Simulated effect of meter class.} On a simulated 55-customer
  feeder, a voltage-based theft detector's accuracy fell from above 0.9
  with class 0.1S meters to about 0.6 with class 2
  \citep{pengwah2023modelless}.
\item
  \emph{Clocks.} On a real 525-home microgrid in Haiti with simulated
  jitter, readings aligned within 30 s gave a median state error of
  0.29\%, and readings spread over 15 minutes gave 1.18\%
  \citep{buevich2016microgrid}. Typical smart-meter clock offsets are
  ``in the order of few seconds'' and larger with infrequent
  synchronisation \citep{antonios2019approximation}.
\end{itemize}

\textbf{Smallest NTL detected, by level of the grid.}

\begin{itemize}
\tightlist
\item
  \emph{Distribution transformer and low-voltage feeder.}

  \begin{itemize}
  \tightlist
  \item
    \citet{sau2023identification}, on ``a real feeder'': ``the minimum
    theft power to be identified is around 400 W. However, the location
    has good performance for power of 1 kW and above.'' This is from the
    abstract only. The method is a hypothesis test on state-estimation
    errors (the mismatch between measured values and a fitted model of
    the network), not a plain energy balance.
  \item
    \citet{buevich2016microgrid} took real data from a 525-home
    microgrid, removed meters to create artificial NTL, and found they
    could ``reliably detect NTL on the order of 1.6\% (40W out of 2.5kW
    load)\ldots{} with a false positive rate lower than 2\%''. The
    system load is tiny and the meters are synchronised, so this is a
    best case.
  \item
    On a simulated 30-customer transformer, theft ``can be detected
    100\% of times when the power theft percentage is greater than 4\%''
    of the group's use \citep[Mitsubishi Electric Research
    Laboratories]{sahoo2015electricity}.
  \end{itemize}
\item
  \emph{Transformer district (China).} A \begin{CJK}{UTF8}{gbsn}台区\end{CJK} (transformer district) is
  the area served by one distribution transformer, and its line-loss
  rate is the transformer's metered supply minus the customers' metered
  use, as a share of supply. State Grid counts a district's monthly
  line-loss rate of 0--7\% as qualified (\begin{CJK}{UTF8}{gbsn}合格\end{CJK}). Above 7\% is high loss
  (\begin{CJK}{UTF8}{gbsn}高损\end{CJK}) and below 0 is negative loss (\begin{CJK}{UTF8}{gbsn}负损\end{CJK}), and both are singled out
  for management and scoring. This is reported in a State Grid Zhejiang
  patent's background section \citep[background
  section]{stategridzj2020patent}, not in the rule itself. A vendor
  article gives the synchronous line-loss system's bands as 0--10\%
  daily and 0--7\% monthly \citep{fx3612022lineloss}. State Grid Jiangsu
  tightened its qualified band from 0--10\% in 2013 to 0--7\% in 2014
  and 0--5\% in 2015, and from 2016 set it from each district's
  theoretical loss \citep{ncdqh2017jiangsu}.

  \begin{itemize}
  \tightlist
  \item
    For an island, our arithmetic: a stolen 19,000 kWh a month adds
    19,000 ÷ S to a district's loss rate, where S is its monthly supply.
    With a normal loss of 3\%, the district crosses 7\% when 19,000 ÷ S
    \textgreater{} 4\%, i.e.~S below about 475 MWh a month (about 650 kW
    average). Under a dynamic band set near theoretical loss, the margin
    is smaller and larger districts would also cross. We found no
    published distribution of district sizes to check this against.
  \end{itemize}
\item
  \emph{Medium-voltage feeder and above.}

  \begin{itemize}
  \tightlist
  \item
    We found no measured detection floor for NTL on a utility feeder.
    The nearest measurement is our own energy balance on a 2.4 kV campus
    network (Appendix A.5), which measured the residual scatter a step
    must clear but found no NTL.
  \item
    \citet{tang2019inference}, in a simulated six-load network, act only
    ``once the discrepancy is larger than 20\% of the normal power
    consumption''. This is a trigger the authors chose so that switching
    is not wasted, not a floor they measured.
  \item
    The only feeder-level case in D.2 with a size attached is Dagestan's
    3.2 MW, and that was a deliberate test (an internet cut), not a
    find. BC Hydro's pilots balanced at the primary or secondary level
    and found 22 thefts, but gave no sizes.
  \item
    At system level, the uncertainty of GB annual loss accounting
    (0.15--0.3\% of units distributed) is far larger than any one site.
  \end{itemize}
\end{itemize}

\textbf{Thresholds for opening an investigation.} We found no utility or
regulator rule of the form ``investigate a transformer whose loss
exceeds X\%'' outside China's \begin{CJK}{UTF8}{gbsn}台区\end{CJK} bands.

\begin{itemize}
\tightlist
\item
  \emph{India.} The Revamped Distribution Sector Scheme (RDSS) targets
  pan-India AT\&C losses (aggregate technical and commercial losses,
  which include unpaid bills) of 12--15\% by 2024-25. It sends smart
  metering first to electricity divisions with AT\&C losses above 15\%
  (urban) or 25\% (others). These are division-level priorities, not
  site triggers \citep{mop2023rdss}.
\item
  \emph{Brazil.} ANEEL sets each distributor's regulatory NTL by
  comparison with distributors in comparable or more socioeconomically
  complex areas \citep{aneel2015proret26, aneel2025ren1114}.\footnote{The
    second reference approves the version now in force.} NTL above the
  regulatory level cannot be passed through to tariffs
  \citep{aneel2024perdas}. In the 2015 rules, distributors below 7.5\%
  (large) or 2.5\% (small) of the measured low-voltage market were set
  no reduction path; these are floors, not targets \citep[footnote
  16]{aneel2015proret26, aneel2024perdas}. We read the 2015 text; the
  2025 version in force may differ.
\item
  \emph{Customer meters.} SMUD's police referral threshold fell from
  7,000 to 2,800 kWh a month (about 3.8 kW); the figure comes from court
  filings \citep{theregister2025smud}. A Kentucky co-op examines
  accounts using under 50\% or over 150\% of the same period a year
  earlier \citep{jpec2014response}.
\end{itemize}

\textbf{How a loss is tracked down.} The chain runs from a system or
feeder loss to the transformer meter, then to analytics on customer
meters, then to a field visit.

\begin{itemize}
\tightlist
\item
  \emph{India.} Under RDSS a ``feeder and DT level automated online
  energy accounting system shall be put in place'' (DT: distribution
  transformer). Meters at feeders and distribution transformers are
  meant ``to enable proper energy accounting every month for
  identification of theft prone pockets and high loss areas''
  \citep{mop2023rdss}.
\item
  \emph{China.} District loss is supply minus use, computed online from
  the meter-data collection system. The Hubei pilot's workflow runs
  «\begin{CJK}{UTF8}{gbsn}精选数据、排查异常\end{CJK}\ldots \begin{CJK}{UTF8}{gbsn}勘验现场\end{CJK}\ldots \begin{CJK}{UTF8}{gbsn}获取证据、追缴损失\end{CJK}» (select
  data, find anomalies, inspect on site, gather evidence, recover
  losses) \citep{ncdqh2018lineloss}.
\item
  \emph{BC Hydro.} Each feeder's measurement points are compared with
  the customer smart meters below them, and analytics rank the
  discrepancies; this feeder-level balance came on line from 2015, after
  meter and transformer analytics
  \citep{bchydro2016smiqr23, emc2016bchydro}. Then ``field
  investigators\ldots{} go out into the field, and use the solution to
  confirm theft is occurring'', before police are brought in
  \citep{kellydetwiler2016bchydro}. The balance flags a section, and
  confirmation is local: the regulation separately requires
  ``investigation devices \ldots{} to identify the location of the
  unmetered loads'' \citep[s. 4(1)(b)]{bcreg3682010}. Before smart
  meters, BC Hydro found ``over 2,600 electricity thefts over the past
  five years'' by ``a time-consuming, inefficient and expensive manual
  process'' \citep[p.~6]{bchydro2011businesscase}.
\item
  \emph{TNB (Malaysia).} TNB combines demand-spike and tamper alarms
  from 5.56 million smart meters with a transformer-meter pilot, drones
  and handheld thermal devices
  \citep{thestar2025tnb, thestar2026smartmeters}.
\item
  \emph{Stedin (Netherlands).} As of 2018 its finds were mostly reactive
  and tip-driven. It was piloting devices ``in voltage rooms which
  collect electricity usage data for research'' in Rotterdam, and said
  ``We are still in the test phase'' \citep{mollen2018bitcoin}. Its 2016
  station-measurement plan was never reported on afterwards.
\item
  \emph{Britain.} A respondent to Ofgem's 2025 consultation said ``only
  37\% of ETTOs leads are investigated, with a conversion rate of 22\%
  to Confirmed Energy Thefts'' \citep{ofgem2025r0173}. This is one
  respondent's figure, not Ofgem's. ETTOS is the Energy Theft Tip-Off
  Service.
\item
  \emph{China, a recent case.} A State Grid unit found
  ``\begin{CJK}{UTF8}{gbsn}某台区线损数据异常偏高\end{CJK}'' (abnormally high line loss in one
  transformer district) from the meter-data and synchronous line-loss
  platforms. It then compared customers' historical load curves and
  phases, and sent a field team with ``\begin{CJK}{UTF8}{gbsn}红外测温、电气检测\end{CJK}'' (infrared
  thermometry, electrical testing) to check ``\begin{CJK}{UTF8}{gbsn}对照线路布局逐一排查\end{CJK}''
  (one by one against the line layout) \citep{chen2026changtai}.
\item
  \emph{CenterPoint (Texas).} It ``investigates approximately 3,000
  potential theft cases each month'' with ``an approximately 50\% to
  60\% truck roll success rate'' (the share of field visits that confirm
  a case). A few transformer monitors pinpoint ``theft at a premise
  attached to the monitored transformer''. Alarm windows are 24 h (no
  communication), 48 h (tamper) and 30 days \citep{neel2015revenue}.
\item
  \emph{Light (Rio de Janeiro).} Light runs an energy balance, an
  inspection programme and a monthly risk matrix ``finalizada até o dia
  15 do mês posterior'' (finished by the 15th of the following month).
  Its triggers include ``Queda de consumo trimestral 20\%'' (a 20\%
  quarterly consumption drop) \citep{riberaneto2019light}.
\item
  \emph{Tata Power-DDL (Delhi).} Automated meter reading on distribution
  transformers and a regular ``energy audit of all DT'' identify ``areas
  in terms of power mismatch'', which go to enforcement teams
  \citep{mpost2014tpddl}.
\item
  \emph{Hit rates per inspection.}

  \begin{itemize}
  \tightlist
  \item
    Endesa (Spain): theft ``en una de cada dos inspecciones'' (in one of
    two inspections) with AI targeting \citep{endesa2021inspecciones}.
    In 2024, about 428,000 inspections found 71,000 cases, about 17\%
    (our arithmetic). Citizen tip-offs gave 17,500 inspections and over
    3,500 cases, about 20\% \citep{endesa2025fraud}.
  \item
    HEDNO (Greece's distribution operator): 1,155 checks of large
    consumers in tourist areas found at least 200 thefts
    \citep{skai2025deddie}.
  \item
    BC Hydro's analytics vendor: ``more than 80 percent of the leads
    actioned have identified theft, unmetered consumption, or errors in
    company data'' \citep{emc2016bchydro}, a vendor claim. Data errors
    count as hits, so this is not a theft hit rate.
  \end{itemize}
\item
  \emph{Speed.} No source gives the time from a high-loss transformer to
  a confirmed site. The nearest are Light's monthly risk-matrix cycle
  and CenterPoint's alarm windows.
\end{itemize}

\subsection*{Appendix E. Data sources}\label{app-data-sources}
\addcontentsline{toc}{subsection}{Appendix E. Data sources}

The counts and screens in this paper rest on the following public
sources; none was collected for this work. Electrical power is given
either as real power in watts (W) or as apparent power in volt-amperes
(VA); the two differ by the power factor. For counting buildings and
hosts by their average load the difference is well within our modelling
uncertainty, and we treat the two units as interchangeable.\footnote{The
  difference matters when detecting the island's swing on a meter that
  records VA: the island's current does not run exactly in step with the
  rest of the load on the line, so only part of its swing shows.
  Appendix B.3 allows for this by counting 90\% of the swing, a power
  factor of 0.90. Losses in converting AC to DC at the rack are a
  different matter: they are real power, not a W/VA difference, and they
  are already inside the island's load at the wall (about 1 kW per GPU).}

\begin{itemize}
\tightlist
\item
  \textbf{EIA-861 (2025 release)} \citep{eia2025form861}. The Energy
  Information Administration's annual census of US utilities: customers,
  meters and advanced (smart) meters by sector and utility. Gives the
  20.7 million commercial meters, about 85\% of them smart meters (as
  compiled in the Electric Power Annual's Table 10.05
  \citep{eia2025epatable1005}), and the ratio of about 1.06 meters per
  commercial account. Its sector definitions place crypto mining and
  data centers in the commercial class.
\item
  \textbf{EIA CBECS 2018} \citep{eia2018cbecs}. The Commercial Buildings
  Energy Consumption Survey public-use microdata: 6,436 sampled
  buildings, weighted to represent 5.9 million. Annual electricity ÷
  8,760 h gives each building's average load. This is the source of the
  count of buildings averaging more than a bare island's load (the
  population whose meters a bare island must blend into), and of the
  commercial part of the candidate-host pool (the sites large enough to
  cloak an island). The survey's jackknife replicate weights give the
  sampling error. The 2018 survey has no data-center building type and
  predates most of the data-center build-out, so its upper tail (8,000
  buildings above 1 MW, the largest at 13 MW) understates today's stock
  of very large sites; those are declared compute and outside this
  paper's pool in any case.
\item
  \textbf{Census County Business Patterns, Economic Census and EIA MECS
  (2022)} \citep{census2022cbp, census2022economic, eia2022mecs}.
  Manufacturing establishments by industry and employee-size class
  (County Business Patterns), purchased electricity per employee by
  industry (Economic Census), and purchased electricity by
  employment-size class (the Manufacturing Energy Consumption Survey,
  Table 11.2). A size-split model combines them: it multiplies each
  class's employees by its industry's electricity per employee, scales
  each size class to the electricity MECS reports for it, and spreads
  sites around that mean, to count manufacturing sites above each load
  threshold. Within the candidate-host pool this model is the only
  source for sites above the largest CBECS building (13 MW), where the
  count is modelled rather than measured. The separate estimate of sites
  averaging 150 MW or more (Figure 4) is built instead from named
  facilities, MECS 2022 sector totals (Table 11.1), EIA-860 and EIA-923
  \citep{eia923} records of on-site generation, and USGS data; it too is
  an estimate, not a measured count.
\item
  \textbf{Industrial Training and Assessment Centers (ITAC) database}
  (Smegal and Muller, 2014; Open Energy Data Initiative; CC BY 4.0).
  Energy assessments of US manufacturing plants, as of 2026-10-26. Gives
  the load factors of manufacturing sites, which stand in for
  manufacturing hosts in the step screen and the pool it leaves for
  random visits (B.4), and the spread of electricity per employee used
  in the size-split model above.
\item
  \textbf{Building Data Genome Project 2} \citep{miller2020bdg2}.
  \protect\phantomsection\label{src-bdgp2}{} Hourly electricity from
  1,578 meters at 19 sites in 2016--17 (the electricity meters among the
  set's 1,636 buildings), of which 796 average 50 kW or more; mostly
  university and public buildings. Gives the measured load factors
  (average load divided by peak); the year-over-year changes in energy
  and load factor behind the step screen's false-positive and
  false-negative rates (B.4); and the flat share, the fraction of
  buildings averaging 50 kW or more whose median monthly load factor
  exceeds 0.9 (6 of 796, 0.8\%), which estimates how many ordinary
  buildings would look like a bare island to a load-factor screen. A
  convenience sample, not a random draw of US buildings.
\item
  \textbf{NREL ComStock (2025 release 3)} \citep{nrel2025comstock}.
  \protect\phantomsection\label{src-comstock}{} Simulated hourly load
  for the US commercial stock. Weighted to the national stock (4.67
  million buildings from 149,272 models; 491,676 of them averaging 50 kW
  or more) but built on DOE prototype schedules, so it is too peaky
  (median annual load factor 0.33 against 0.46 measured, both among
  buildings averaging 50 kW or more) and produces essentially no flat
  buildings. Used only as the comparison for load factor, not as the
  source of the flat share.
\item
  \textbf{LBNL California customer profiles} \citep{smith2024phase4}.
  Load shapes for clusters of California commercial and industrial
  customers: each shape is the summed load of the cluster's members,
  with the number of customers as its weight. Used twice. First,
  together with utility counts of large customers quoted from utility
  filings (ComEd's count of accounts peaking at 100 kW or more), to
  convert sites into utility accounts: the main text's 1.17 accounts per
  site. Second, as a check on the BDG2 and ITAC load factors: with
  LBNL's in their place, the pool of hosts left for random visits after
  the 20\% energy-step screen comes out 7--15\% smaller. Because each
  shape is summed over many customers, its load factor is higher than a
  typical member's own.
\item
  \textbf{Portland General Electric feeder data} \citep{pge2026feeders}.
  Peak load for 559 distribution feeders (median 6.5 MVA), and their
  daytime minimum load; the source of the island's share of a feeder
  phase (Figure 1).
\item
  \textbf{Distributed-training simulator} \citep{rahman2026distributed}.
  Software rather than a dataset, used with its author's permission. A
  compute and communication model of training split across islands. We
  run it with two changes (documented in the reproduction package): a
  fix to its search over configurations, and its penalty for averaging
  many drifting model copies applied to the reducible part of the loss
  only, rather than to the whole loss. It gives the number of sites per
  run by island size, link speed and evader scenario (Table 1).
\item
  \textbf{Logged GPU jobs and campus recordings} (Appendix B): NLR
  \citep{vercellino2026measurement} training-job power traces and
  \citet{elsayed2026characterization} single-node logs; Caltech campus
  meters at 1 s \citep{xie2025socal}; LBNL micro-PMU at 120 samples per
  second \citep{stewart2016openupmu}.
\end{itemize}

\end{document}